\documentclass[superscriptaddress,showpacs,preprintnumbers,amsmath,amssymb,aip,citeautoscript,twocolumn,reprint,nofootinbib,floatfix]{revtex4-2}
\usepackage{color}
\usepackage{graphicx}
\usepackage{caption}
\usepackage{subcaption}
\usepackage{tabularx} 
\usepackage{amsmath}  
\usepackage{scrextend}
\usepackage{hyperref}
\usepackage{dcolumn}
\usepackage{bm}
\usepackage{IEEEtrantools}
\usepackage[misc,clock,geometry]{ifsym}
\usepackage{pifont}
\usepackage{adjustbox}
\usepackage[dvipsnames]{xcolor}

\newcommand{\overbar}[1]{\mkern 1.5mu\overline{\mkern-1.5mu#1\mkern-1.5mu}\mkern 1.5mu}

\begin{document}
	

\title{A study of DC electrical breakdown in liquid helium through analysis of the empirical breakdown field distributions}
	
\author{N.~S.~Phan}
\email{nphan@lanl.gov}
\affiliation{Los Alamos National Laboratory, Los Alamos, New Mexico 87545, USA}

\author{W.~Wei}
\affiliation{Los Alamos National Laboratory, Los Alamos, New Mexico 87545, USA}

\author{B.~Beaumont}
\affiliation{Department of Physics, North Carolina State University, Raleigh, NC 27695}
	
\author{N.~Bouman} 
\affiliation{Department of Physics and Astronomy, Valparaiso University, Valparaiso, IN 46383}
	
\author{S.~M.~Clayton}
\affiliation{Los Alamos National Laboratory, Los Alamos, New Mexico 87545, USA}
	
\author{S.~A.~Currie}
\affiliation{Los Alamos National Laboratory, Los Alamos, New Mexico 87545, USA}
	
\author{T.~M.~Ito}
\email{ito@lanl.gov}
\affiliation{Los Alamos National Laboratory, Los Alamos, New Mexico 87545, USA}
	
\author{J.~C.~Ramsey}
\affiliation{Los Alamos National Laboratory, Los Alamos, New Mexico 87545, USA}
	
\author{G.~M.~Seidel}
\email{george\_seidel@brown.edu}
\affiliation{Department of Physics, Brown University, Providence, Rhode Island 02912}
	
\date{\today}


\begin{abstract}
We report results from a study on electrical breakdown in liquid
helium using near-uniform-field stainless steel electrodes with a stressed area of
$\sim$0.725~cm$^2$. The distribution of the breakdown field is obtained
for temperatures between 1.7~K and 4.0~K, pressures between the
saturated vapor pressure and 626~Torr, and with electrodes of different surface polishes.  A data-based approach for determining the electrode-surface-area scaling of the breakdown field is presented. The dependence of the breakdown probability on the field strength as extracted from the breakdown field distribution data is used to show that breakdown is a surface phenomenon closely correlated with Fowler-Nordheim field emission from asperities on the cathode.  We show that the results from this analysis provides an explanation for the supposed electrode gap-size effect and also allows for a determination of the breakdown-field distribution for arbitrary shaped electrodes. Most importantly, the analysis method presented in this work can be extended to other noble liquids to explore the dependencies for electrical breakdown in those media.

\end{abstract}

	
	
\maketitle

\section{Introduction} \label{sec:intro}

There has been a long-standing interest in electrical breakdown in
liquid helium (LHe), motivated by practical applications as well as
fundamental interest. LHe may be used as electric insulator and
thermal conductor at cryogenic temperatures, which may simplify the
design of superconducting magnets\cite{GER98}. Studying the behavior of
LHe under a strong electric field may be of interest in
understanding electronic properties of condensed helium\cite{BEL15}.
More recently, the study of electrical breakdown of LHe is motivated
by the importance of superfluid LHe as a medium in which experiments
in nuclear, particle, and astroparticle physics are performed, along
with other noble liquids, in particular LAr and LXe.\cite{REB14} For some experiments, the
application of a high electric field is necessary. These include
experiments to search for the electric dipole
moment of the neutron\cite{GOL94,GRI09,ITO12,ITO16,AHM19} and experiments to look for
light dark matter particles.\cite{GUO13,ITO13,KNA17,HER18} For these
experiments, breakdown of LHe in response to DC electric fields (as
opposed to AC or pulsed electric fields) is particularly relevant. For
a new method to generate a large electric potential ($\sim$1~MV) inside LHe, see
Ref.~\onlinecite{CLA18}.
	
The study of electrical breakdown in LHe dates back to the late
1950s.~\cite{BLA59,BLA60} Since then, active investigations have been
conducted by many authors. These studies employed various electrode
geometries, electrode materials, and temperature and pressure of
LHe. The electrode geometries employed include sphere to sphere,
sphere to plane, and uniform field electrodes. The materials used
include stainless steel, tungsten, brass, copper, bronze, lead, niobium, and others. Data exist for
the temperature range of $1.2-4.2$~K mostly at the saturated vapor
pressure, with the bulk of the data taken at 4.2~K\footnote{Ref.~\onlinecite{ITO16} reports measurements at 0.4 K}. A review of
earlier work can be found in e.g. Ref.~\onlinecite{GER98}. While the
existing data show little consistency in general, several parameters
have been identified as potentially affecting breakdown, including 
(1)~electrode material and surface finish, (2)~gap size and electrode size, 
and (3)~temperature and pressure of the liquid. 

Based on a detailed theoretical study of electron multiplication induced by an electric field in LHe starting from the kinetic Boltzmann equation~\cite{BEL93}, the intrinsic breakdown field of LHe is expected to be at least in the low MV/cm region. On the other hand, breakdown fields around or below a few 100's kV/cm have been commonly reported in the literature for LHe. These observations are therefore presumed to be the result of the following process: (1)~Field emission of electrons occurs on a rough surface because of extremely high local fields at sharp asperities. (2)~This causes local heating, which in turn results in vapor bubble formation leading to electron multiplication and vapor growth proceeding together. (3)~Breakdown is then the result of a vapor column extending to the anode. Note that in other dielectric liquids, formation and growth of vapor bubbles associated with electrical breakdown have been observed experimentally\cite{KAT89,KEL81,FOR90}.

This picture is at least qualitatively consistent with the observed dependence of the breakdown field on the electrode material and surface finish, the electrode size, and the pressure of liquid helium. However, there is no model that can quantitatively describe the dependencies of the breakdown field on these parameters. 

The study reported in this paper was performed as part of the research and development (R\&D) for the SNS nEDM experiment.\cite{SNSnEDM,ITO07,AHM19} This experiment, currently under development to be mounted at the Spallation Neutron Source (SNS) at the Oak Ridge National Laboratory, will search for the permanent electric dipole moment of the neutron (nEDM) using a method proposed by Golub and Lamoreaux\cite{GOL94}.  The experiment will be performed inside a bath of LHe at approximately 0.4~K, and large-area electrodes (61~cm $\times$ 30~cm) will provide an electric field of up to $\sim$75~kV/cm inside the experimental volume.  Because the sensitivity of the experiment is directly proportional to the strength of this applied field, it is of critical importance to establish the following: (1)~the highest electric field that can be sustained stably; (2)~how the breakdown depends on factors such as the pressure and temperature of LHe, the electrode material and surface condition, and the size of the system (the electrode area and/or the size of the gap between electrodes); and (3)~what distribution describes the breakdown field (or voltage).

However, studying all these aspects of electrical breakdown in LHe
using a full-scale SNS nEDM system would be a rather daunting task,
considering that it would require 1500~l of LHe to be cooled to
$\sim$~0.4~K for each set of measurements requiring a change in the 
cryogenic apparatus. While the ultimate performance test of the electrodes 
to be used for the SNS nEDM experiment will need to be conducted utilizing 
a full scale system, it is possible to obtain information on the dependence 
of the breakdown properties on the electrode materials and LHe temperature
and pressure using a much smaller system, which is significantly easier to 
operate. In addition, since the breakdown is a stochastic phenomenon, how the 
breakdown properties scale with the size of the system depends on and 
can be inferred from the breakdown-field distribution for a small system, 
as discussed in this paper. Therefore significant insights can be gained 
from a detailed study using a small-scale system. 

To this end, we have performed a comprehensive study of electrical breakdown in
LHe using an apparatus with small electrodes.  We refer to this apparatus
as the ``Small Scale High Voltage Test Apparatus'' (SSHV apparatus), which can accommodate electrodes with a high field area of 
$\sim$0.725~cm$^2$.  Using this apparatus, we studied (1)~the dependence of the breakdown field on 
the LHe pressure, LHe temperature, and electrode surface finish, (2)~the 
distribution of the breakdown field with a constant field ramp rate, 
and (3) distribution of the time it takes before a breakdown occurs when a
field is set at some predetermined value. The dependence of the
probability of breakdown on the field strength, extracted from the
data, closely resembles that of the field emission, giving a strong
indication that the initial process involves the field emission from
the cathode. Also, having the dependence of the probability of
breakdown on the field strength determined for one electrode size
naturally allowed us to predict how the breakdown field scales with
the surface area of the electrode. Our prediction agrees well with
data taken with our ``Large Scale High Voltage Test Apparatus'' (LSHV apparatus)
with an electrode area of $\sim$1100~cm$^2$, which had previously been
constructed in our laboratory\cite{LON06} as well as those obtained by
other investigators. In addition, our study shows that what was
thought be the gap size dependence can be explained by the surface
area scaling when the change in the stressed area caused by the change
in the gap size is properly taken into consideration.

In this paper, we describe the SSHV apparatus, the measurements performed 
using it, the analysis of the data, and the above-mentioned findings
and their implications, as well as our findings on the dependencies of
the breakdown field and distributions on various parameters.

The paper is organized as follows. In Sec.~\ref{sec:apparatus} we describe 
the design of the SSHV apparatus. In addition, we give a brief description of 
the LSHV system. We describe the experimental procedure and present the 
data in Sec.~\ref{sec:experiment}. Our analysis of the data is 
presented in Sec.~\ref{sec:analysis_discussion}, where we also discuss the implications of our findings.

\section{Apparatuses} \label{sec:apparatus}

\subsection{Small scale high voltage test apparatus} \label{sec:SSHV} 
The design of the SSHV apparatus was guided by the following requirements: (1)~the
electrodes have a finite uniform field region with good electrode alignment,
(2)~the temperature can be varied between 1.7~K and 4.0~K, (3)~the pressure can be 
varied, (4)~the pressure can be measured reliably, and (4)~the apparatus can fit in the 
existing Dewar (Janis Model 10 CNDT Research Dewar\cite{JANIS}).

\begin{figure*}
\centering
\includegraphics[width=\textwidth]{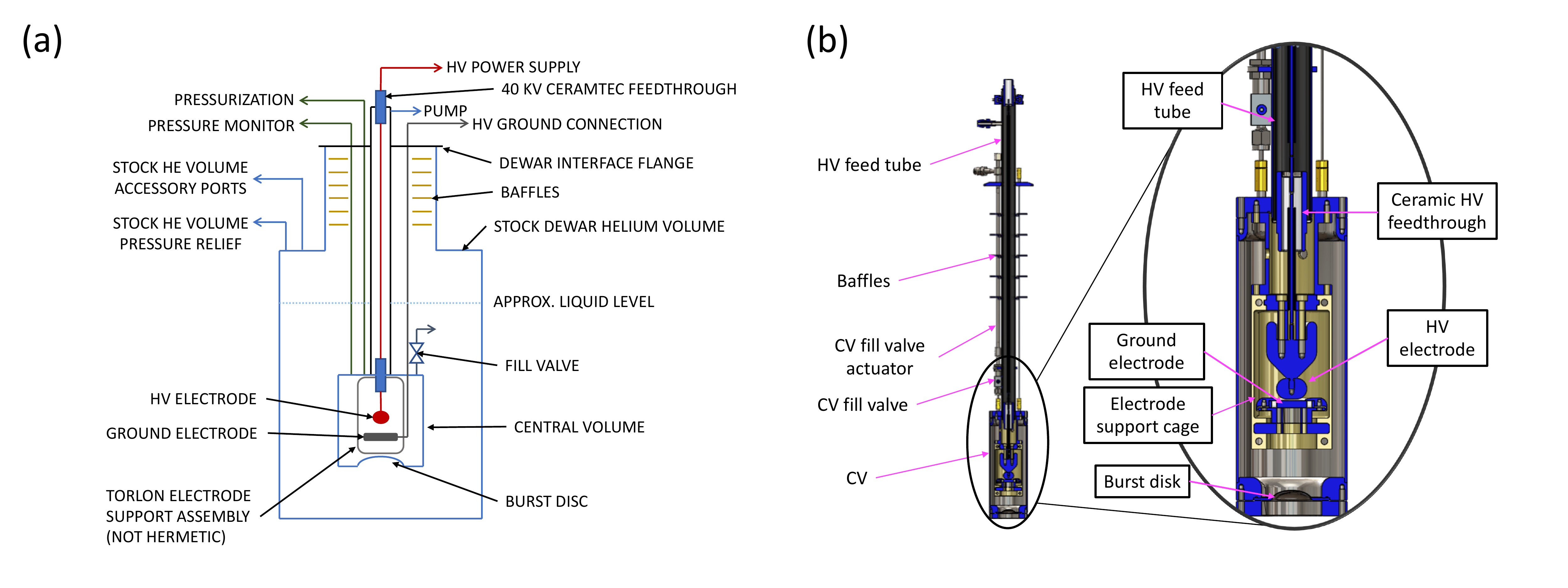}
\caption{Schematics of the SSHV system. (a)~Flow diagram, showing the Dewar LHe volume, CV, various lines, and the high voltage (HV) feed line. (b)~Scale drawing of the SSHV insert along with expanded view of the CV }
\label{fig:SSHVschematic}
\end{figure*}

\begin{figure}
\includegraphics[width=0.49\textwidth]{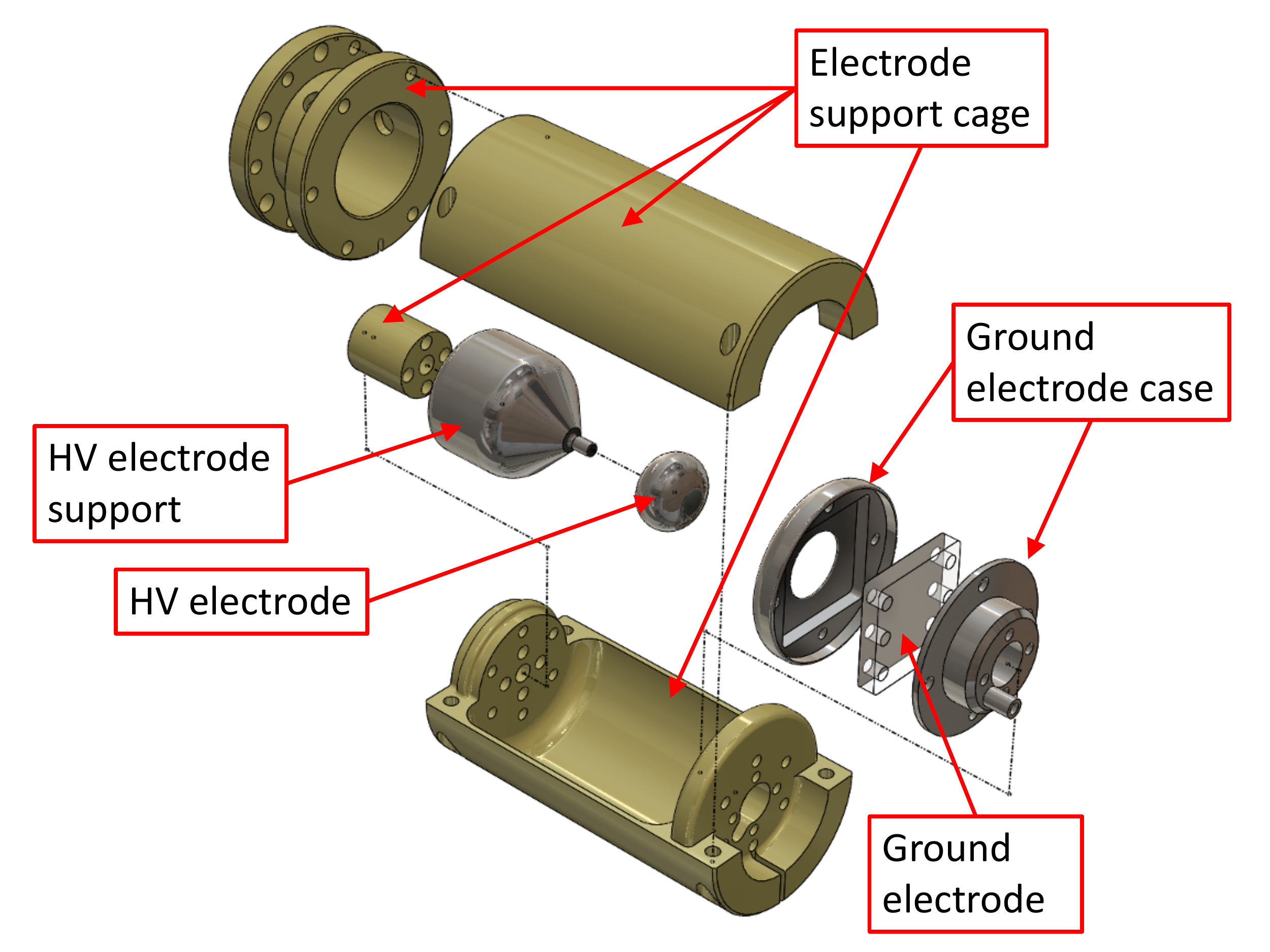}
\caption{Schematic of the SSHV electrode assembly}
\label{fig:SSHVschematic3}
\end{figure}

Schematic diagrams of the apparatus are shown in
Fig.~\ref{fig:SSHVschematic}. 
The electrodes were supported by a
structure made of
polyamide-imide, which has a similar coefficient of thermal expansion
to stainless steel, the electrode material, and is a good electrical
insulator (see Fig.~\ref{fig:SSHVschematic3}). The electrode system was then placed inside the ``Central Volume'' (CV) enclosure, made of stainless steel. The CV was inserted
into the LHe volume of the Janis Dewar. LHe was introduced into the CV through the CV fill
valve. The CV was cooled by evaporatively cooling the LHe outside the
CV. Leaving the CV fill valve open ensured that the pressure of the
LHe inside the CV was at the saturated vapor pressure (SVP). When the
LHe inside the CV needed to be pressurized to a pressure higher than
the SVP, the CV fill valve was closed. The CV was connected to an
external source of high purity helium gas through a small diameter
tubing. The pressure of the LHe inside the CV was controlled by
adjusting the pressure of the external helium gas source with the CV
fill valve closed. In this way, the temperature ($T$) and pressure ($P$)
were varied independently in the ranges of $1.7<T<4.0$~K and
SVP~$<P<626$~Torr.
	
The high voltage (HV) feed line was designed with the following three design
considerations in mind:
	
(1) The heat transmitted to the HV electrode through the HV feed line
needs to be minimized because a small amount of heat imparted to the HV
electrode can boil LHe on the electrode surface due to the rather
small heat of vaporization of LHe, creating bubbles and significantly
reducing the breakdown field.
	
(2) The conductor carrying HV should not be directly exposed to the
volume above the LHe surface, which is filled with helium vapor, as it
would cause electrical breakdown at relatively low voltages.
	
(3) The capacitance of the HV feed line needs to be
minimized. Electrical breakdown causes the energy stored capacitively
in the HV feed line to be released, significantly altering the surface
condition of the electrodes or even damaging the electrodes.
	
In order to meet these considerations, the HV was brought in from the
outside of the cryostat to the inside of the CV using a feed line that
consisted of a stainless steel wire placed inside an evacuated
tube. Using vacuum as electrical insulation (instead of dielectric)
kept the capacitance low. At the same time, it prevented the HV conductor
from being exposed to the helium vapor above the LHe surface. At the
ends of the evacuated tube were 40-kV ceramic HV feedthroughs,
providing seals between the atmosphere and the vacuum inside the tube on the room temperature end,
and the vacuum inside the tube and the LHe inside the CV on the other end. The HV
feedthrough on the cold end was mounted on the top wall of the CV,
providing thermal anchoring to the HV wire. The wire
diameter was chosen to minimize the heat flow from the room
temperature while meeting the electrostatic requirement that the field
on the surface of the wire was sufficiently low when the full 40~kV was
applied. The heat through the feed tube wall was removed by the LHe
outside the CV, as the tube wall was directly in contact with the LHe.

The electrodes were made of stainless steel. The shape of the electrodes was chosen so that the electric field on the surface of each electrode was near uniform in a region around the center and that the field rapidly decreased outside this region. Such an arrangement simplified the data analysis reported in Sec.~\ref{sec:analysis_discussion}.

\begin{figure}
\includegraphics[width=0.49\textwidth]{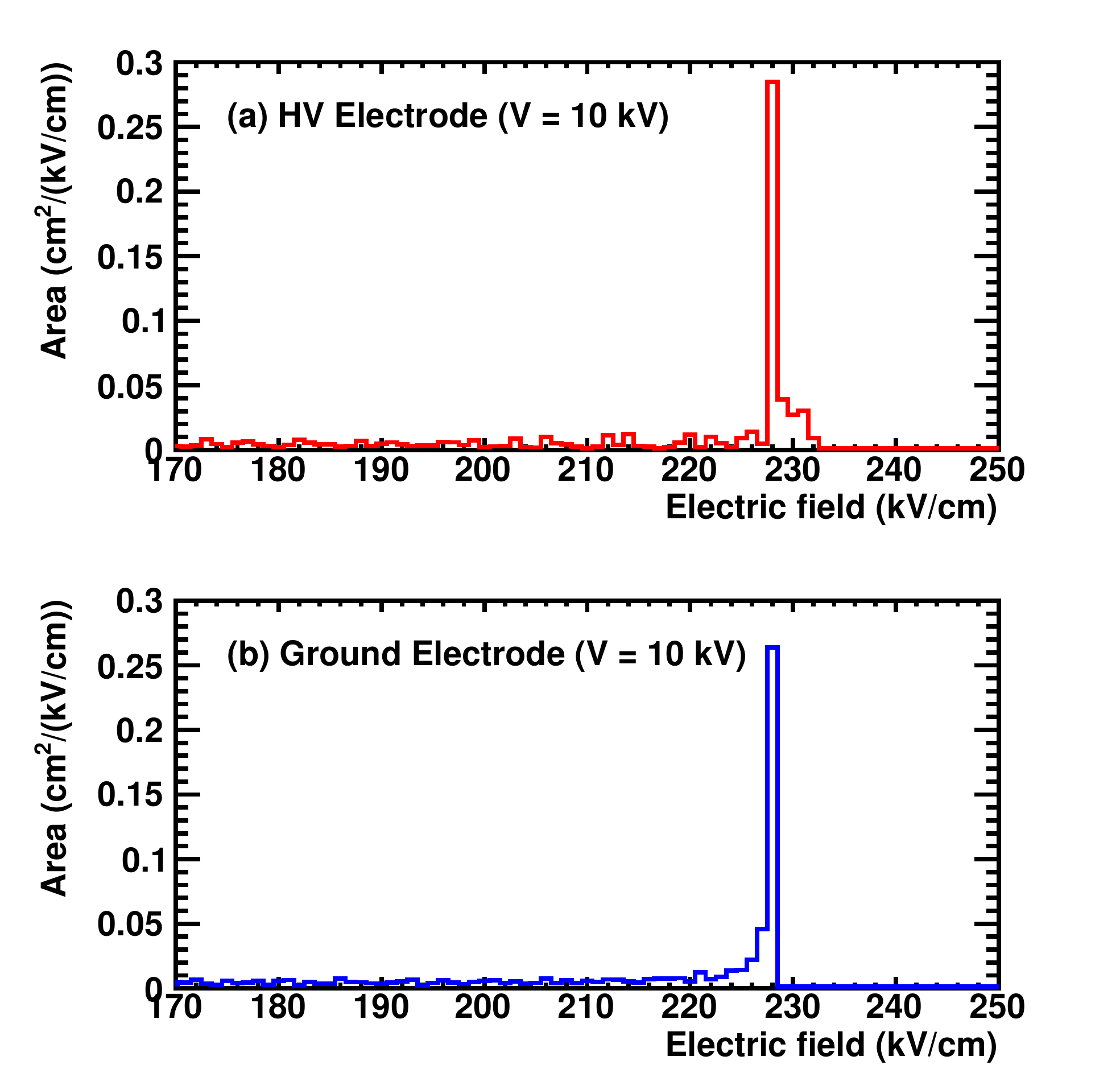}
\caption{Calculated distribution of the electric field strength on (a)~the HV electrode and (b)~the ground electrode for a potential difference of 10 kV and a gap distance of 0.44~mm.}
\label{fig:sshv_field_distribution}
\end{figure}

A field distribution, calculated using the finite element method\cite{COMSOL}, is shown in Fig.~\ref{fig:sshv_field_distribution} for both the HV and ground electrode for a potential difference of 10 kV and a gap distance of 0.44~mm.  We measured the average gap distance to be 0.52~mm at room temperature. Based on the known coefficients of thermal expansion for the materials used for the electrode assembly, we estimate that the gap distance shrinks by 15\% when the system is cooled, giving an average gap distance of 0.44~mm. The estimated overall uncertainty on this quantity is 20\% due to variation in electrode thicknesses.

When the gap was set to be 2~mm and the pressure was set to be 600~Torr, no breakdown was observed even at 40~kV. This confirms that all the HV components were able to withstand potentials up to 40~kV and that the breakdowns observed were indeed in the gap between the two electrodes. 

\begin{figure}
\includegraphics[width=0.47\textwidth]{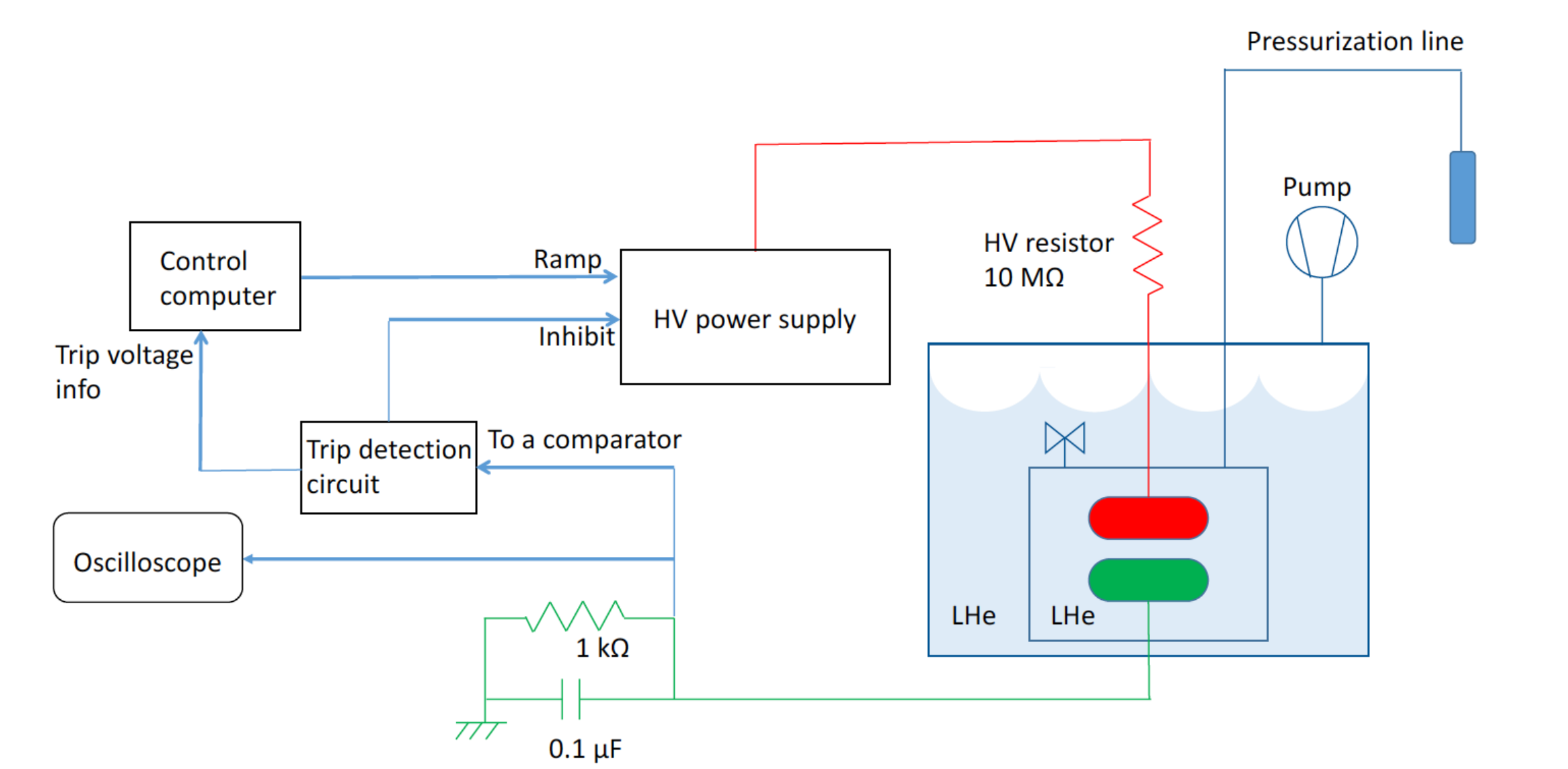}
\caption{A schematic showing electrical connections for the SSHV system.}
\label{fig:sshv_circuit}
\end{figure}

The HV power supply for the SSHV system was computer controlled. The ground electrode was connected to a trip circuit, which shuts off the HV power supply as soon as it detected a current flowing out of the ground electrode. The circuit had a comparator that tripped and latched when the current exceeded 100 $\mu$A with the 1-k$\Omega$ load resistor.  This shut off a relay switch to the interlock of the power supply and caused the stoppage of the voltage ramp.  The breakdown voltage was then recorded and a wait time of a few seconds passed before the latched interlock was reset.  Then a new voltage ramp sequence was started. Refer to Fig.~\ref{fig:sshv_circuit}.

\subsection{Large scale high voltage test apparatus}\label{sec:LSHV}  
\begin{figure}
    \centering
    \includegraphics[width=0.49\textwidth]{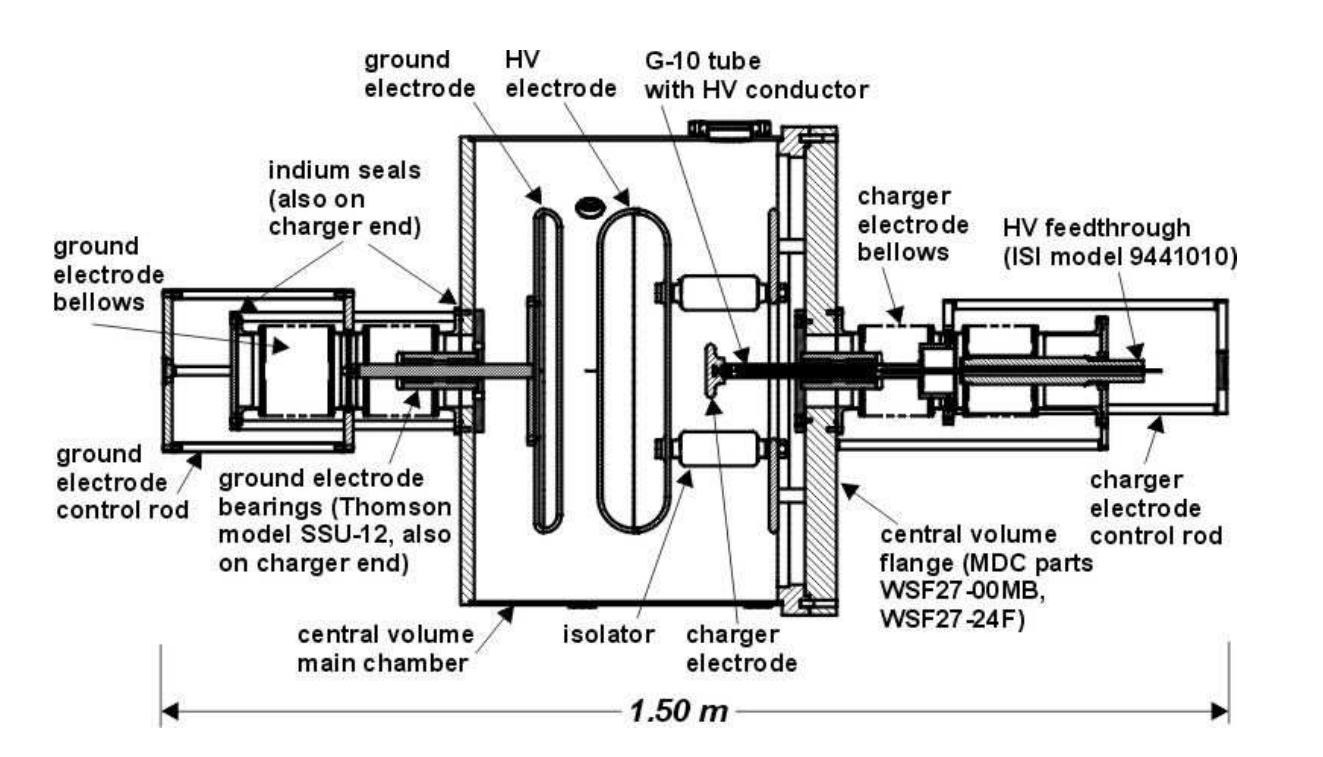}
    \caption{A schematic of the LSHV CV (Central Volume, the volume filled with LHe).}
    \label{fig:lshv_schematic}
\end{figure}

A schematic of the LSHV apparatus is shown in
Fig.~\ref{fig:lshv_schematic}. Details of its design and operation can
be found in Ref.~\onlinecite{LON06}. Here we give a brief description
of the apparatus relevant for the measurements reported in this paper. 

The CV (= Central Volume, a volume filled with LHe) was a cylinder with an inner diameter of 66~cm and an inner length of 44~cm. The CV housed the HV, ground, and charging electrodes. Both the HV and ground electrodes had a diameter of 45.8~cm. However, since the HV electrode had a thickness of 10.2~cm and the edge was rounded with a radius of curvature of 5.1~cm, the flat area had a diameter of 35.6~cm. Using the finite element method~\cite{COMSOL} we calculated the effective stress area to be 1,100~cm$^2$. The ground electrode was designed to be movable and the charging electrode was designed to be disconnectable from the HV electrode, because one of the purposes of this apparatus was to demonstrate the concept of HV amplification based on a variable capacitor as described in Ref.~\onlinecite{LON06}. However, for acquiring the data reported in this paper, these features were not used. The position of the ground electrode was fixed to 3~mm and the charging electrode remained connected to the HV electrode. 

The temperature and pressure of the LHe were varied by directly pumping on the LHe in the CV. Because of this arrangement, the pressure was always the saturated vapor pressure at a given temperature. In order to avoid the bubbling of LHe affecting the breakdown field measurements, the measurements were made with the pumping turned off and the volume closed off after the desired pressure and temperature were reached. 

\section{Experimental procedure and results}\label{sec:experiment}

\subsection{Procedure}\label{sec:expprocedure}

Data were acquired for three different electrode surface polish configurations under varying pressures and temperatures. These configurations were: electropolished (EP), mechanically-polished (MP) and hybrid-polish (HP), where the latter configuration consisted of an electropolished electrode in conjunction with a mechanically-polished electrode. For a given set of operating conditions (temperature, pressure, and electrode polish), the voltage on the HV electrode was ramped up from zero at a constant rate until a breakdown occurred and which caused the power supply to trip (or in the case of no breakdown until 40~kV was reached).  The voltage at which the breakdown was observed and the operating conditions were recorded.  This procedure was repeated many times to obtain a set of breakdown voltages for the given operating conditions. The parameters and conditions under which the datasets were acquired are summarized in Tabs.~\ref{tab:sshvdatasets} and \ref{tab:lshvdatasets} for the SSHV and LSHV apparatuses, respectively.  Henceforth, for brevity a dataset will be referenced by the polish of the electrodes used in the measurement and its set number (e.g., EP:3, HP:24, or MP:32).

The majority of the datasets were acquired at both fixed temperature and pressure and with a constant ramp rate, but several datasets were acquired under slightly different procedure and conditions.  Dataset EP:1 was acquired at a fixed temperature of 1.92~K but with variable pressure.  Dataset MP:50 was acquired at a fixed pressure of 480 Torr but with the temperature increasing from $2.2 - 4.0$~K as the system was allowed to warm up.  The same procedure was applied to acquire dataset MP:38.  Dataset MP:53 was acquired at a fixed temperature and pressure but with three different voltage ramp rates of 50 V/s, 100 V/s, and 200 V/s.  The dependence of the breakdown on the ramp rate is discussed in further details in Sec.~\ref{sec:ramprate}. Finally, datasets $55-58$ involved measurements of the time to breakdown and the procedure used in acquiring these data are discussed in more detail in Sec.~\ref{sec:analysis_discussion}.

\subsection{Results}\label{sec:expresults}

\subsubsection{Distribution of breakdown field} \label{sec:bfielddistr}

\begin{figure*}[]
	\captionsetup[subfigure]{justification=centering}
	\centering
	\begin{subfigure}[]{0.32\textwidth}
		\includegraphics[width=\textwidth]{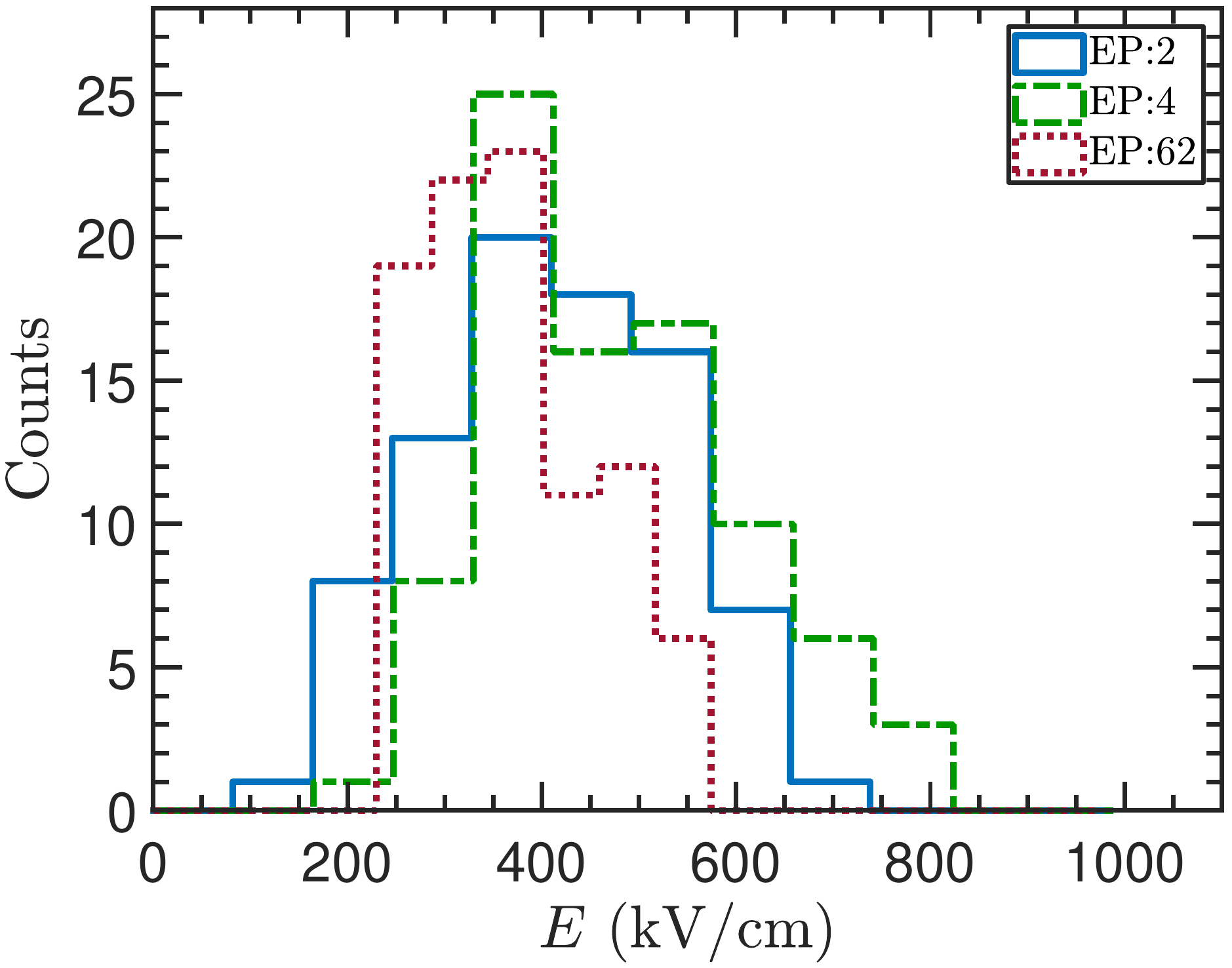}
		\caption{EP, $< 20$ Torr}
		\label{fig:EP-04-62}
	\end{subfigure}
	\hfill
	\begin{subfigure}[]{0.32\textwidth}
		\includegraphics[width=\textwidth]{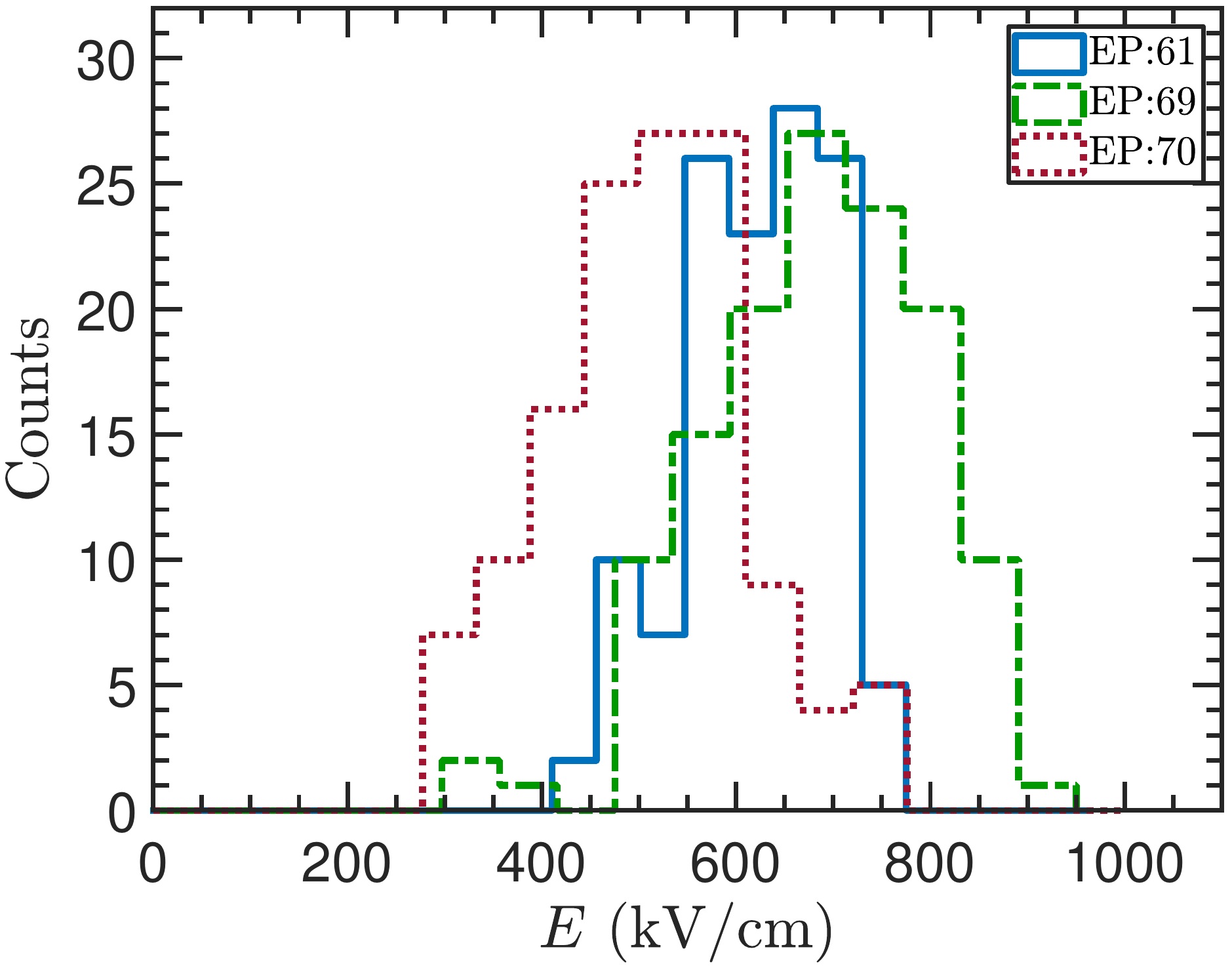}
		\caption{EP, $100-200$ Torr}
		\label{fig:EP-61-70}
	\end{subfigure}
	\hfill
	\begin{subfigure}[]{0.32\textwidth}
		\includegraphics[width=\textwidth]{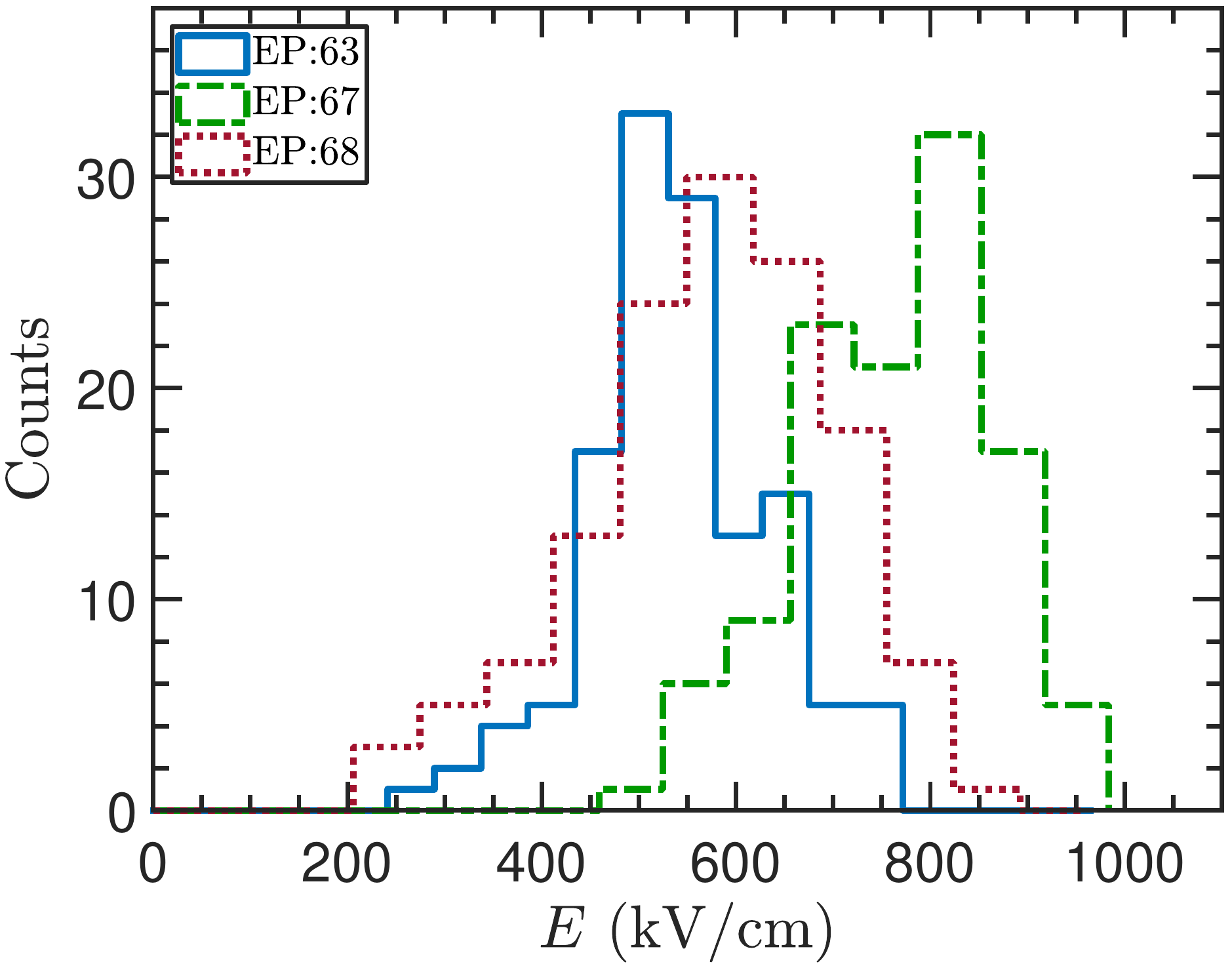}
		\caption{EP, $> 200$ Torr}
		\label{fig:EP-63-68}
	\end{subfigure}
\hfill
	\begin{subfigure}[]{0.32\textwidth}
	\includegraphics[width=\textwidth]{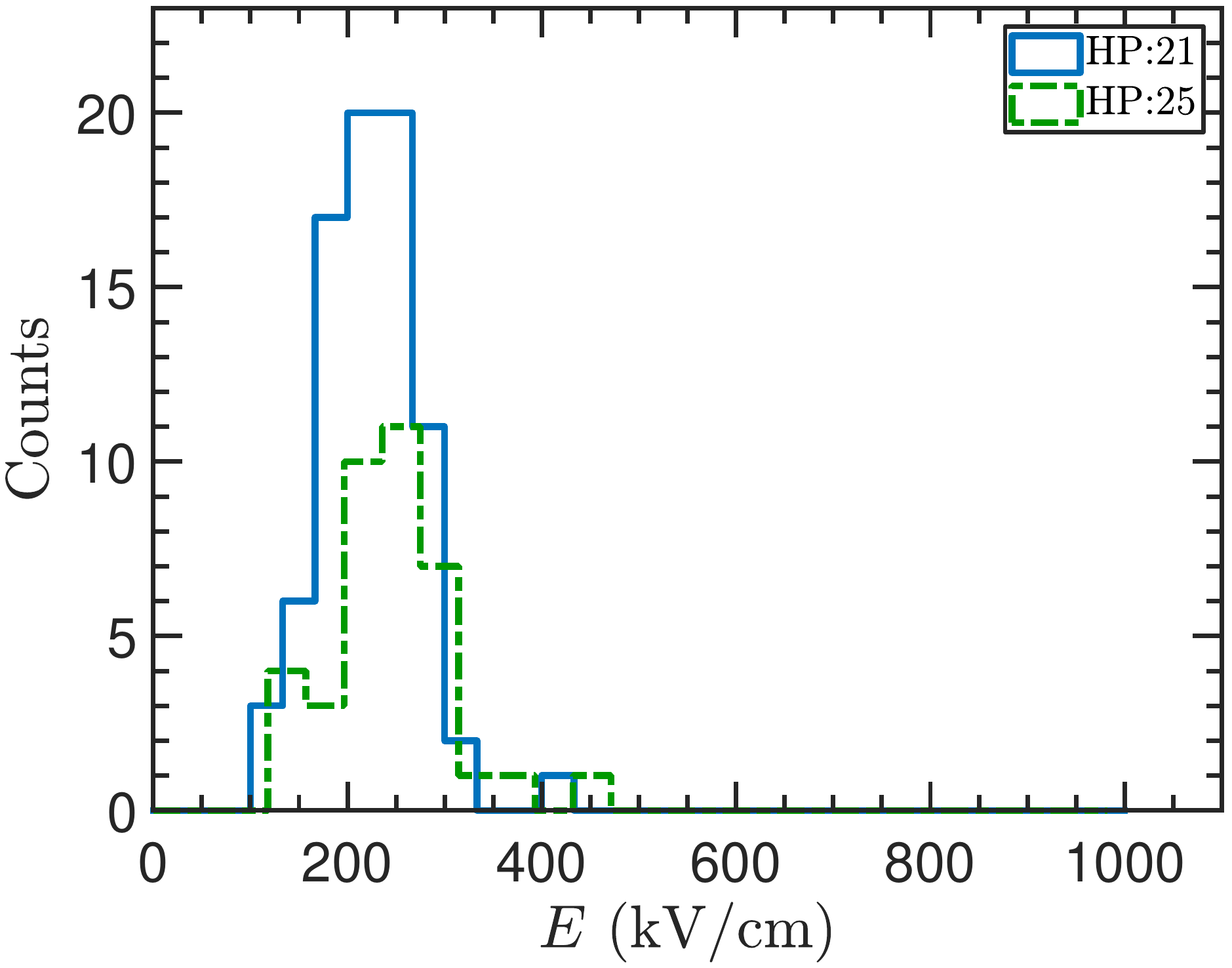}
	\caption{HP, $< 20$ Torr}
	\label{fig:HP-21-25}
\end{subfigure}
\hfill
\begin{subfigure}[]{0.32\textwidth}
	\includegraphics[width=\textwidth]{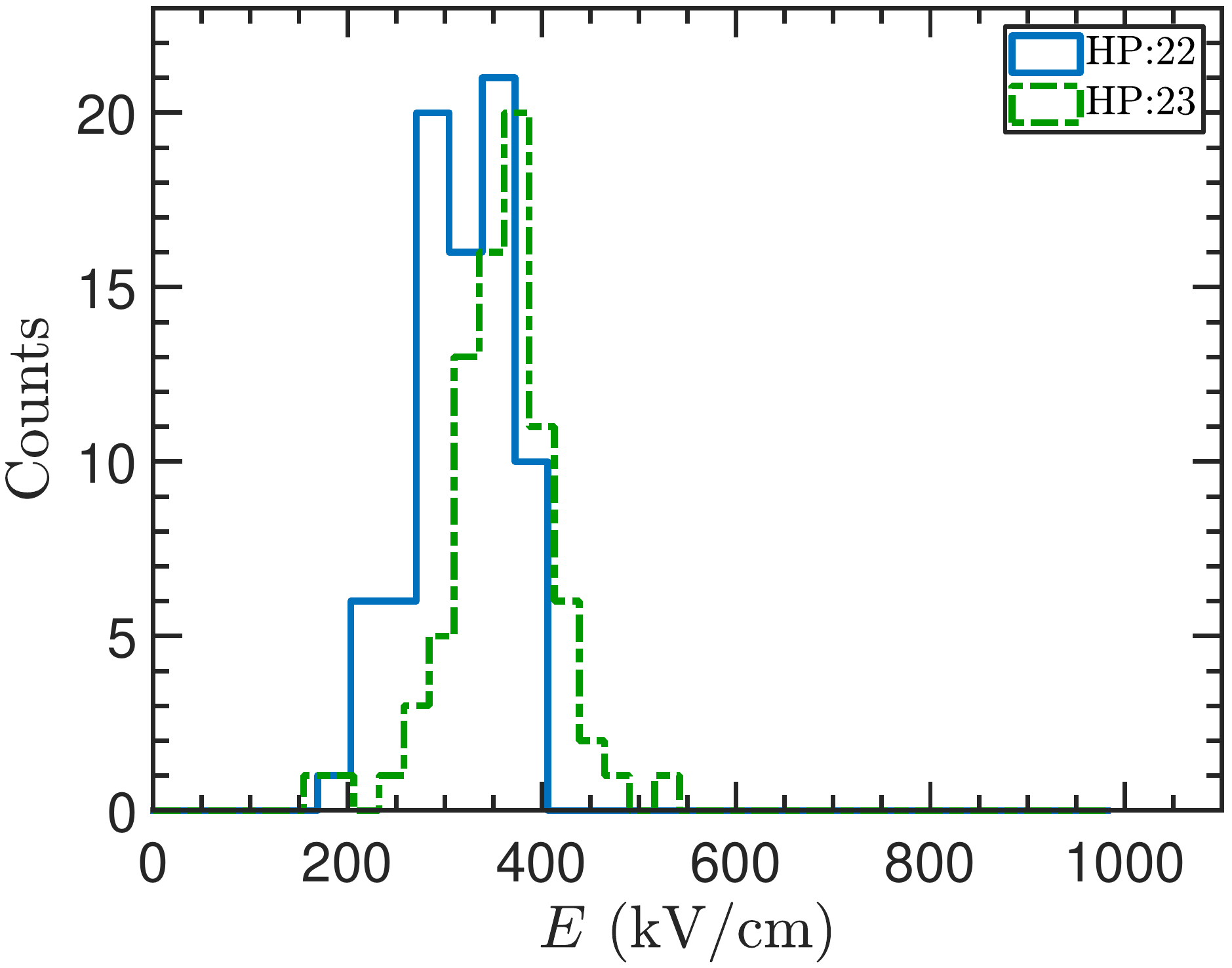}
	\caption{HP, $200-400$ Torr}
	\label{fig:HP-22-23}
\end{subfigure}
\hfill
\begin{subfigure}[]{0.32\textwidth}
	\includegraphics[width=\textwidth]{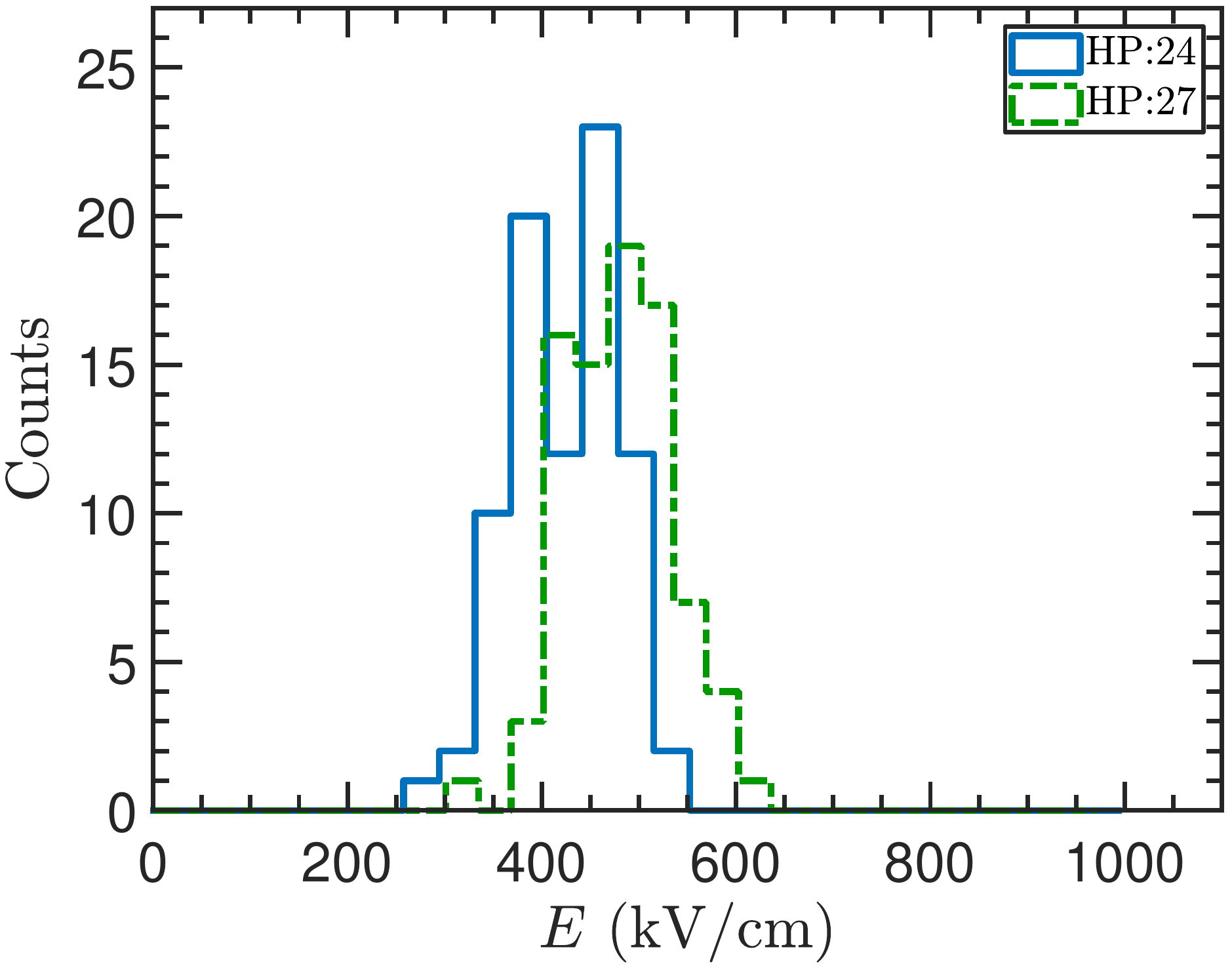}
	\caption{HP, $> 500$ Torr}
	\label{fig:HP-24-27}
\end{subfigure}
\hfill
\begin{subfigure}[]{0.32\textwidth}
	\includegraphics[width=\textwidth]{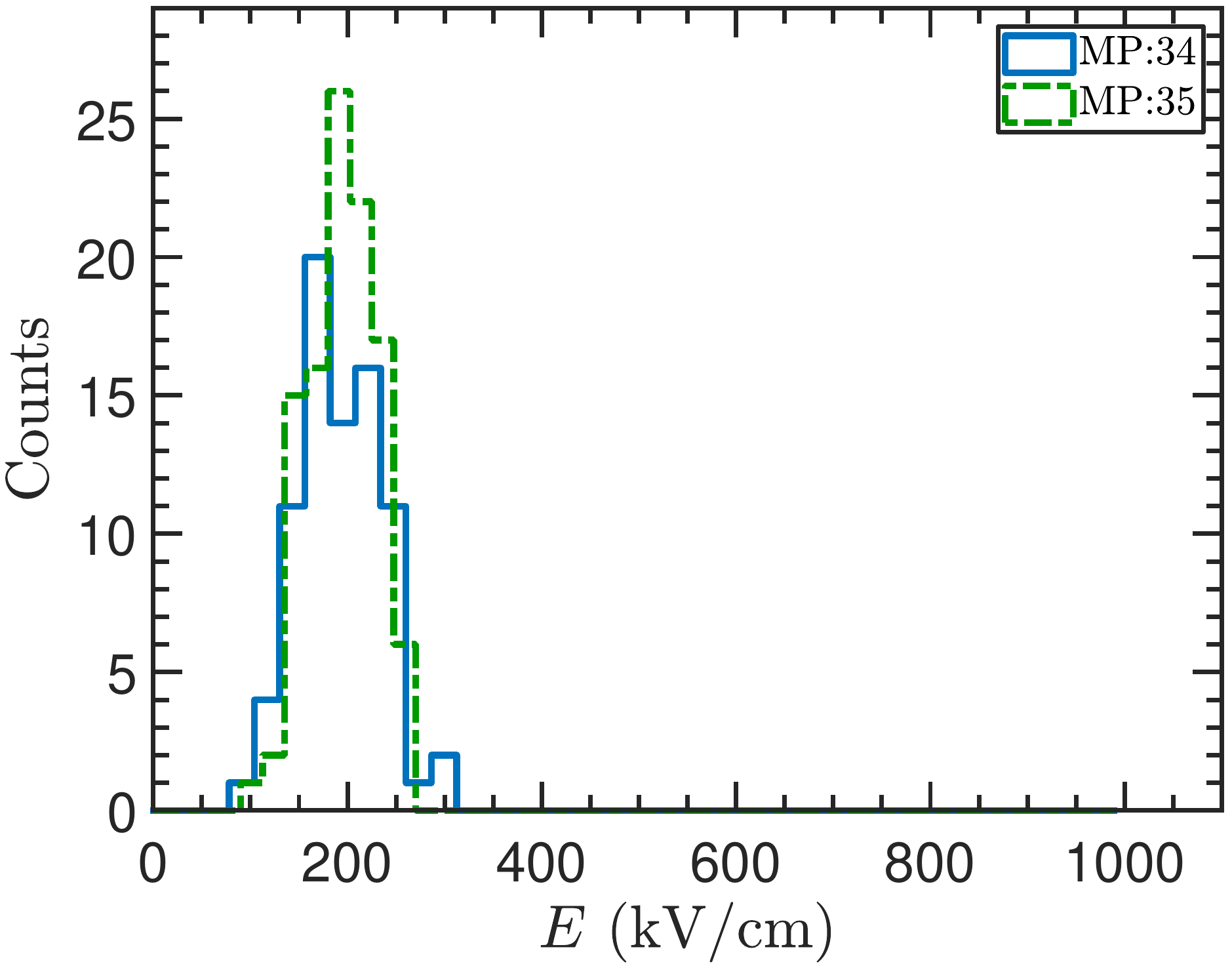}
	\caption{MP, $< 20$ Torr}
	\label{fig:MP-34-35}
\end{subfigure}
\hfill
\begin{subfigure}[]{0.32\textwidth}
	\includegraphics[width=\textwidth]{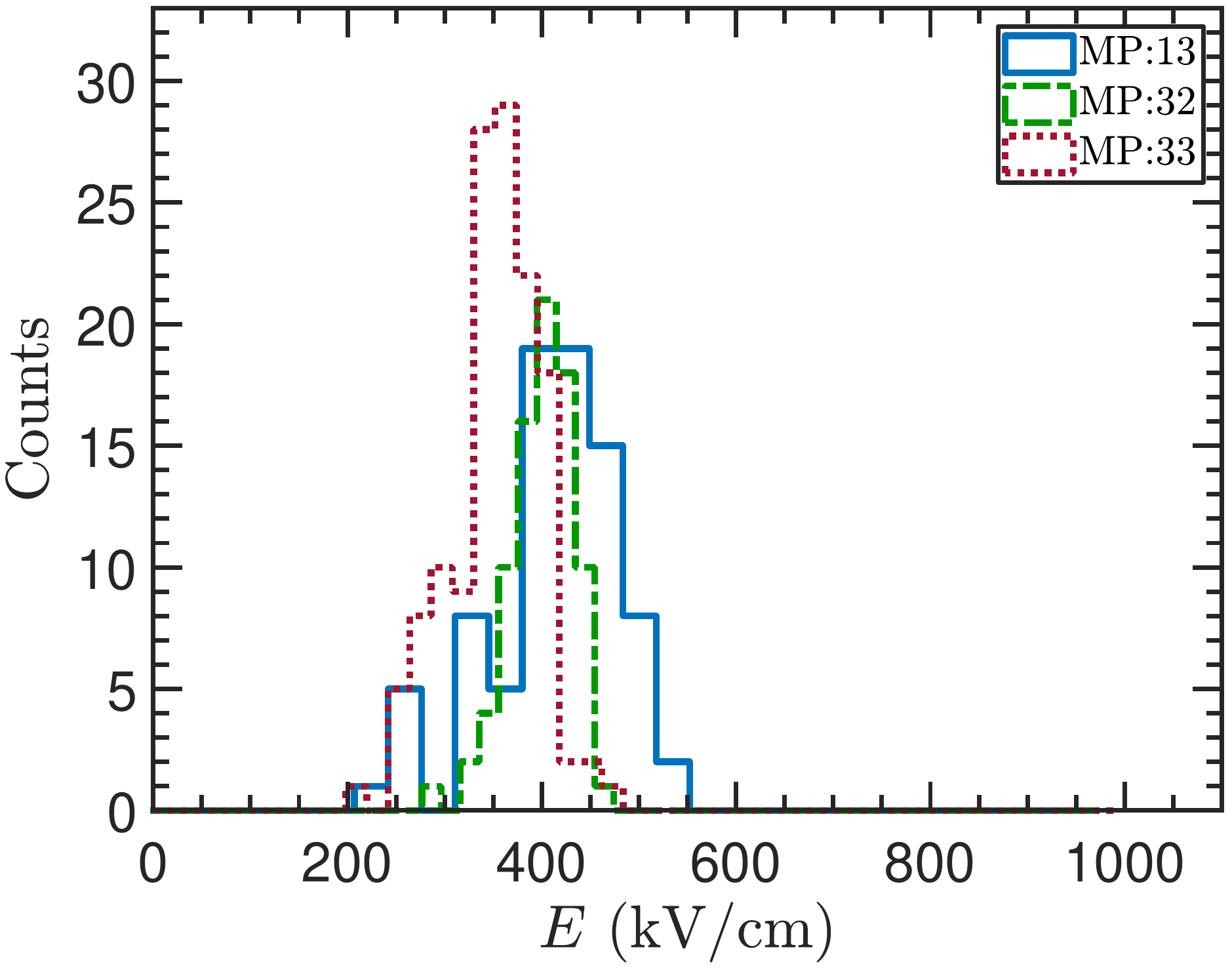}
	\caption{MP, $300-400$ Torr}
	\label{fig:MP-13-32-33}
\end{subfigure}
\hfill
\begin{subfigure}[]{0.32\textwidth}
	\includegraphics[width=\textwidth]{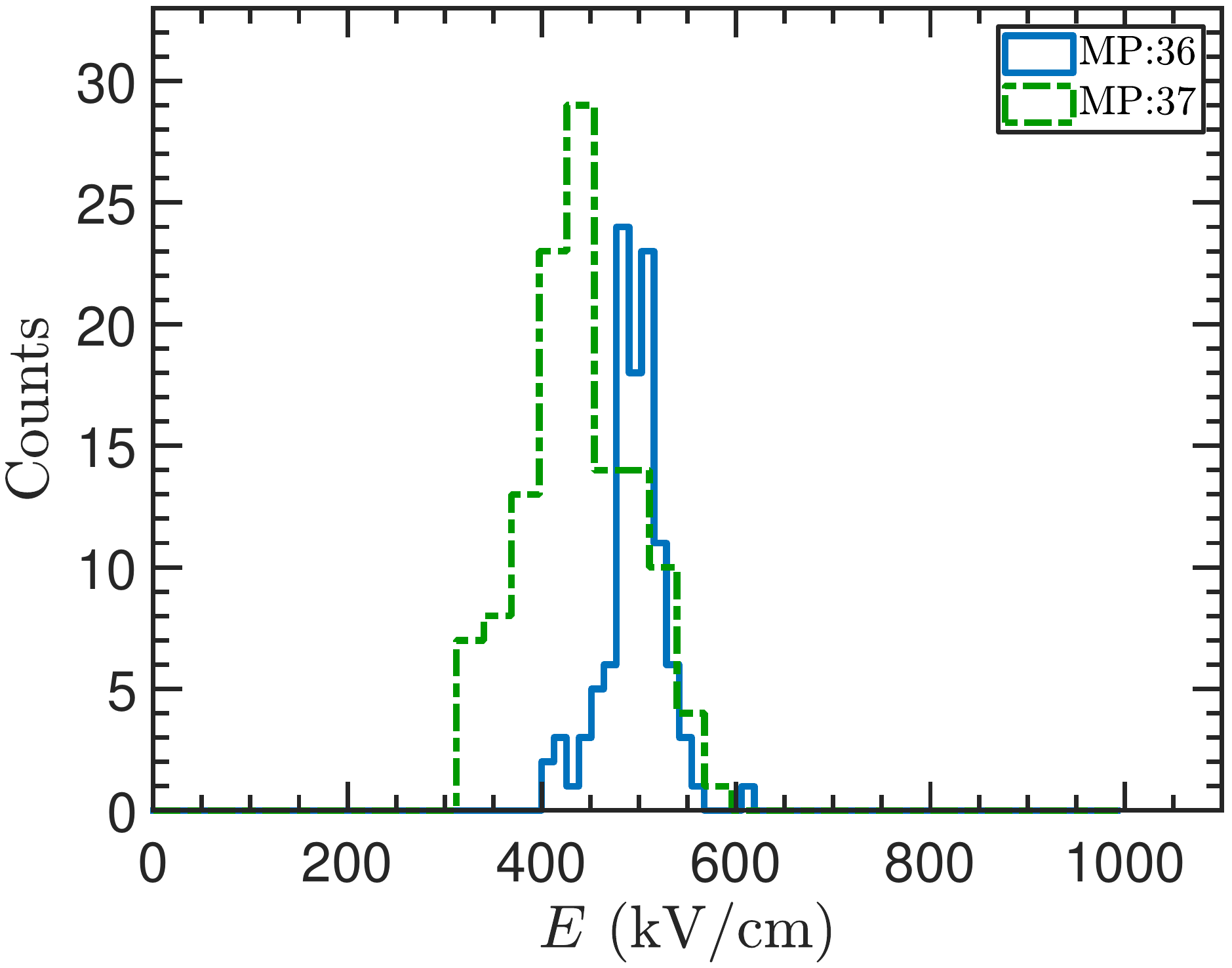}
	\caption{MP, $\sim600$ Torr}
	\label{fig:MP-36-37}
\end{subfigure}
\hfill
\begin{subfigure}[]{0.32\textwidth}
	\includegraphics[width=\textwidth]{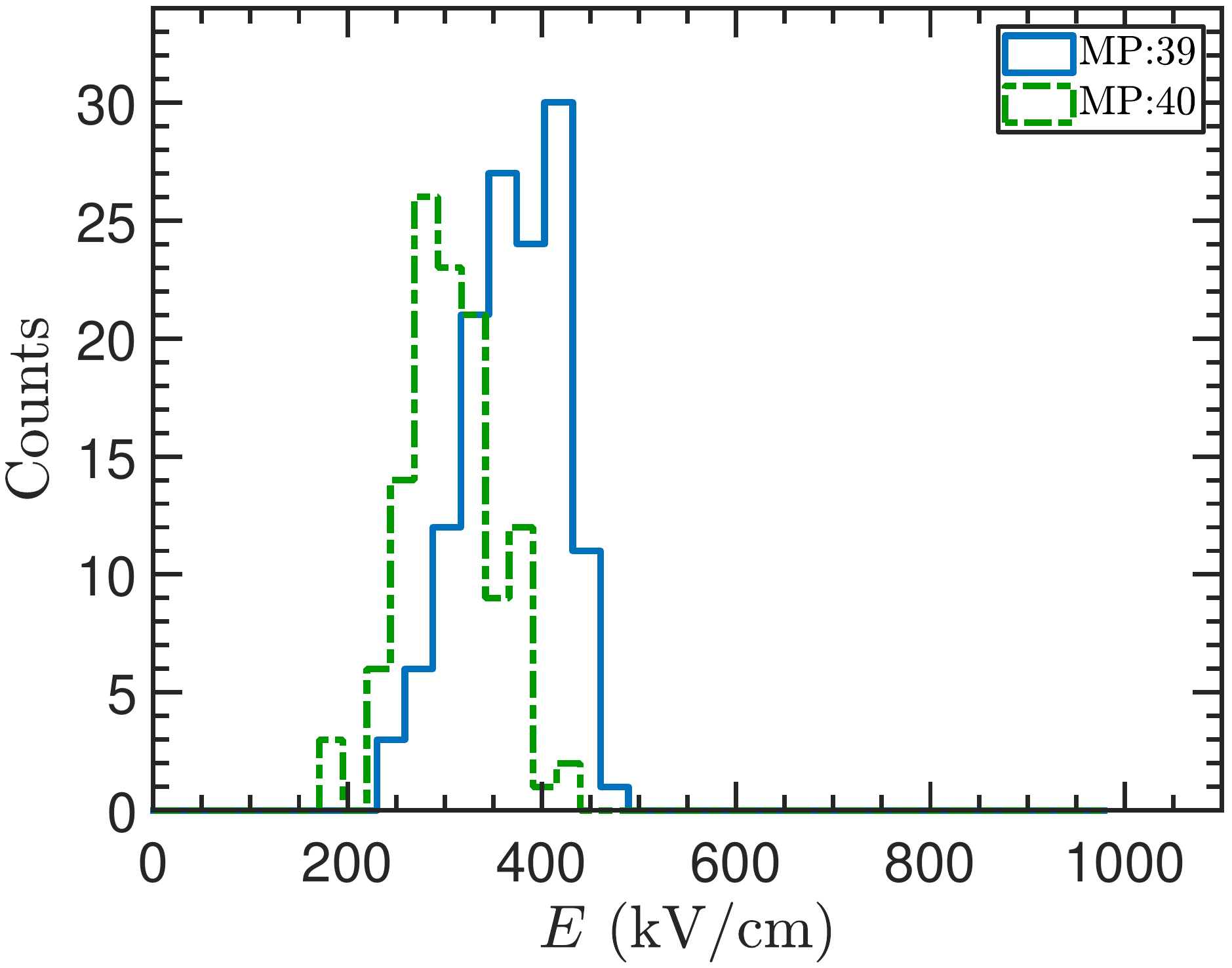}
	\caption{MP, $130-140$ Torr}
	\label{fig:MP-39-40}
\end{subfigure}
\hfill
\begin{subfigure}[]{0.32\textwidth}
	\includegraphics[width=\textwidth]{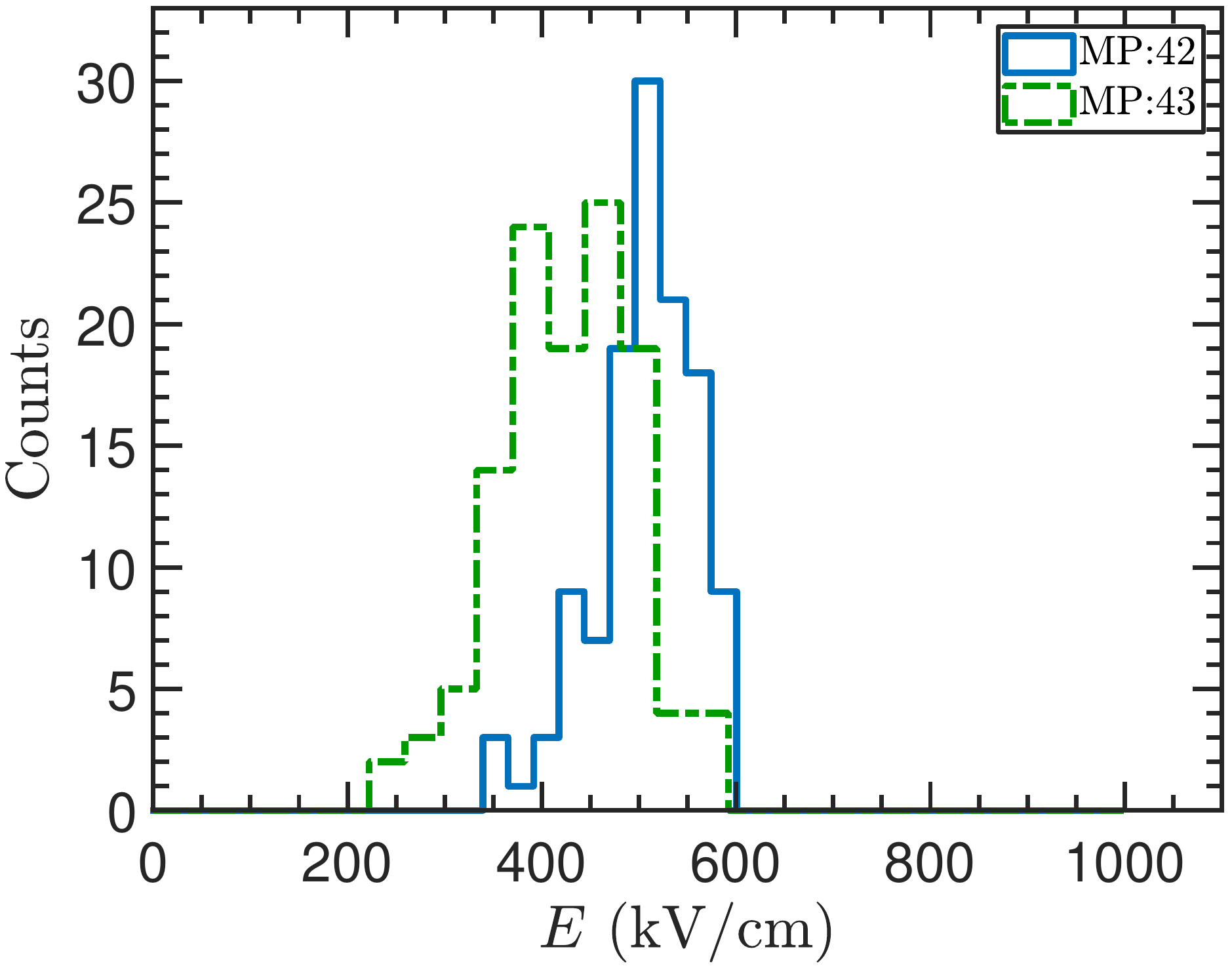}
	\caption{MP, $\sim480$ Torr}
	\label{fig:MP-42-43}
\end{subfigure}
\hfill
\begin{subfigure}[]{0.32\textwidth}
	\includegraphics[width=\textwidth]{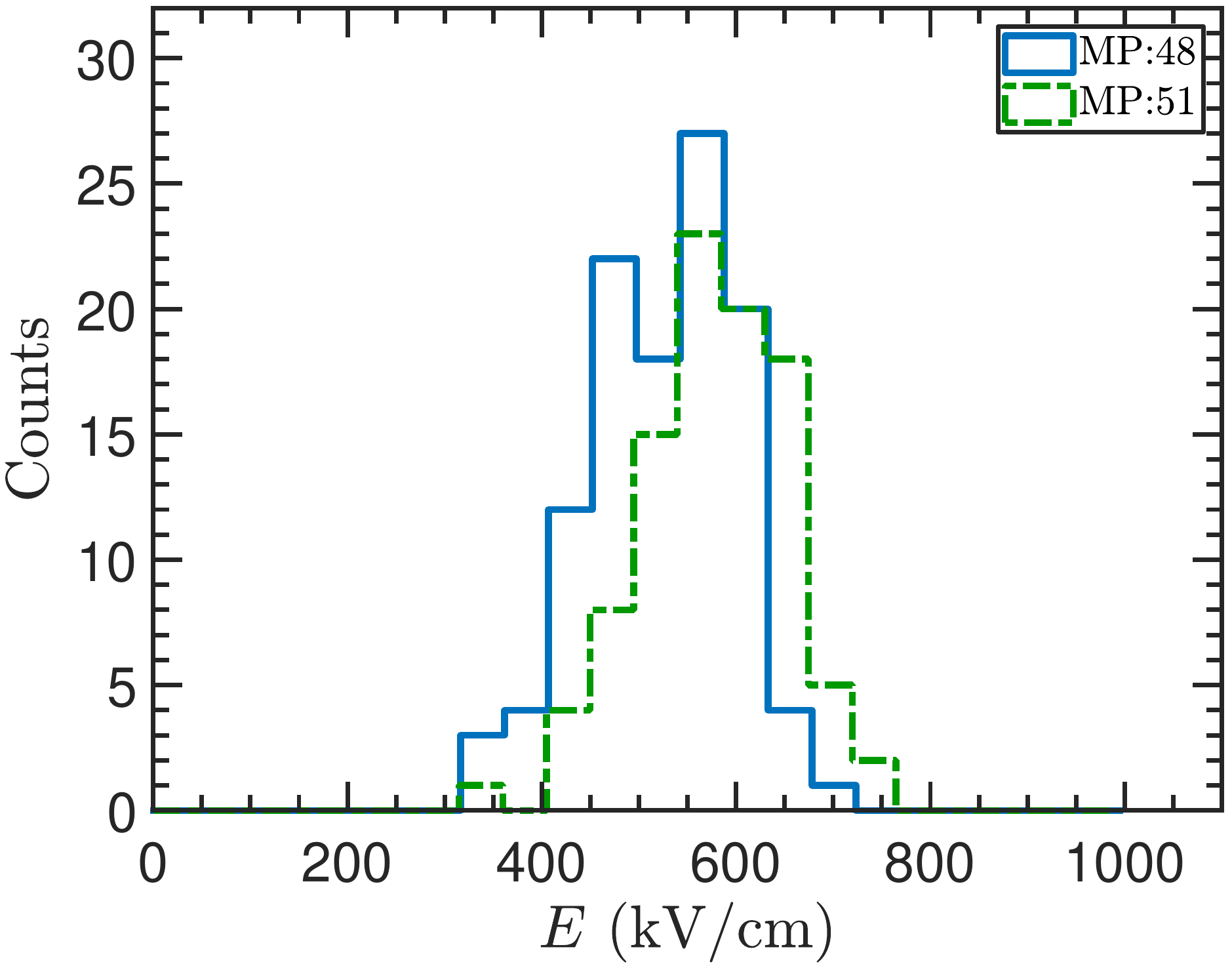}
	\caption{MP, $\sim620$ Torr}
	\label{fig:MP-48-51}
\end{subfigure}

	\caption{(a)-(l) Distributions of breakdown field for select SSHV datasets.}
	\label{fig:hist_datasets}
\end{figure*}

The breakdown voltages were converted to breakdown fields through, $E = V/d$, where $d$ was the separation between the HV and ground electrodes and $V$ was the voltage at which a breakdown was observed.  The uncertainty on this distance is estimated to be 20\% for the SSHV data and 15\% for the LSHV data.  The breakdown field distribution for selected SSHV datasets acquired with the electropolished, mechanically-polished, and hybrid electrodes are shown in Fig.~\ref{fig:hist_datasets}. The distributions for the LSHV datasets are not shown because each dataset contained only a few breakdown measurements.

From Fig.~\ref{fig:hist_datasets}, the mean values of the distributions are seen to be shifted to higher fields with increasing pressure for all types of electrode polish.  Interestingly, the widths of the electropolished distributions are much broader than the distributions obtained with the other electrode polishes.  We will discuss a possible explanation for this observation in Sec.~\ref{sec:analysis_discussion}.  Furthermore, a quite intriguing feature is the absence of any significant differences between the distributions acquired with mechanically-polished electrodes as compared to those acquired with the hybrid electrodes.  This has important implications regarding whether breakdown is initiated on the cathode or anode surface.  We will discuss this in more detail in Sec.~\ref{sec:analysis_discussion}.

\begin{table*}
	\caption{\label{tab:table3}SSHV datasets and the operational conditions in which they were acquired.  The surface polish of the electrodes is specified by electropolished(EP), hybrid-polish(HP), and mechanically-polished(MP).}
	\begin{ruledtabular}
		\begin{tabular}{lccccccc}
			Set & Electrode polish & No. breakdowns & P (Torr) & T (K)
			& Ramp rate (V/s) & $\overbar{E}$ (kV/cm) & $\sigma_{\text{br}}$ (kV/cm)\\ \hline
			1\footnote[7]{Measurements made during cooling with $P$ between 12.6 - 35.2 Torr at SVP.} & EP  & 104 & 19.0 & 1.93  & 100  &  349 & 83 \\
			2 & EP & 84 & 12.6 & 1.81  & 100  & 415  & 121 \\
			3 & EP & 93 & 12.7 & 1.81  & 100  & 436  & 124   \\
			4 & EP & 86 & 12.9 & 1.81  & 100  & 473  & 127   \\
			5 & EP & 91 & 12.9 & 1.81  & 100  & 430  & 139    \\
			6 & EP & 109 & 7.4 & 1.67  & 100  & 425  & 134    \\
			11 & MP & 80 & 8.8 & 1.75  & 100  & 213  & 47    \\
			12 & MP & 81 & 193.5 & 3.01  & 100  & 310  & 51    \\
			13 & MP & 82 & 402.0 & 3.61  & 100  & 414  & 66    \\
			14 & MP & 61 & 617.2 & 4.01  & 100  & 525  & 86    \\
			15 & MP & 40 & 621.4 & 4.01  & 100  & 499  & 82    \\
			16 & MP & 40 & 193.4 & 3.05  & 100  & 365  & 56   \\
			17 & MP & 40 & 9.0 & 1.72  & 100  & 253  & 41    \\		
			21 & HP & 80 & 9.6 & 1.73  & 100  & 222  & 50 \\
			22 & HP & 80 & 202.9 & 3.08  & 100  & 316  & 50  \\
			23 & HP & 81 & 395.9 & 3.60  & 100  & 358  & 55   \\
			24 & HP & 82 & 618.6 & 4.01  & 100  & 425  & 54   \\
			25 & HP & 38 & 8.0 & 1.69  & 100  & 242  & 63   \\
			26 & HP & 30 & 469.5 & 3.75  & 100  & 442  & 81   \\
			27 & HP & 83 & 617.5 & 4.01  & 100  & 479  & 56 \\	
			31 & MP & 100 & 9.1 & 1.72  & 100  &  162 &  32 \\
			32 & MP & 83 & 330.2 & 1.75  & 100  & 399  &  31 \\
			33 & MP & 135 & 324.1 & 3.43  & 100  & 353  &  46 \\
			34 & MP & 80 & 9.8 & 1.74  & 100  & 191  &  41 \\
			35\footnote[14]{Negative polarity on HV electrode.} & MP & 105 & 9.8 & 1.74  & 100  & 196  & 34    \\
			36 & MP & 107 & 606.9 & 1.76  & 100  & 495  & 32     \\
			37 & MP & 123 & 625.9 & 4.02  & 100  & 438  & 56    \\		
			38\footnote[16]{$P$ fixed with $T$ increasing from 2.2 to 2.9~K as a natural warm-up.}  & MP & 40 & 202.2 & 2.56  & 100  & 370  & 41   \\
			39 & MP & 135 & 131.0 & 1.76  & 100  & 370  & 51   \\
			40 & MP & 117 & 140.5 & 2.84  & 100  & 305  & 46   \\
			41 & MP & 116 & 10.5 & 1.76  & 100  & 211  & 50   \\
			42 & MP & 120 & 484.2 & 1.77  & 100  & 507  & 53   \\
			43 & MP & 118 & 487.1 & 3.78  & 100  & 427  & 69   \\
			44 & MP & 91 & 319.2 & 1.77  & 100  & 486  & 56   \\
			45 & MP & 102 & 328.3 & 3.44  & 100  & 412  & 63   \\
			46 & MP & 26 & 612.3 & 1.77  & 100  & 551  & 70   \\
			47 & MP & 96 & 609.8 & 1.76  & 100  & 575  & 62   \\		
			48 & MP & 111 & 619.3 & 4.01  & 100  & 525  & 76  \\
			49 & MP & 100 & 11.1 & 1.77  & 100  & 248  & 48   \\
			50\footnote[20]{$P$ fixed with $T$ increasing from 2.2 - 4.0~K as a natural warm-up.} & MP & 72 & 622.3 & 3.30  & 100  & 547  & 86   \\
			51 & MP & 96 & 593.7 & 2.16  & 100  & 576  & 71   \\
			52 & MP & 94 & 589.7 & 3.13  & 100  & 565  & 69   \\
			53\footnote[18]{Three different voltage ramp rates: 50, 100, and 200 V/s.} & MP & 974 & 10.8 & 1.76  & 50, 100, 200   & 271  &  72  \\
			54 & MP & 148 & 10.7 & 1.76  & 200  & 303  &  83  \\	
			55\footnote[8]{Holding time measurements. Refer to text for further details.} & MP & 1364 & 10.6 & 1.76  & 200  & -  & -   \\
			56\footnotemark[8] & MP & 1149  & 10.5 & 1.76  & 200  &  - & -   \\
			57\footnotemark[8] & MP & 1168  & 10.5 & 1.76  & 200  & -  & -   \\
			58\footnotemark[8] & MP & 994 & 10.5 & 1.76  & 200  & -  & -   \\	
			61 & EP & 127 & 223.0 & 1.79  & 200  & 621  & 77   \\
			62 & EP & 93 & 11.8 & 1.79  & 200  & 368  &  86  \\
			63 & EP & 129 & 225.8 & 3.16  & 200  &  545  & 93  \\
			64\footnote[19]{Additional 26 holding time measurements.} & EP & 97 & 422.7 & 1.79  & 200  & 814  & 63  \\
			65 & EP & 129 & 104.2 & 1.79  & 200  & 684  &  87   \\
			66 & EP & 39 & 231.1 & 1.80  & 200  &  771 &  78   \\
			67 & EP & 114 & 225.9 & 1.79  & 200  & 767 &  100 \\		
			68 & EP & 134 & 229.6 & 3.17  & 200  & 574  &  128   \\
			69 & EP & 131 & 115.8 & 1.80  & 200  & 688  &  116   \\
			70 & EP & 130 & 119.1 & 2.74  & 200  & 510  &  104   \\
		\end{tabular}
	\end{ruledtabular}
\label{tab:sshvdatasets}
\end{table*}

\begin{table}
 	\caption{\label{tab:lshvdatasets} Summary of LSHV datasets. }
	\begin{ruledtabular}
		\begin{tabular}{lcccc}
			Set & Electrode polish & P (Torr) & T (K)
			& $\overbar{E}$ (kV/cm) \\ \hline
			1 & EP & 600.0 & 3.98 & 112.5     \\
			2 & EP & 210.0 & 3.10 & 100.0     \\
			3 & EP & 117.5 & 2.73 & 100.0      \\
			4 & EP & 29.0 & 2.07 & 75.0      \\
			5 & EP & 18.0 & 1.91 & 60.0     \\
			6\footnote[16]{Pressurized.} & EP & 58.0  & 2.07 & 90.0      \\
			7 & EP & 600.0 & 3.98 & 100.0     \\
			8 & EP & 28.8 & 2.07 & 73.0      \\
			9 & EP & 16.2 & 1.88 & 67.0      \\
			10 & EP & 6.9 & 1.65 & 57.0      \\
		\end{tabular}
	\end{ruledtabular}
\end{table}

\subsubsection{Ramp rate dependence}\label{sec:ramprate}

\begin{figure}
	\centering
	\includegraphics[width=0.475\textwidth]{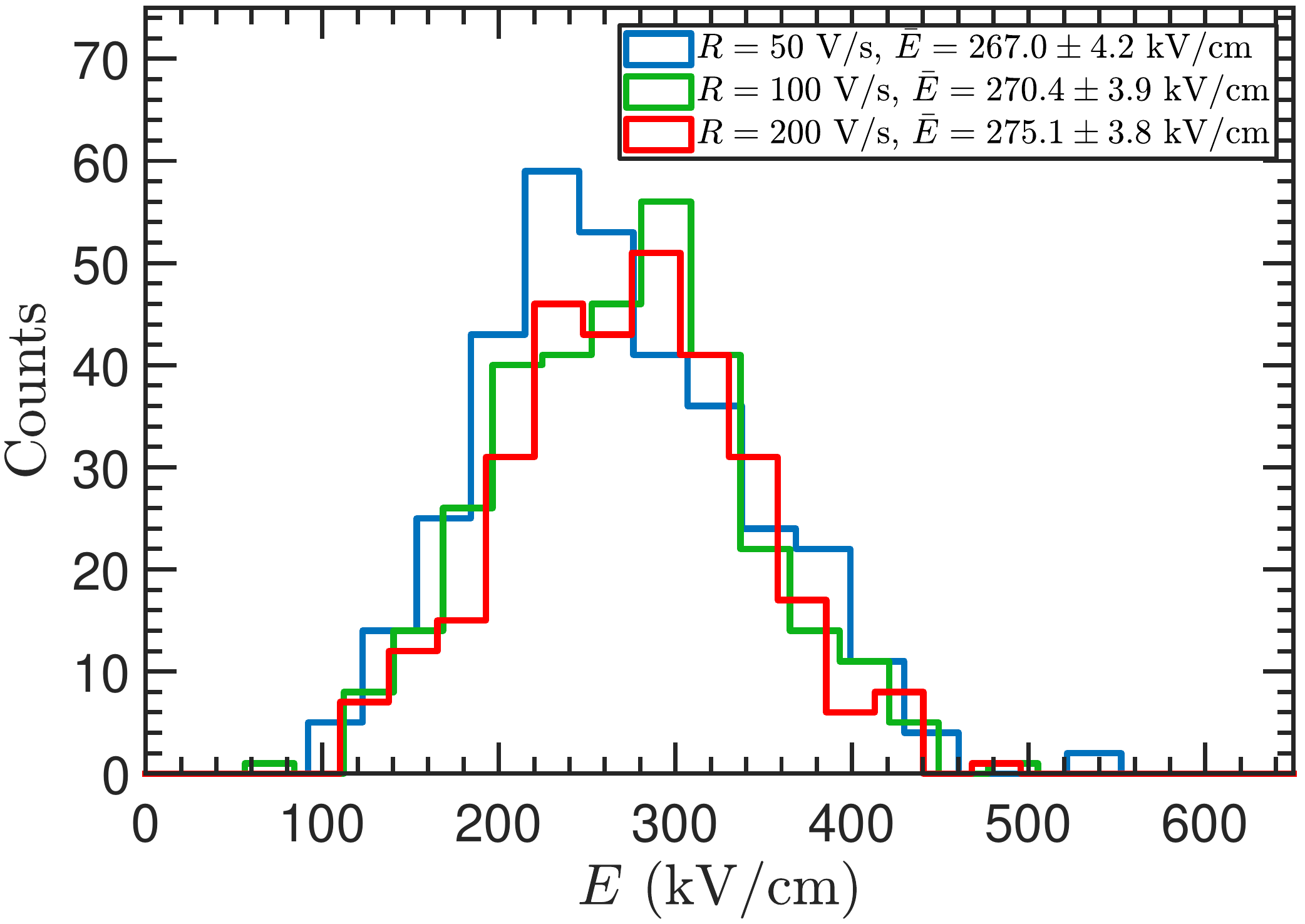}
	\caption{Distribution of breakdown fields for three different voltage ramp rates from SSHV dataset MP:53, showing the absence of any statistically significant differences due to a variable ramp rate.}
	\label{fig:ramp_rates}
\end{figure}

It is natural to ask if there are time scales involved in DC electrical breakdown and if so, what they are. For example, Ref.~\onlinecite{KUP02} considers a probability density function $\mu(E)$ of breakdown initiation per unit time interval per unit area on an electrode surface. Introducing such a quantity to attempt to describe the statistical aspect of breakdown data is in fact quite reasonable if the electrode surface properties are completely identical everywhere on the electrode surface. In this case, the inception probability becomes a function of the electric field strength only and the probability of breakdown inception on a small surface element is independent of what happened previously and what happens in other parts of the electrode. If this is the case, and if $\mu(E)$ is a smooth function of $E$, then the DC breakdown field distribution should depend on the ramp rate. (For example, if we increase the ramp rate, we can get to higher voltage before breakdown occurs.) Motivated by this, we measured the breakdown field distribution for different ramp rates. We chose ramp rates of 50 V/s, 100 V/s, and 200 V/s. The results are shown in Fig.~\ref{fig:ramp_rates}.

The data do not show a clear ramp rate dependence. Rather, the mean breakdown voltages of the three data sets are consistent with each other within the uncertainties of the measurements.


\subsubsection{Distribution of time to breakdown}\label{sec:timedistr}
\begin{figure}
	\centering
	\includegraphics[width=0.49\textwidth]{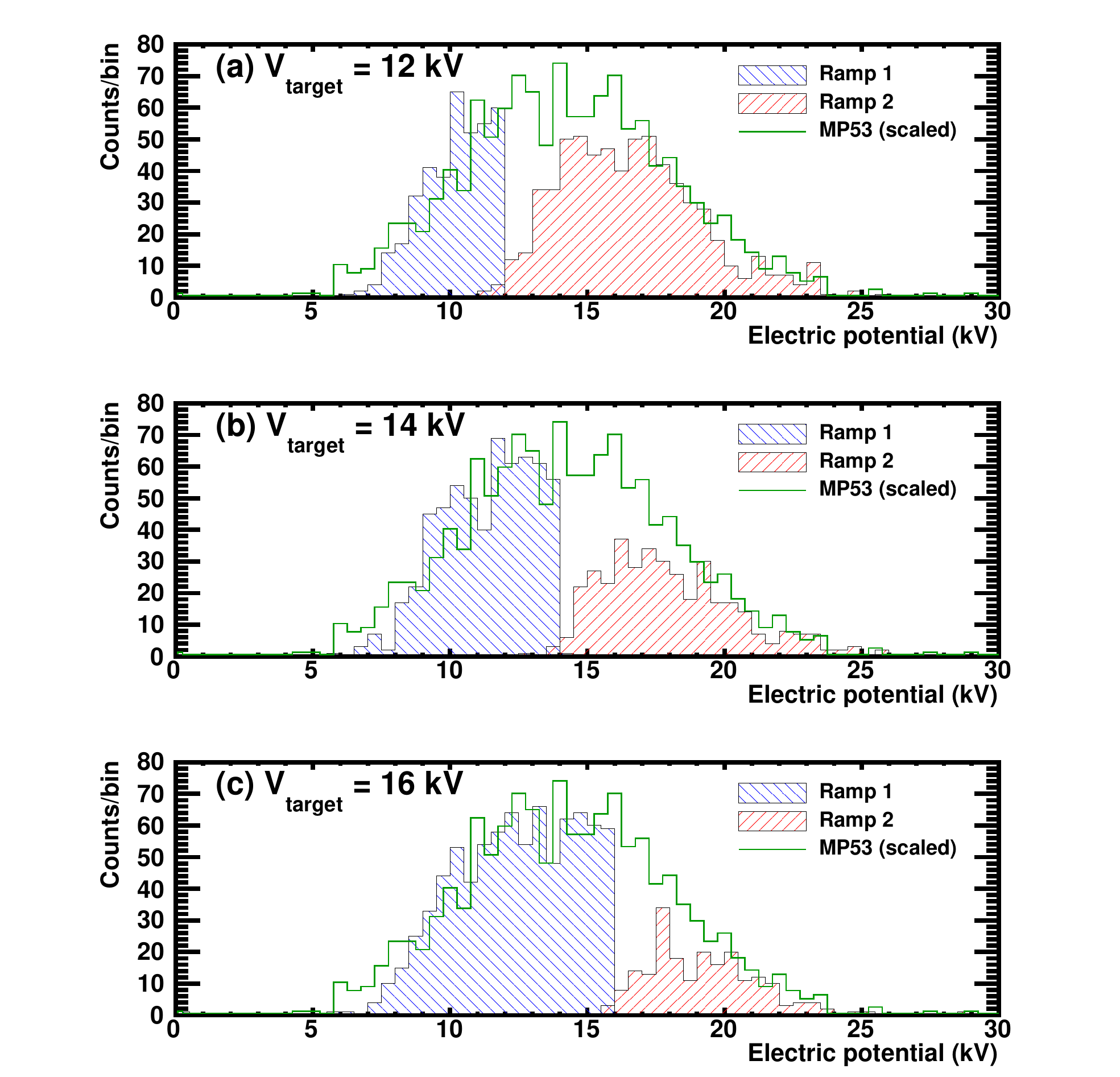}
	\caption{Distribution of breakdown fields for events in which breakdown occurred during the 1st ramp before reaching the preset target value (blue) and for events in which breakdown occurred during the 2nd ramp (red). The three panels correspond to three target voltages: (a)~12~kV, (b)~14~kV, and (c)~16~kV.}
	\label{fig:holding_time1}
\end{figure}

\begin{figure}
	\centering
	\includegraphics[width=0.49\textwidth]{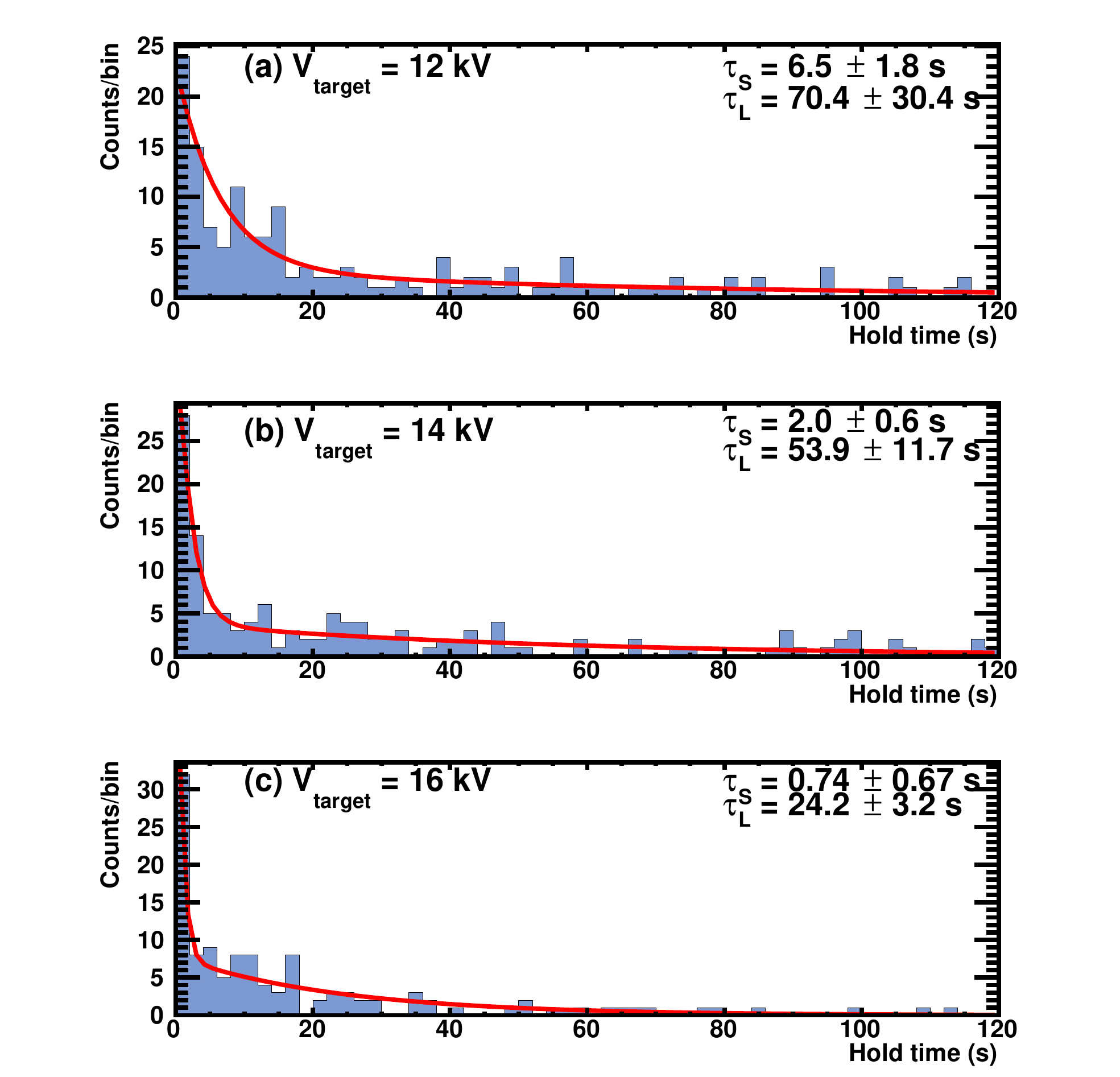}
	\caption{Distribution of time between setting the field to be a preset value and the occurrence of breakdown. The three panels correspond to three target voltages: (a)~12~kV, (b)~14~kV, and (c)~16~kV. A single exponential function gives a poor fit to the data. A sum of two exponential functions fit the distribution much better. The obtained two time constants are shown in each plot.  }
	\label{fig:holding_time2}
\end{figure}

In order to further address the question of the time scales are involved in DC breakdown, we performed a measurement of the time between applying an electric field to be a certain predetermined value and the occurrence of breakdown. We followed the following procedure for this measurement:
\begin{enumerate}
    \item Ramp the voltage at a constant rate towards a preset target value (first ramp).\label{step:ramp1}
    \item If a breakdown occurs before reaching the target value, then record the voltage at which the breakdown occurred and go back to step~\ref{step:ramp1}. The distribution of the breakdown voltage for these events are indicated in blue in Fig.~\ref{fig:holding_time1}.
    \item If the target voltage is reached without a breakdown, then hold the voltage there until a breakdown occurs or until a predetermined amount of time (2 min) elapses.\label{step:hold} If a breakdown occurs, then the time between the target voltage reached and the occurrence of breakdown is recorded. The distribution of this time is plotted in Fig.~\ref{fig:holding_time2}.
    \item If a breakdown does not occur in Step~\ref{step:hold}, then ramp the voltage down to zero, and ramp up again until a breakdown occurs (2nd ramp). Record the voltage at which breakdown occurs. The distribution of the breakdown voltage for these events are indicated in red in Fig.~\ref{fig:holding_time1}.\label{step:ramp2}
    \item Repeat these steps.
\end{enumerate}
We performed such a measurement for target voltages of 12, 14, and 16 kV, corresponding to electric field strengths of 273, 318, and 364 kV/cm. These data correspond to datasets MP:55, MP56, MP:57, and MP:58. The three target voltages were randomly interspersed in these data sets to avoid the possibility of a ``conditioning" effect causing a trend that would depend on the target voltage. In Fig.~\ref{fig:holding_time1}, a scaled breakdown distribution from dataset MP:53, taken under a similar condition, is superimposed. 

We make the following observations:

(1)~When a breakdown occurs in the first ramp, the breakdown voltage distribution below the target voltage agrees with that from MP:53 (scaled).

(2)~Once the target voltage is reached, in a small fraction of cases a breakdown follows with a relatively small amount of time ($\sim$20~s). In the majority of cases, a breakdown is not observed within the observation time.

(3)~In the 2nd ramp, which happens for cases where a breakdown is not observed during the hold (step~\ref{step:hold}), a breakdown is not observed below the target voltage except for a few cases. The breakdown distribution of the 2nd ramp agrees with that from dataset MP:53 in the high end of the distribution. The discrepancy between the second ramp and dataset MP:53 in the part above the target voltage can be attributed to the events in which a breakdown is observed during the hold (step~\ref{step:hold}).

As shown in Sec.~\ref{sec:correlations}, the breakdown voltages among successive ramps have little or no correlations for the majority of the datasets. And yet, the results presented here appear to indicate that the outcome of a ramp can depend on the outcome of the previous one, for which there is no breakdown observed, in that the breakdown voltage in the 2nd ramp is above the voltage up to which a breakdown was not observed in the 1st ramp.  It is not entirely clear what these observations mean to the underlying mechanism of electrical breakdown. This is a subject of future studies along with the time scales involved in the phenomenon.


\subsubsection{Correlation and conditioning}\label{sec:correlations}

\begin{figure}
	\centering
	\includegraphics[width=0.49\textwidth]{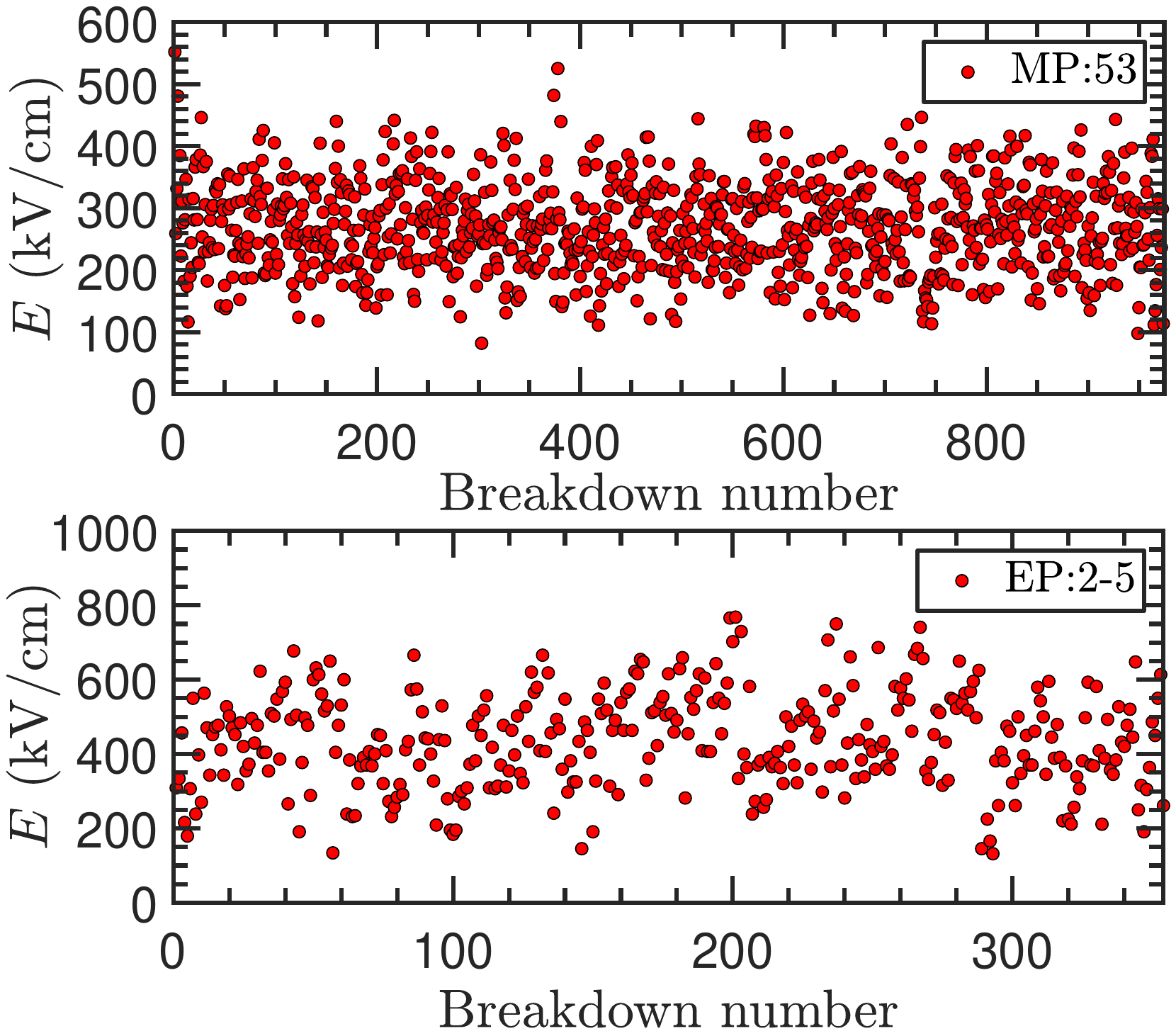}
	\caption{Breakdown sequences for SSHV datasets MP:53 and EP:2-5.}
	\label{fig:breakdown_seq}
\end{figure}

In Fig.~\ref{fig:breakdown_seq} the breakdown sequences for SSHV datasets MP:53 and EP:2-5 are shown.  These two datasets will be the focus of more detailed statistical analysis and discussion in Sec.~\ref{sec:analysis_discussion}.  Importantly, such an analysis is made much more tractable when the data meet certain conditions.  One such condition is the stationarity of the data, which requires the mean, variance, and autocorrelation to be constant over time.  The time series plot in Fig.~\ref{fig:breakdown_seq} show that both these datasets have constant mean and variance.  For instance, the mean of the first 100(50) breakdowns differ only by a few percentages from the mean of the last 100(50) breakdowns for dataset MP:53(EP:2-5), and this difference is within the statistical uncertainty.

Correlation tests can be performed on each of dataset in Table~\ref{tab:sshvdatasets}.  Many statistical tests exists for such a task, including two relatively simple ones in the Durbin-Watson test\cite{DWtest} and the Ljung-Box Q-test\cite{LBtest}.  These correlation tests do have some difficulties.  In particular, both tests assess the residuals of the breakdown field time series for correlation, and so the results have a strong dependence on the regression model used in the analysis. Furthermore, any temperature or pressure drifts inside the experimental volume can complicate the analysis. The result of applying those tests shows that about half of the datasets exhibit no correlations, a quarter are statistically inconclusive, and the remaining quarter display some correlations. For instance, the tests hint at possible correlations for datasets MP:53 and MP:54 but no correlations for other datasets (e.g., MP:31, MP:34, MP:35, MP:41, MP:49) acquired under the same conditions. The consensus result of all the datasets does not show a definitive presence of correlations, and the reason for the lack of complete consistency across datasets is not clear.  However, the stability of operating conditions, particularly the pressure (see Sec.~\ref{sec:pressure and temperature dependence} ), during the acquisition of each dataset could be playing a significant role in determining the presence or absence of correlations.  
 
 Conditioning effects on electrodes are observed in vacuum systems as well as those containing dielectric media.  The presence and magnitude of the effect depend on both the intrinsic properties of the electrode, including its composition and surface roughness, as well as extrinsic factors such as the capacitance and stored energy inside the system, that is ultimately released during a breakdown. For the datasets shown in Fig.~\ref{fig:breakdown_seq}, no signs of conditioning effects are seen up to nearly 1000 breakdown sequences for datasets MP:53 and over 350 breakdown sequences for dataset EP:2-5. Based on the results described above, we assume stationarity of these datasets for the analysis presented in this paper.
 
 Finally, the measurements in Sec.~\ref{sec:timedistr} together with the findings in this section suggest the possibility that breakdown in LHe may be described by a Markov chain process, in which the future breakdown is determined only by the present state of the electrode, and not by previous breakdowns.  However, a better understanding of the effects of conditioning and how changes to the electrode surface condition resulting from a breakdown or an applied stress over a time period will require more dedicated measurements.

\subsubsection{Pressure and temperature dependence}\label{sec:pressure and temperature dependence} 

The pressure and temperature dependence of breakdown in LHe have been previously explored by other experimenters.  Meats\cite{MEATS72} investigated breakdown for temperatures above 4.2~K and pressures up to 7500~Torr and showed a clear dependence of breakdown field on the applied pressure.  Similarly, Chigusa \emph{et al.} \cite{Chigusa1999} observed a significant increase in the breakdown voltage corresponding to an increased pressure between 750 to 1500~Torr for both uniform and non-uniform field configurations.  With sphere-sphere electrodes, Gerhold\cite{Gerhold1989} measured a rise in the breakdown field strength with pressure above one atmosphere.  The same trend is observed by Burnier \emph{et al.}\cite{Burnier1970} who also utilized sphere-sphere electrodes and performed measurements above one atmosphere.  

For measurements conducted at pressures below an atmosphere, Hara \emph{et al.}\cite{Hara1993} and Davidson\cite{Davidson2011} showed that the trend persists, with a reduction in breakdown strength with decreasing pressure.  A pressure dependence is also seen for measurements made with AC power by Fallou \emph{et al.}\cite{Fallou1970} and pulsed power by Yoshino \emph{et al.}\cite{Yoshino1982}.  Altogether, these observations of the pressure dependence may be interpreted as related to gas discharge within a vapour bubble, and we will revisit this point in Sec.~\ref{sec:discuss_temp_pressure_dependence}.

\begin{figure*}[]
	\captionsetup[subfigure]{justification=centering}
	\centering
	\begin{subfigure}[]{0.475\textwidth}
		\includegraphics[width=\textwidth]{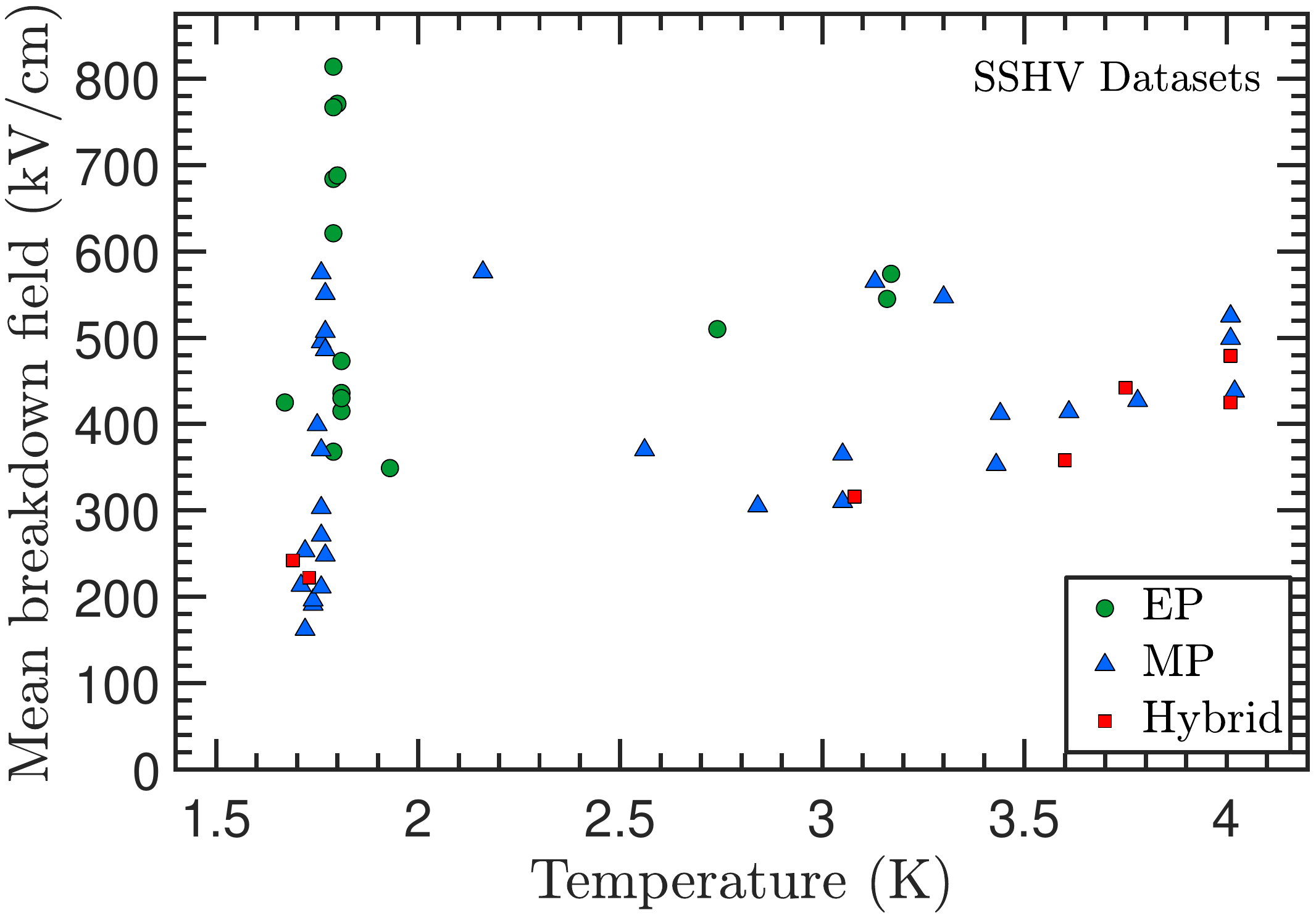}
		\caption{Temperature dependence}
		\label{fig:sshv_temperature_dependence}
	\end{subfigure}
	\hfill
	\begin{subfigure}[]{0.485\textwidth}
		\includegraphics[width=\textwidth]{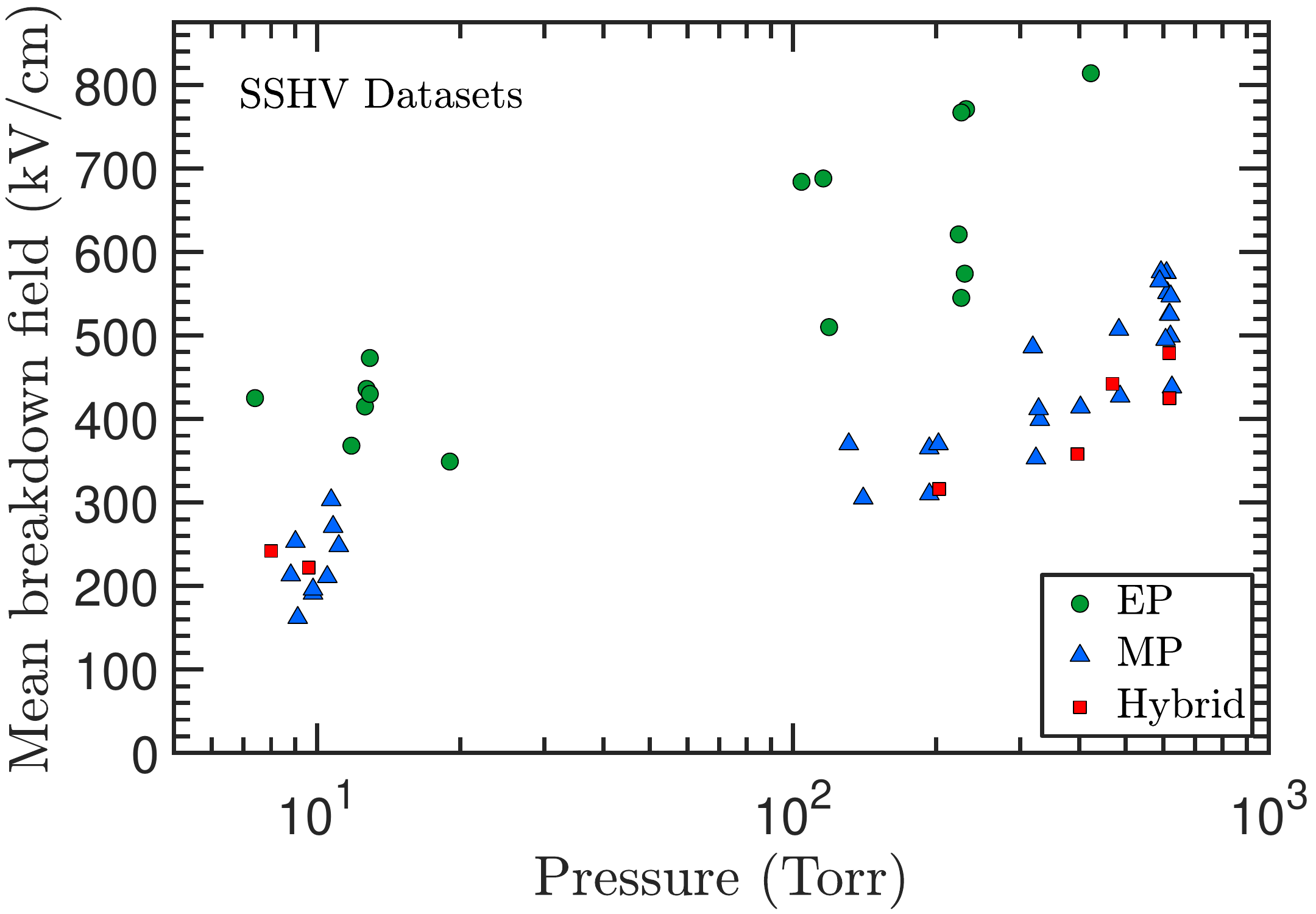}
		\caption{Pressure dependence}
		\label{fig:sshv_pressure_dependence}
	\end{subfigure}
	\caption{The mean breakdown field of SSHV datasets as a function of (a) the temperature and (b) the pressure.  Data were acquired at both SVP and with pressurization to allow for separation of the temperature and pressure dependences.}
	\label{fig:sshv_temperature_pressure_dependence}
\end{figure*}

\begin{figure*}[]
	\captionsetup[subfigure]{justification=centering}
	\centering
	\begin{subfigure}[]{0.475\textwidth}
		\includegraphics[width=\textwidth]{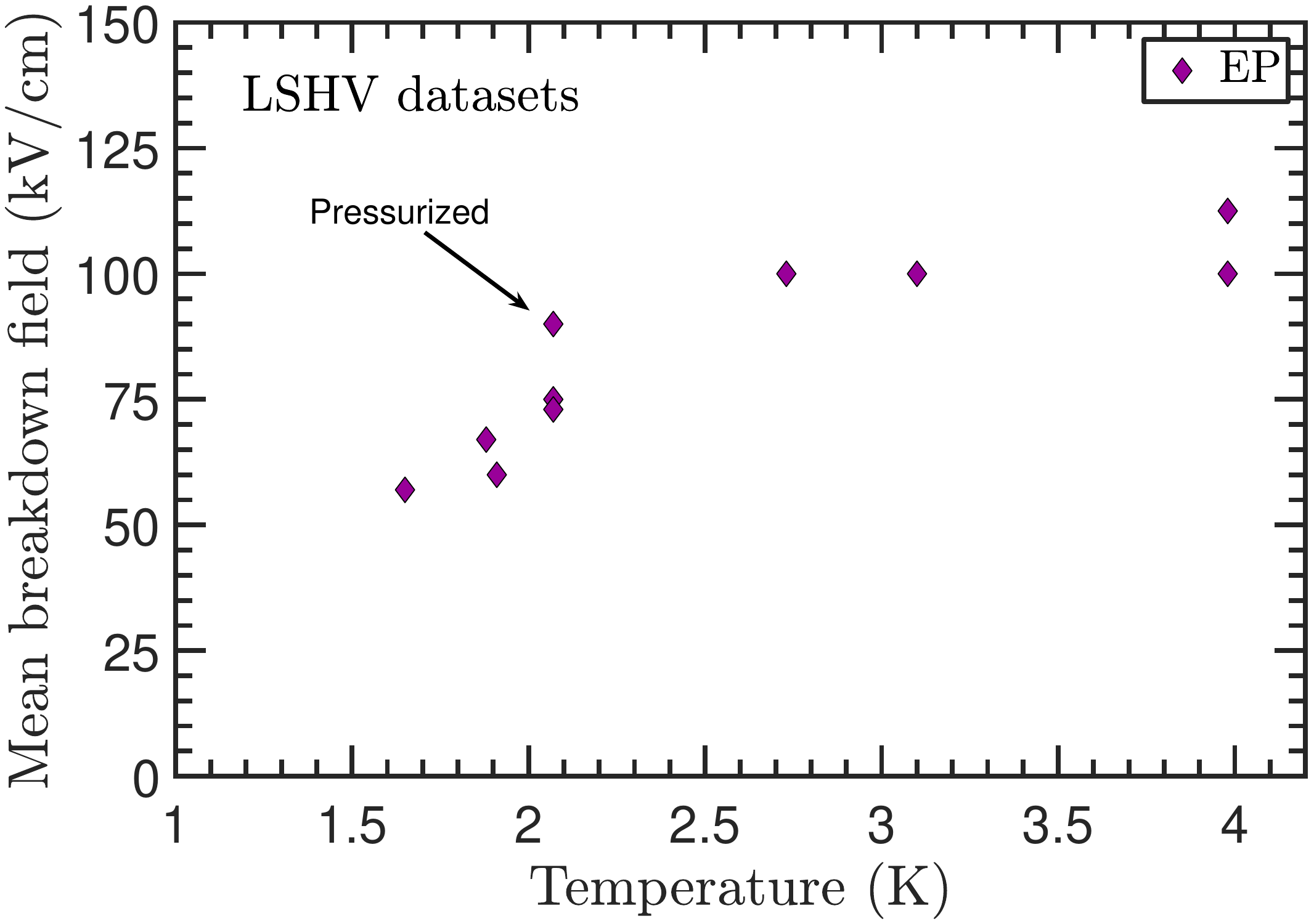}
		\caption{Temperature dependence}
		\label{fig:lshv_temperature_dependence}
	\end{subfigure}
	\hfill
	\begin{subfigure}[]{0.49\textwidth}
		\includegraphics[width=\textwidth]{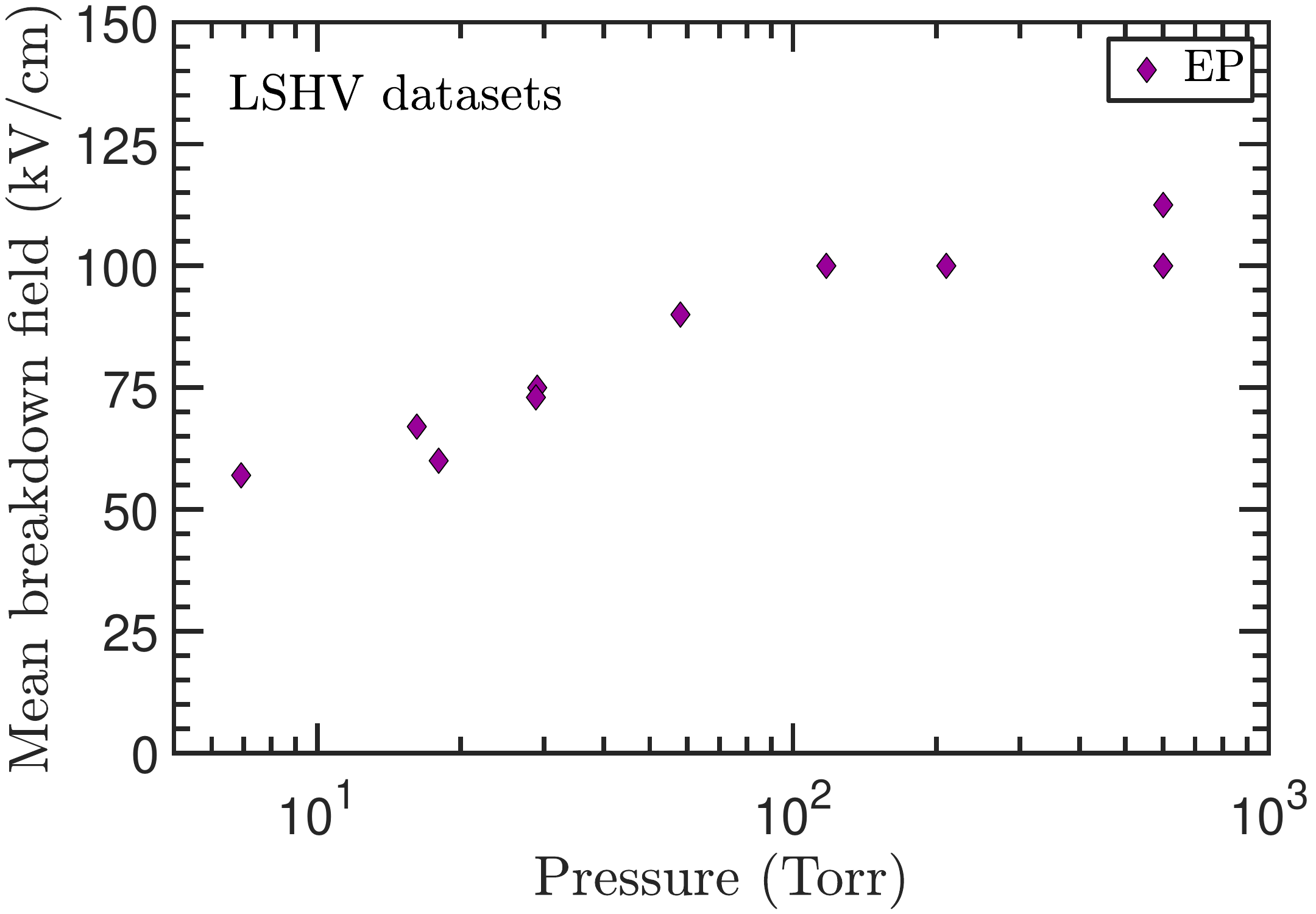}
		\caption{Pressure dependence}
		\label{fig:lshv_pressure_dependence}
	\end{subfigure}
	\caption{The mean breakdown field of LSHV datasets as a function of (a) the temperature (b) the pressure.  All data were acquired at SVP with the exception of the single point at 2.07 K.}
	\label{fig:lshv_temperature_pressure_dependence}
\end{figure*}

At 1.7~K and between $\sim$10 and 600~Torr, we observed a factor of approximately 2.5 increase in the mean breakdown field for the MP datasets as shown in Fig.~\ref{fig:sshv_temperature_pressure_dependence}.  These results are in close agreement with those of Davidson\cite{Davidson2011} acquired at the same temperature and pressure range.  For the SVP data, the factor is slightly lower at approximately 2.1 and 1.9 for the SSHV and LSHV data, respectively (Figs.\ref{fig:sshv_temperature_pressure_dependence} and \ref{fig:lshv_temperature_pressure_dependence}).  This may be due to a small temperature dependence but could also be indicative of the effects of boiling liquid at 4~K which is known to reduce the breakdown strength.

\subsubsection{Electrode surface dependence} \label{sec:surfacedepend} 

The dependence of the breakdown field on the electrode surface finish is evident from Fig.~\ref{fig:sshv_pressure_dependence}.  Here, the mean breakdown field at a given pressure is about a factor of two higher for the electropolished datasets as compared to the mechanically-polished datasets.  For the mechanically-polished electrodes utilized in our measurements, the surface was polished using a lapping film with average particle size in the range of $1-10$~$\mu$m.  The electropolished electrode surfaces should have much smaller feature sizes, however, the surface profile was not measured. 

It must be noted that attributing the difference between these two types of electrodes solely to surface roughness is not necessarily accurate.  Rather, it is more proper to characterize the surface by its ``condition" rather than just its topographical roughness.  To qualify the meaning of the term ``surface condition", we utilize the definition proposed by May and Krauth\cite{MAY81}, in which surface condition is characterized by other aspects in addition to surface roughness such as the electrode material, oxidation layer (thickness and composition), and the presence of microcracks.  The oxidation layer can depend on a variety of factors, including the exact chemical composition of the electrode, the chemical treatment the surface has undergone, the amount of time and environment the electrode has been exposed for oxidation, etc.  Furthermore, various processes such as machining, stress-corrosion from exposure to the ambient atmosphere, and local temperature gradients created during the cooling process can generate microcracks on the electrode surface.  Together, all of these factors can complicate any comparison between the surfaces of different electrodes.

The mechanical polishing process, beyond the likelihood of embedding abrasives during the polishing process, can lead to changes in the material properties such as a local increase in the mechanical strength of the surface.  The electropolishing process, even more so than mechanical polishing, can cause changes to surface composition.  In this chemical surface finishing technique, metal is electrolytically removed and an undisturbed, metallurgical clean surface is exposed.  The metals that make up the stainless steel alloy are removed at different rates during the process.  Chromium atoms are removed at a much lower rate than nickel and iron, resulting in a surface rich in chromium.  When exposed to oxygen, a very thin passive oxide layer is formed on the surface, but this thin layer prevents further diffusion of oxygen into the base material and thus prevents corrosion.  These surface composition changes can potentially lead to a corresponding change in the surface work function, which can ultimately affect charge emission from the surface.  Depending on the nature of the change to the work function, an enhancement or suppression of the breakdown initiating process is possible.

In this regard, it is worth noting that the work function for typical metals tend to be in the range of $4.0 - 5.0$~eV \cite{CRCHandbook14} whereas the difference in work function between a metal and its oxide can be as high as 2.0~eV\cite{VIJH76}.  Moreover, just as the local microscopic field on the surface of the electrode can be significantly different than that of the applied macroscopic field as a result of micro-protrusions that vary in both size and shape on the surface, the work function is not necessarily uniform and constant across the electrode surface or with the externally applied field\cite{Qui2015}.  For fields of $10^8$ V/cm or greater, work function changes above 0.5~eV are possible\cite{Tsong1969}.


\subsubsection{Scale dependence} \label{sec:scale dependence} 

It is well known that electrical breakdown in dielectric liquids has a dependence on the size of the electrodes~\cite{Weber1956, Gerhold1994, Gerhold1998, Acciarri2014, Auger2016, Tvrznikova2019}.  This dependence is often referred to as the ``area-effect" and the ``volume-effect" in which the breakdown field decreases with increasing electrode area and volume of the liquid under high electric fields.  For uniform field electrode geometries and constant electrode areas, the latter can be viewed as an effect due to the inter-electrode spacing or the so-called ``gap size" effect.  As such, the electrode geometry is an important consideration when endeavoring to independently study these effects in a systematic manner.  For example, in a sphere-plane or sphere-sphere electrode geometry, both the surface area of the electrodes and the volume of the medium subjected to high fields change when the distance between the two electrodes changes.  Consequently, attributing the contribution of each effect to the observed breakdown behavior can be challenging. 

This issue has lead to some inconsistencies in published results regarding the existence of a gap size effect for electrical breakdown in LHe.  Several previous studies\cite{MATHES67,GALAND68,LEHMANN70,MEATS72,MAY81,MEYERHOFF95} have shown that the breakdown voltage increases approximately linearly with gap size, whereas the results from others\cite{FALLOU69,SCHWENTERLY74} have suggested a possible square-root law as found in vacuum breakdown.  At the same time, it has also been suggested that whenever spherical electrodes are used, a hidden area effect could emerge\cite{Sharbaugh1955, Gallagher75}.  Hence, for such scaling studies, the area of the electrode subjected to high fields must be carefully quantified, particularly in the case of spherical and point geometries.

But even in the situation of a uniform field geometry where the volume can be changed independently of the area, the definition of the ``stressed" or ``active" area as the region in which a breakdown can occur is not necessarily straight-forward given that breakdown is initiated over a range of electric fields and is not solely restricted to the highest field region.  Typically in the literature, an ``effective" stressed area or volume is defined as the region where the electric field intensity is greater than 90\%, or some other percentage, of the maximum value.  However, the width of a breakdown field distribution at a given pressure can be quite broad as shown in Fig.~\ref{fig:hist_datasets}.  For instance, in SSHV dataset EP:02 acquired with a electropolished electrode, the ratio between the highest and lowest breakdown field is about a factor of eight.  This would imply that when a breakdown occurs at the maximum of the distribution, the area of the electrode with a field higher than the minimum of the distribution is larger by some amount than when a breakdown occurs at lower fields.  The stressed area, hence, is a function of the applied field, and the effect would be most pronounced for geometries involving spherical or point electrodes.  Given that the probability for breakdown is field dependent, the contribution of this additional area is not purely geometric.  The shape of the electrodes will play a critical role in determining the importance of this effect, with the implication being that the shape of the breakdown field distribution may be influenced by the geometry of the electrodes.

In an effort to account for this effect, some investigators\cite{GOSHIMA95} have defined a statistical stressed electrode area/volume when the breakdown field distribution is described by a Weibull\cite{Weibull51} distribution.  But even this definition is not entirely general considering that the shape of the breakdown distribution is not known a priori.  Rather, it is more appropriate to define the statistical stressed area of the electrode in reference to the empirically acquired breakdown field distribution.  This implies that a proper quantification of the stressed area necessitates a sufficiently well-measured breakdown distribution.  From this consideration, the stressed area should be viewed as a measured quantity rather than a calculated geometric value.  Therefore, our aim is to point out that caution is warranted when studying the scaling behavior and, in particular, when comparing and interpreting data from different experiments possessing dissimilar electrode geometries.  

For near uniform-field electrodes, the stressed area may be determined by examining the distribution of craters on the electrode surface created by the breakdowns.  An analysis of the spatial distribution of craters created by breakdowns on the surface of a mechanically-polished electrode used in our measurements finds that nearly all of the craters are contained within a circular region of $\sim$5~mm radius.  This region gives approximately the same stressed area as a region with $\geq$70\% of the maximum field value.  Therefore, to facilitate comparison of results from other experiments, we will adopt this working definition of the electrode stressed area.  For the SSHV HV and ground electrodes, the stressed areas are 0.74~cm$^2$ and 0.725~cm$^2$, respectively.  Similarly, for the LSHV electrodes, the stressed areas are 1126~cm$^2$ and 1125~cm$^2$.  

A comparison of the SSHV and LSHV results shows the effect of electrode area scaling on electric breakdown in LHe.  As shown in Figs.\ref{fig:sshv_temperature_pressure_dependence} and \ref{fig:lshv_temperature_pressure_dependence}, the mean breakdown field strength of the LSHV data is about a factor of four lower than that of the SSHV data at the same pressure.  This observation suggests that breakdown is a surface phenomenon that is dependent on the size of the electrode.  The lower breakdown fields measured in the LSHV apparatus, which has an electrode $\sim$1500 times larger than that of the SSHV electrode, is consistent with the trend observed in previous experiments on the scale dependence of breakdown. A more detailed discussion of area scaling will be given in Sec.~\ref{sec:discuss_area_scaling}.  

It should be noted that the electrode spacing in the LSHV apparatus is a about factor of seven times larger than the SSHV apparatus, and the surface conditions of the electrodes employed in the two apparatuses are similar but not necessarily identical. For these reasons, it is not possible to definitively separate out the relative contributions of the area and spacing effects or the surface quality to the observed scaling behavior.  But to the best of our knowledge, there is no conclusive evidence for a spacing effect in uniform field electrodes, and published data indicate that the scaling behavior is dominated by the area effect.  We will discuss a possible explanation for the apparent gap dependence observed by some experimenters in more detail in Sec.~\ref{sec:discuss_gap_dependence}.


\section{Analysis and discussion}\label{sec:analysis_discussion}

\subsection{Statistical approach}\label{sec:discuss_statistical_approach}

In the analysis of experimental data, there are often two approaches that are adopted: a statistical approach and a physical model-based approach.  Each approach has it own merits, and a complete understanding of a problem will necessarily require the union of both approaches.  Nonetheless, when the physical mechanism for the phenomenon under study is not well known, which is indeed the case for electric breakdown in LHe, the statistical approach is more suitable for use in interpretation of the data.  A statistical analysis will necessarily remove us somewhat from the underlying physical mechanism of breakdown, which is the principal interest for experimentalists.  However, it is valuable from the standpoint of identifying patterns and attributes in the data that can assist in constructing models to describe the phenomenon.

The statistical analysis method of choice frequently employed in studies of electrical breakdown is a subset of the field of order statistics known as extreme value statistics.  The most commonly used function in this field is certainly the one originally introduced by Weibull~\cite{WEIBULL51}, which is an empirical distribution function that has been shown to be one of the three allowable forms of the asymptotic extreme value functions \cite{GUMBEL58}.  The second form of these functions most often encountered in applications is called the Gumbel function.  These two functions have been extensively used in studies of breakdown in dielectric media, while a third form, called the Fr\'{e}chet function, is closely related to the Weibull function but differing from it in the range of validity.  The Fr\'{e}chet function, like the Weibull function, has a finite bound whereas the Gumbel function is two-side unbounded.  The choice of which function to use depends on the phenomenon and property being studied.

However, imprudent use of those functions in the analysis of experimental data can have considerable drawbacks.  Because the parent distribution from which the data are drawn cannot be guaranteed to be known a priori, limiting the analysis of the data to using the three extreme value distributions can introduce a bias in the conclusions obtained from such an analysis.  It is clear that any analysis built upon unfounded assumptions at the outset can produce unreliable predictions. But above all, it can be quite detrimental in the quest to further our understanding of the physics behind electrical breakdown.

A statistical analysis of data most likely does not capture all the information contained within it.  For instance, although routinely done, it is generally true that when empirical data are binned there is an inherent loss of information.  This is especially important for small datasets where binning can result in amplified fluctuations and an obscuration of the true shape of the distribution, particularly in the tail regions where the data are sparse.  The loss of information can in some instances be compensated for with knowledge of the underlying distribution, but when the distribution is not known, the loss can be significant. Hence, to retain the maximum amount of information, it is preferable to analyze the data using the cumulative distribution function (cdf) rather than as a binned histogram; the latter being more common in the analysis of experimental data.  

Absent the assumption of a model distribution, the analysis must start with the most general formulation for breakdown statistics. 
Following Choulkov\cite{CHOULKOV05}, for an electrode with surface area, $S_{0}$, composed of small area elements, $dS$, the total number of surface elements on the electrode is $N = S_{0}/dS$.  Let $W$ be the probability density of breakdown initiation on a small element of the electrode, then the probability of breakdown on a single small element is $p = W dS = W S_{0}/N$.  Correspondingly, the probability that a breakdown does not occur (i.e., the survival probability) on the surface element is $1-p$.  According to weakest link theory, the absence of a breakdown on the entire electrode surface depends on the absence of breakdown on all of its elements; that is, no surface element undergoes breakdown.  Thus, the electrode survival function, $P_{s}$, is the product of the probabilities of survival for each surface element,
\begin{eqnarray}
P_{s} = \prod_{i=1}^{N} (1-p) = \left( 1 - p  \right)^{N} =  \left( 1 - S_{0} W/N  \right)^{N},
\label{eq:survprobdef}
\end{eqnarray}
where the second equality follows from the assumption that $p$ does not depend on the surface element (i.e., the uniform case).  In the limit $N \rightarrow \infty$, Eq.~\ref{eq:survprobdef} reduces to 
\begin{eqnarray}
P_{s} = e^{-S_{0}W}.
\label{eq:survprobexp}
\end{eqnarray}
The complement of the survival function is the breakdown cumulative distribution function, \cite{WEIBULL51,CHOULKOV05}
\begin{eqnarray}
P_{b} = 1 - P_{s} = 1 - e^{-S_{0} W}.
\label{eq:bprobuni}
\end{eqnarray}
For the most general case in which $p$ is not uniform but depends on the location of the surface element, the cdf can be shown to be given by
\begin{eqnarray}
P_{b} = 1 - P_{s} = 1 - e^{-\int {W dS}}
\label{eq:bprobgen}
\end{eqnarray}
from substitution of Eq.~\ref{eq:survprobexp} for $p$ in Eq.~\ref{eq:survprobdef} and taking the limit of large $N$.

The quantity of special significance in Eqs.~\ref{eq:bprobuni} and \ref{eq:bprobgen} is $ W $. In general, the cdf can be dependent on a variety of factors such as the electrode surface condition (roughness, work function, micro-particles, surface inclusions, absorbed gas layers, microcracks, etc.), the temperature and pressure of the liquid, and distance between electrodes.  But for a given dataset, these parameters are held fixed, so $W$ is a function of only the applied electric field, $E$.  

In survival analysis, $W$ is often referred to as the hazard function (or failure rate) and can be interpreted as the ``instantaneous rate" of a breakdown(failure) at field $E$, but note that in the strict mathematical sense, the hazard function is not a true probability since it can be greater than 1.  More properly, the hazard function is a conditional probability density that the event has not yet occurred prior to field $E$ and is a non-negative function that uniquely defines the distribution function.  The integral $\int{W dS}$ is known as the cumulative hazard function, and the behavior of the distribution function and its scaling properties are determined by this quantity.  With the appropriate choice for $W$, the cdf can become any one of the three extreme value distributions.  For instance, the three parameter Weibull distribution function\cite{WEIBULL51} is obtained when
\begin{eqnarray}
W(E) \propto \left(\frac{E-E_{T}}{\alpha} \right)^{m},
\label{eq:Weibull_phi}
\end{eqnarray}
where $\alpha > 0 $, $E_T < E$ , and $m > 0$ are known as the scale, location, and the shape parameter, respectively.  If instead, $W$ follows an exponential function,
\begin{eqnarray}
W(E) \propto \exp {\left( \frac{E - \alpha}{\beta}  \right)},
\label{eq:Gumbel_phi}
\end{eqnarray}
the result is the Gumbel distribution function~\cite{GUMBEL58}.

Therefore, the specific form of extreme value distribution appropriate for describing the data depends upon the functional form of the continuous (initial) distribution that the chosen property (e.g., microscopic field intensity) would have if the sample size (electrode area) were infinite. In particular, it depends upon the way in which the probability density decays as the chosen extreme limit is approached.  If this initial distribution density falls off exponentially or more rapidly (e.g., a Gaussian function), then the appropriate form is the Gumbel distribution.  In the case that it falls off as a power law, then the Weibull or Fr\'{e}chet form is the most appropriate form.  However, absent a fundamental reason for breakdown data to follow one of the three extreme value distributions, the initial density can take many other functional forms, and the data should be the final determinant on the appropriate form.  We discuss the method for this determination below.

The cumulative hazard function can be estimated by determining the empirical cumulative distribution function (ecdf), $\hat{P_{b}}$, from the data.  In the absence of censored data, the ecdf can be determined very simply.  But for the case in which the data is censored, several methods exists to address this situation, and the Kaplan-Meier product limit estimate~\cite{KaplanMeier58} and the Nelson-Aalen estimator~\cite{Nelson69, Nelson72, Aalen78} are two of the more versatile methods.  

\begin{figure*}[]
	\captionsetup[subfigure]{justification=centering}
	\centering
	\begin{subfigure}[]{0.49\textwidth}
		\includegraphics[width=\textwidth]{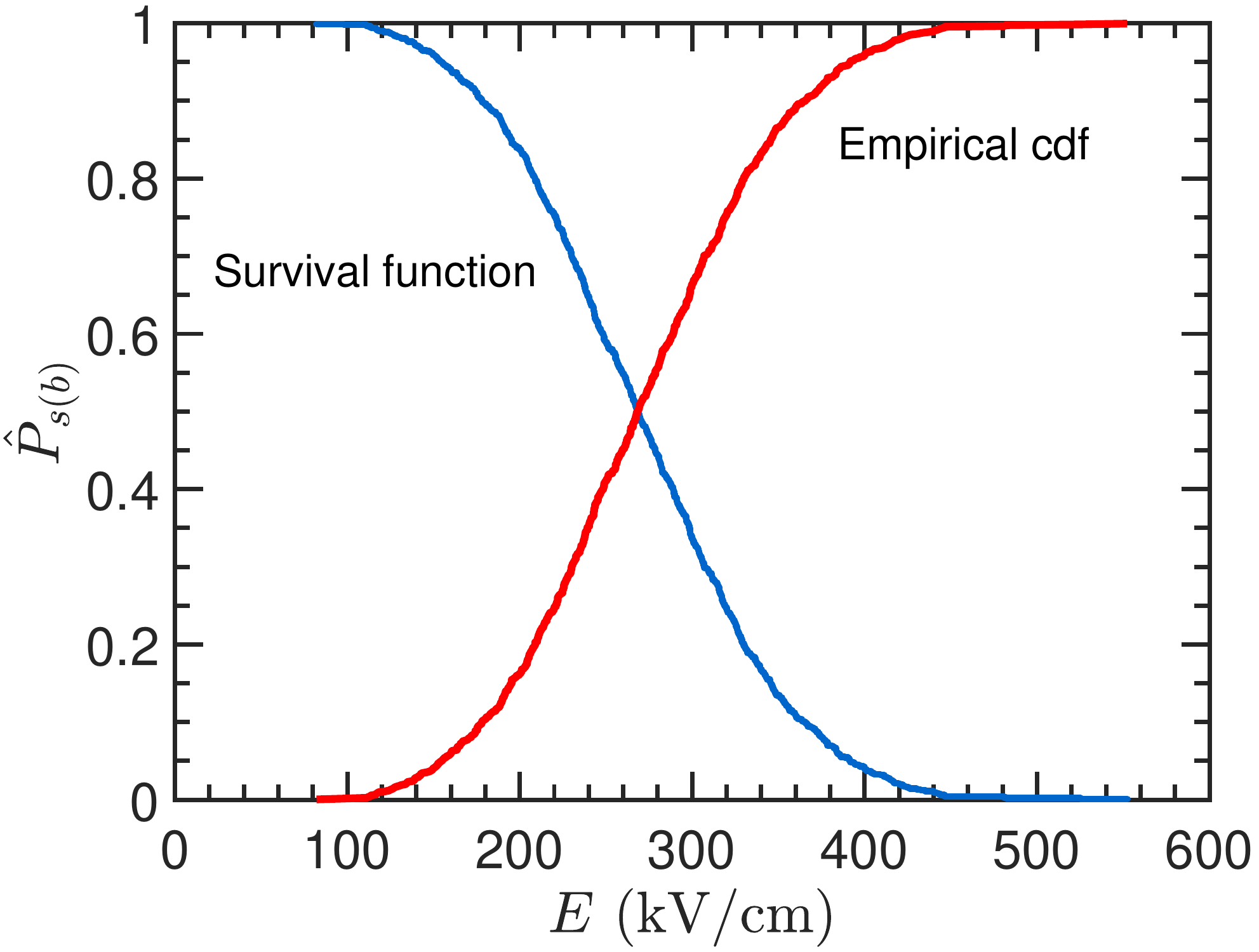}
		\caption{Survival and empirical distribution functions}
		\label{fig:KM_plots53}
	\end{subfigure}
	\hfill
	\begin{subfigure}[]{0.475\textwidth}
		\includegraphics[width=\textwidth]{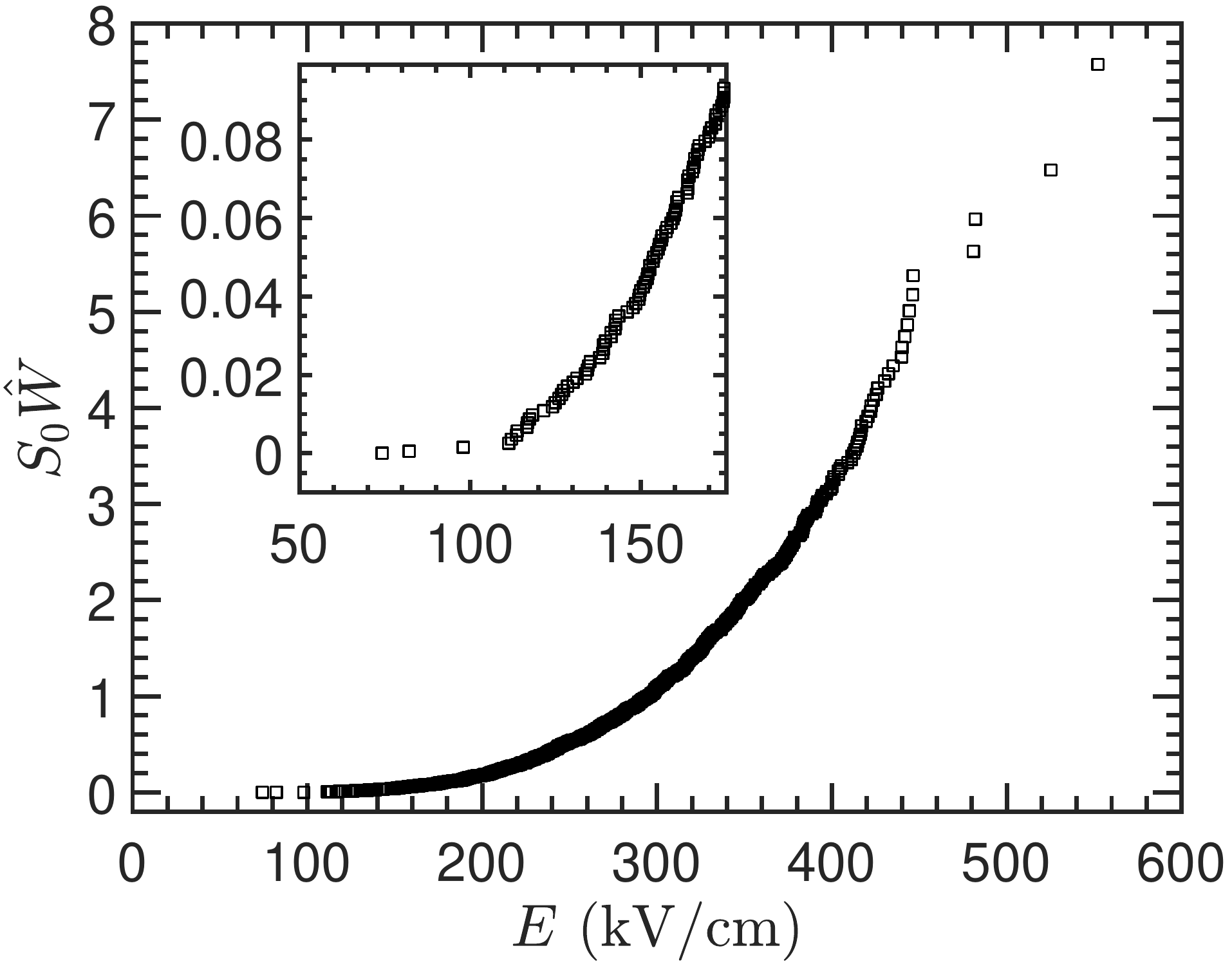}
		\caption{Cumulative hazard function}
		\label{fig:SW53}
	\end{subfigure}
	\caption{(a) The survival and empirical cumulative distribution functions obtained from the Kaplan-Meier product limit estimate for SSHV dataset MP:53.  (b) The estimated cumulative hazard function, $S_{0}\hat{W}$, for the same dataset. Data from all three ramp rates are included in these results. }
	\label{fig:surv_ecdf_hazard_functions}
\end{figure*}

The Kaplan-Meier estimator of the survival function, $\hat{P_{s}}$, is given by
\begin{eqnarray}
\hat{P_{s}}(E) = \prod_{j:E_{j}\leq E} \left( 1- \frac{B_{j}}{N_{j}} \right) ,
\label{eq:KM_sf}
\end{eqnarray}
where $B_{j}$ is the number of breakdowns that occurred at field $E_{j}$ and $N_{j}$ is the number of breakdowns that is yet to occur up to field $E_{j}$.  We take SSHV dataset MP:53 (mechanically polished, 1.76 K, 10.8 Torr), which includes data for all three ramp rates and is the largest dataset taken at one set of conditions, to illustrate the analysis approach.  In Fig.~\ref{fig:KM_plots53}, the survival and empirical cumulative distribution functions derived from this dataset are shown.  Note that the estimator is a step function, hence, only piecewise-continuous.  However, we can define a piecewise linear function with breakpoints at the midpoints of the jumps in the ecdf estimate.  The first and last linear segments must then be extended beyond the data to make the function reach 0 and 1.  After which the estimated cumulative hazard function for a uniform field is determined from Eq.~\ref{eq:bprobuni} to give
\begin{eqnarray}
S_{0}\hat{W} = -\log \left( 1 - \hat{P}_{b} \right).
\label{eq:SW}
\end{eqnarray}

Figure~\ref{fig:SW53} shows the cumulative hazard function obtained from Eq.~\ref{eq:SW}.  Note that $S_{0}$ is the effective stressed area of the SSHV electrode.  We have purposely left the value of this quantity unspecified as it is not necessarily a constant that is independent of the applied electric field.  A more detailed discussion of the surface area was given in Section~\ref{sec:scale dependence}.  Thus far, we have not made any assumptions regarding the form that this function must take, but below we will examine the possibilities that are justified from physical considerations, starting with what the electrode surface condition can reveal about this function.


\subsection{Electrode surface condition}\label{sec:discuss_electrode_surface}

A full characterization of the surface condition of an electrode requires the consideration of many different factors.  However, the one that is the point of focus for most studies on electrical breakdown is the roughness of the electrode surface.  This is not surprising since roughness is a very tangible concept and there is an intuitive notion of its effect on electrical breakdown.  

The roughness of an electrode surface may be characterized by the distribution of surface asperities.  Because each asperity give rise to a local enhancement of the field, it is then natural to consider the distribution of this quantity, $\beta$, which is defined by
\begin{eqnarray}
\beta = \frac{E_{\mu}}{E},
\label{eq:beta}
\end{eqnarray}
where $E_{\mu}$ is the local (i.e., microscopic) field at the emission site and $E$ is the macroscopic applied electric field.  Larger values of the field enhancement factor correspond to asperities on the electrode surface with higher aspect ratios.  

The hazard function, $W$, as previously discussed is a function of the electric field.  But in particular, the field that is of interest is the microscopic field on surface of the electrode rather than the macroscopic applied field.  This interpretation is supported by the results from the mechanically-polished electrode in comparison with those from the electropolished one acquired under the same conditions.  The difference in breakdown field between the two electrode polishes must be a consequence of dissimilar surface conditions.  This then leads to the question of what is the functional form for $W$.

Our qualitative understanding of the physics of breakdown in LHe can be used to provide guidance as to the functional form of the hazard function.  Breakdown is believed to be initiated by electrons tunneling from the cathode electrode at asperities where the field is enhanced. When the heat introduced into the liquid by the tunneling current becomes sufficiently large for a vapor bubble to form at an asperity, breakdown ensues. This physical picture strongly suggests that the hazard function is most appropriately expressed as a function of the tunneling current, $I$, and following Weibull\cite{WEIBULL51}, that it is linearly proportional to the current, $W(I) \propto  I$. We have no measurement of the tunneling current at breakdown, which is estimated be less than than a picoamp, but it can be related to the field through the Fowler-Nordheim equation\cite{FN1928, Good1956} for tunneling of electrons from a metal to vacuum or an insulator.

The current from Fowler-Nordheim field emission, $I$, can be expressed as\cite{ Wang1997}
\begin{IEEEeqnarray}{rCl}
	I & = & A_{e} \frac{1.54 }{\phi}  10^{4.52 \phi^{-1/2}}  (\beta E)^{2}
	\nonumber\\
	&&  \times  \exp \left( \frac{-6.53 \times 10^{4} \phi^{3/2}}{\beta E} \right) \textrm{A}.
 	\label{eq:FN_eq}
\end{IEEEeqnarray}
Here, $A_e$ is the effective emission area in cm$^2$, $\beta$ is the local field enhancement factor, $\phi$ is the electrode work function in eV, and $E$ is the applied electric field in kV/cm.

In fitting the cumulative hazard function with Eq.~\ref{eq:FN_eq}, the value of the parameter $\phi$ is known and can be fixed.  We set $\phi = 5.4$~eV, motivated by two considerations: the work function for stainless steel is about 4.4~eV and the energy of a non-localized electron in LHe is $\sim$ 1.0~eV higher than in vacuum \cite{Broomall1976}.  These factors effectively bring the work function of the surface to 5.4~eV.

\begin{figure*}[]
	\captionsetup[subfigure]{justification=centering}
	\centering
	\begin{subfigure}[]{0.50\textwidth}
		\includegraphics[width=\textwidth]{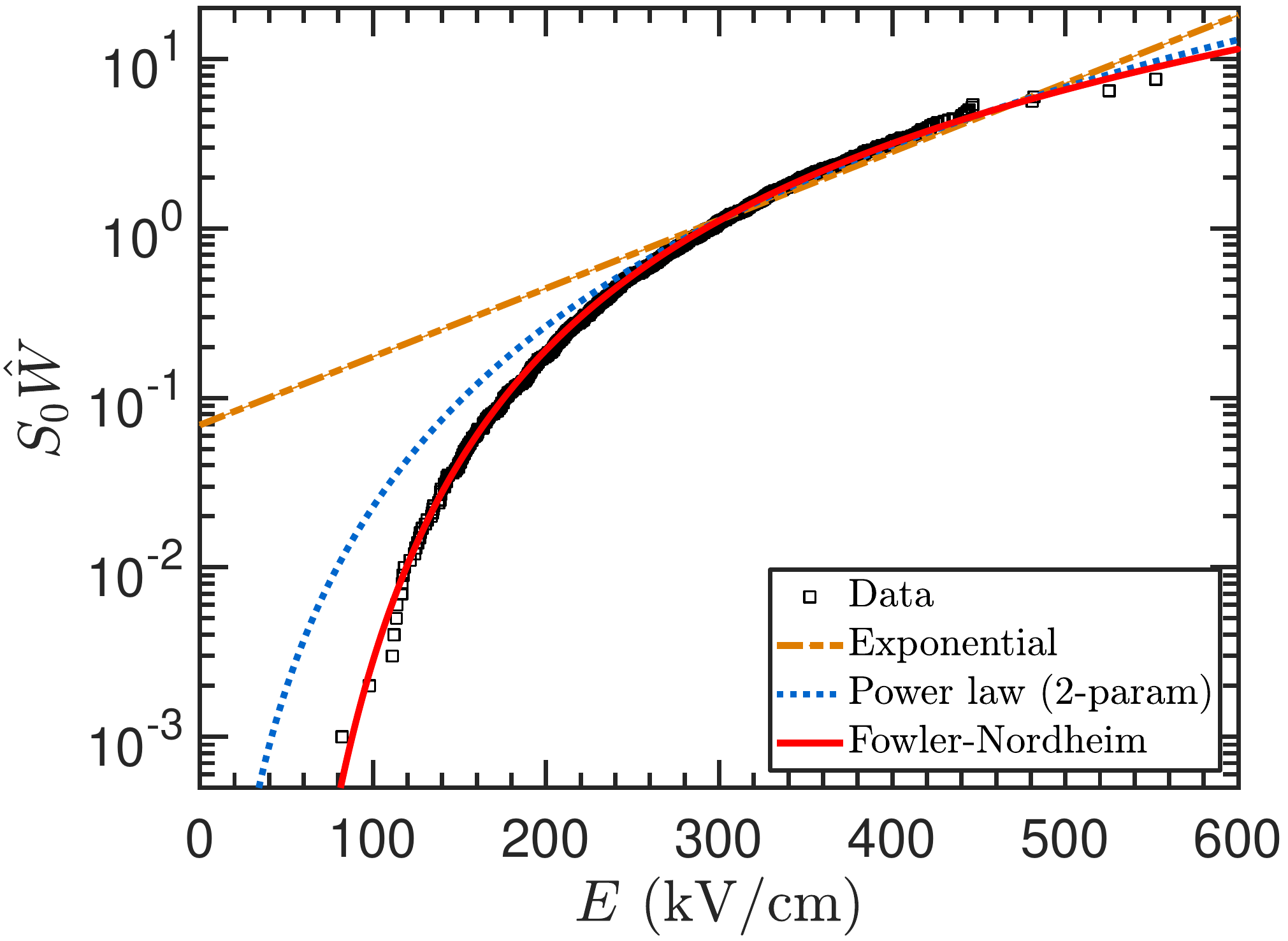}
		\caption{Fits of the cumulative hazard function}
		\label{fig:multiple_fits_WS}
	\end{subfigure}
	\hfill
	\begin{subfigure}[]{0.48\textwidth}
		\includegraphics[width=\textwidth]{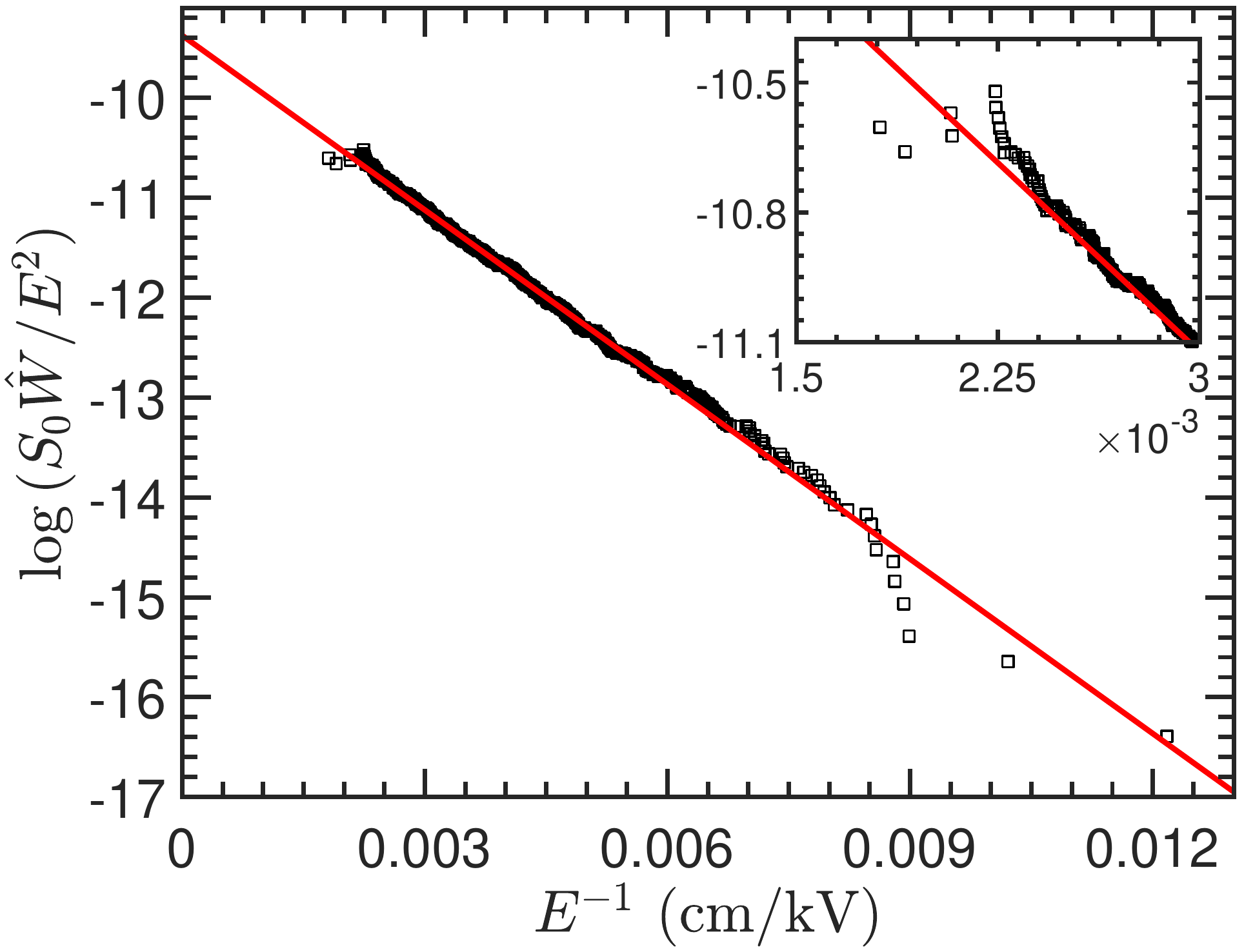}
		\caption{Fowler-Nordheim plot}
		\label{fig:FN_plot_MP53}
	\end{subfigure}
	\hfill
	\begin{subfigure}[]{0.50\textwidth}
		\includegraphics[width=\textwidth]{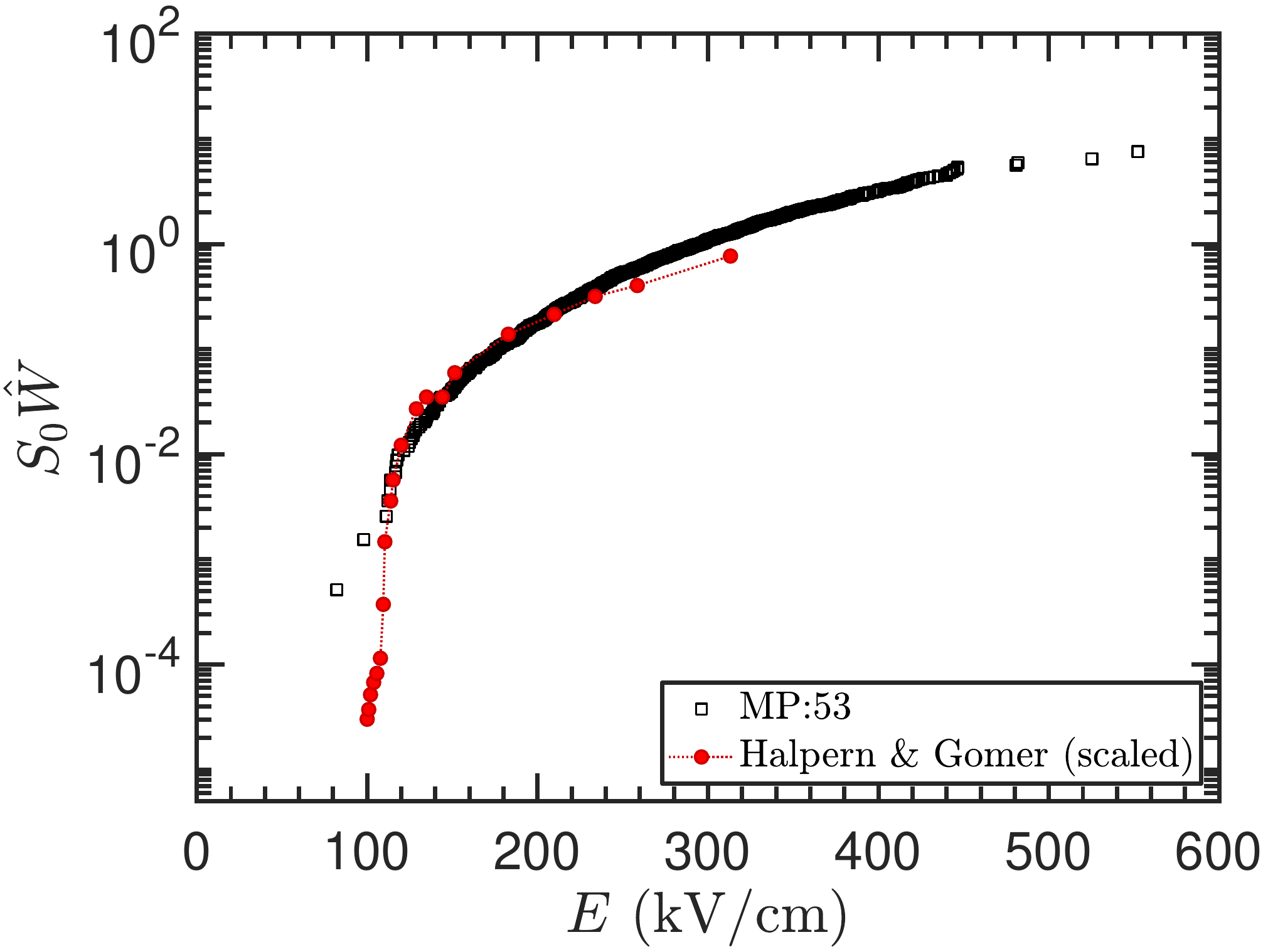}
		\caption{Comparison with field emission in LHe}
		\label{fig:SW_vs_E_HalpernGomer_scaled_current}
	\end{subfigure}
	\hfill
	\begin{subfigure}[]{0.49\textwidth}
		\includegraphics[width=\textwidth]{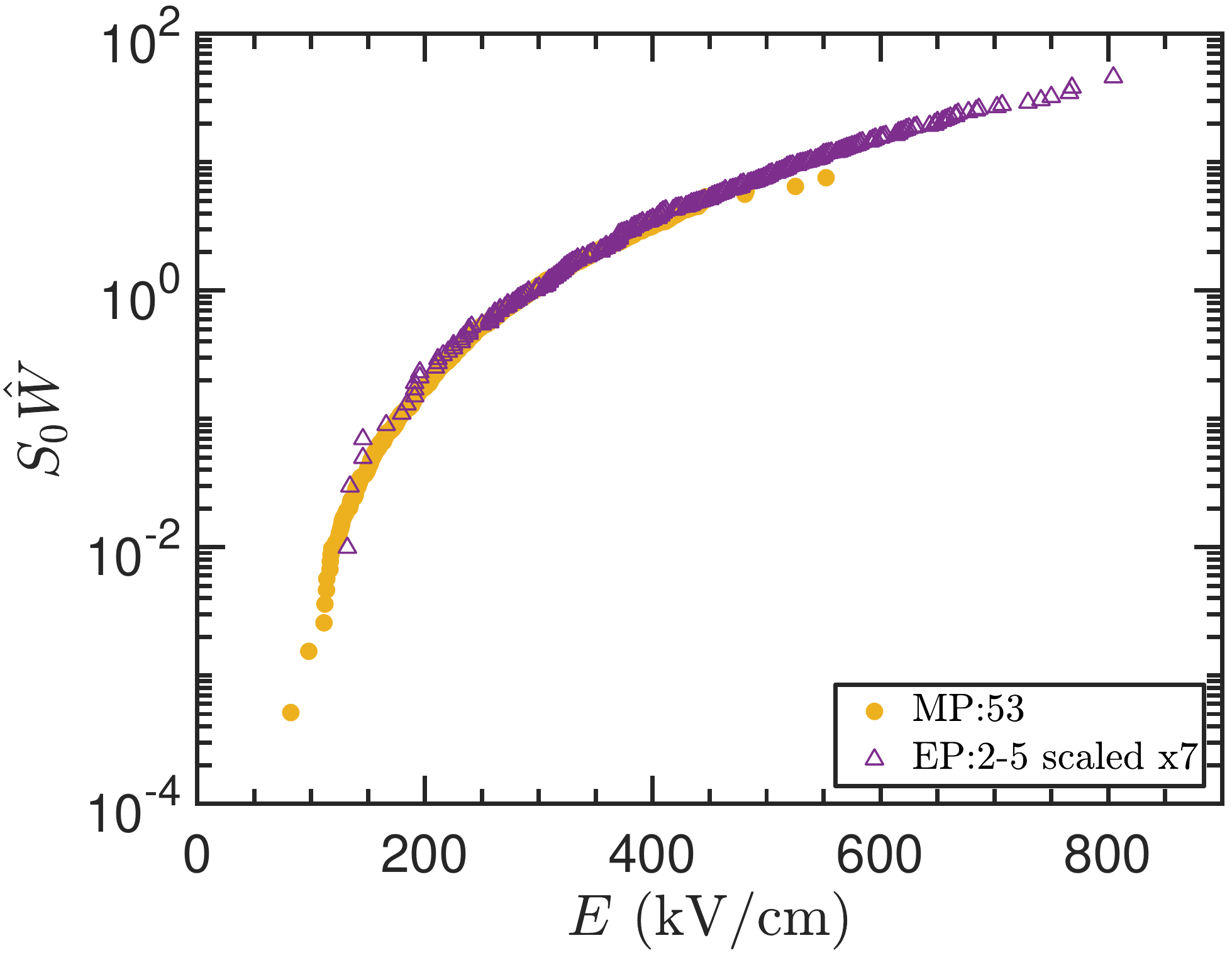}
		\caption{Comparison of electrode surface polish}
		\label{fig:SW_vs_E_scaled_EP_MP}
	\end{subfigure}
	\caption{(a) Multiple functional fits to the estimated cumulative hazard function, $S_{0}\hat{W}$, for dataset MP:53. (b) The Fowler-Nordheim plot of SSHV dataset MP:53 obtained by treating $S_{0}\hat{W}$ as a ``current".  (c) The estimated cumulative hazard function of SSHV dataset MP:53 overlayed with measurements of field emission currents in LHe from Halpern and Gomer~\cite{HalpernGomer1969A} that have been scaled to best match the breakdown data.  Refer to text for more details.  (d) Comparison of the estimated cumulative hazard functions for SSHV mechanically-polished dataset MP:53 and the SSHV electropolished datasets EP:(2-5).}
	\label{fig:SW_plots}
\end{figure*}

The fit of the Fowler-Nordheim equation in Eq.~\ref{eq:FN_eq} to the empirically obtained cumulative hazard function is shown in Fig.~\ref{fig:multiple_fits_WS}.  The goodness of fit of the Fowler-Nordheim equation to data is much greater than either the exponential function or the two parameter power function, and returns fit parameters of $\beta = 1444$ and $A_{e} = 1.6 \times 10^{-12}$~cm$^2$.  The field enhancement factor obtained here is similar to what is found in vacuum field emission from stainless steel electrodes of $\beta \sim 200-1000$\cite{BastaniNejad2015}.  The interpretation of the emission area obtained from the fit is, however, not straightforward due to the proportional relationship between the hazard function and the current\footnote{The current obtained from Eq.~\ref{eq:FN_eq} at the mean breakdown field is the order amps whereas the actual current is below a picoamp. The constant of proportionality must be quite small.} Consequently, any comparison to literature values is at best tenuous, but values of $ \sim 10^{-16} - 10^{-14}$~cm$^2$ have been obtained in a previous experiment\cite{BastaniNejad2015}.

If the quantity, $S_{0}\hat{W}$, is treated as a ``current" and plotted in the Fowler-Nordheim plot as shown in Fig.~\ref{fig:FN_plot_MP53}, the data is observed to follow a linear trend accordant with the behavior for vacuum field emission from metal surfaces\cite{Fursey2005}.  This result and the fit of the data to the Fower-Nordheim equation provide strong support to the hypothesis that field emission is involved in the breakdown initiating process.

Thus far, several pieces of evidence point to the connection between the field emission current and the cumulative hazard function.  This connection is more substantial when we compare our empirical function with measurements of field emission in LHe made by Halpern and Gomer~\cite{HalpernGomer1969A}.  In Fig.~\ref{fig:SW_vs_E_HalpernGomer_scaled_current}, the cumulative hazard function derived from SSHV dataset MP:53 is plotted along with the current($I_{HG}$)~vs.~voltage($V_{HG}$) measurements of Halpern and Gomer~\cite{HalpernGomer1969A}.  Their measurements have been scaled in both variables to best match our data, that is, $I_{HG} \rightarrow qI_{HG}$ and $V_{HG} \rightarrow kV_{HG}$, where $q$ and $k$ are the scaling factors with a value of $5.2 \times 10^{6}$ and $9.3 \times 10^{-2}$, respectively.  The resemblance between the shape of their current vs. voltage measurement curve and the cumulative hazard function further indicates a strong connection between the two.

A comparison can also be made between the cumulative hazard functions for different electrode surface polishes as shown in Fig.~\ref{fig:SW_vs_E_scaled_EP_MP}.  The comparison shows that the mechanically-polished electrode can be viewed approximately as an area-scaled electropolished electrode.  This suggests a better representation of the size parameter is $\sigma_d S_{0}$ with $\sigma_d$ as the density of breakdown initiation sites.  So the relevant size parameter to consider is the number of initiating sites rather than the surface area, and such an interpretation is reasonable from the point of view of surface roughness.  A consequence of this size effect is seen in the difference of the breakdown distribution widths shown in Fig.~\ref{fig:hist_datasets}, but a more detailed discussion of this is given in Sec.~\ref{sec:discuss_area_scaling}.

\begin{figure}[]
	\centering
	\begin{adjustbox}{center}
		\includegraphics[width=\columnwidth]{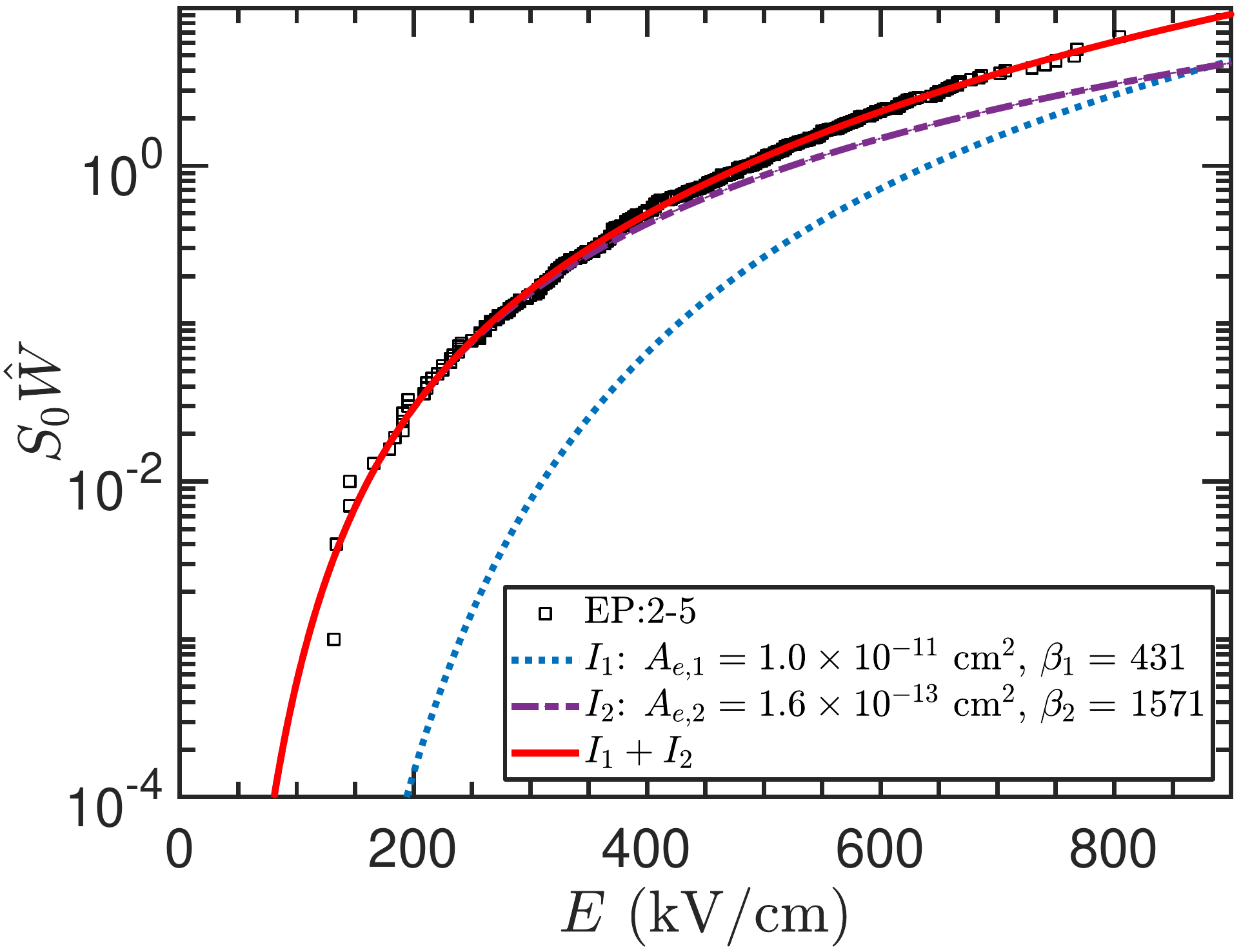}
	\end{adjustbox}
	\caption{A fit of electropolished electrode dataset EP:2-5 to a two term Fowler-Nordheim equation.}
	\label{fig:DFN_fit_EP2-5}
\end{figure}

Although Fig.~\ref{fig:SW_vs_E_scaled_EP_MP} shows considerable similarity in the shape of the cumulative hazard function between the mechanically-polished and electropolished datasets acquired under similar conditions, there are notable differences.  These are apparent from analyzing the fit to dataset EP:2-5 which is shown in Fig.~\ref{fig:DFN_fit_EP2-5}.  It is found that the best fit is obtained by using a sum of two Fowler-Nordheim functions.  The field enhancement factor of $\beta_2 = 1571$ obtained for one of the functions is close to that obtained from dataset MP:53, but the effective emission area, $A_{e,2}$, is about a factor of ten lower, which is consistent with the comparison shown in Fig.~\ref{fig:SW_vs_E_scaled_EP_MP}.  For the data points above 450~kV/cm, the addition of a second function with enhancement factor of $\beta_2 = 431$ provides the best fit.  

The result of this exercise of characterizing the breakdown in LHe as being the result of asperities with two distinct enhancement factors is not intended to provide an accurate representation of the physical reality of the electrode surface. Rather it demonstrates that breakdown is not necessarily limited to sites having the largest field emission current. It illustrates the inability of the statistical approach in determining the microscopic nature of the electrode surface.  Further, it suggests the possible limitations of extreme value theory and of the description of the emission current through the Fowler-Nordheim equation.  
For instance, at sufficiently high fields and high current densities, deviations from the standard Fowler-Nordheim description\cite{Fursey2005} and surface work function changes\cite{Tsong1969} are known to exist.

Based on the results presented, we will adopt the interpretation of the proportionality between the Fowler-Nordheim field emission function and the hazard function as a working hypothesis in the discussion that follows. Our present understanding of the breakdown mechanism in LHe, although still qualitative in nature, provides strong motivation for such an interpretation, and which we showed is supported by our data.  What's more, the hypothesis has broad enough scope with the potential for establishing a unified model to explain breakdown in different dielectric materials and other related studies.  Thus, it serves as a practical starting point to guide exploration in a direction that may lead to a better understanding of the processes behind these phenomena, for up until the present time, the connection between the statistics of breakdown to the underlying physical mechanism has not been clearly established.


\subsection{Area scaling}\label{sec:discuss_area_scaling}

It is common in the literature to explore the scaling behavior of the breakdown field with electrode surface area by fitting measurements of experiments having different electrode areas. However, dissimilarities in setups, conditions and methodologies may obscure underlying scaling laws and other features of a comparison. It is preferable to compare data taken in different experiments whose conditions are similar and well characterized.  But absent the availability of such information, one can also explore the scaling predicted by a single well-characterized experiment.

In the most general case where no assumptions are made on the form of the cumulative hazard function, it is possible to determine the area scaling behavior by utilizing the framework outlined in Sec.~\ref{sec:discuss_statistical_approach}.  Let $n$ be defined as a scaling factor so that the area-scaled empirical cumulative distribution is expressed as
\begin{eqnarray}
\hat{P_{b}}(E,n) =  1 - e^{- n S_{0} \hat{W}},
\label{eq:scaledcdf}
\end{eqnarray}
where $n S_{0} \hat{W}$ is determined from a dataset with an arbitrary choice for $n$.  Here, we will take $n$ to be one so that scaling is performed relative to the surface area of the SSHV electrode.  The $n = 1$ dataset is what we will refer to as the reference dataset, or reference electrode.

\begin{figure}[htb]
	\centering
	\begin{adjustbox}{center}
		\includegraphics[width=\columnwidth]{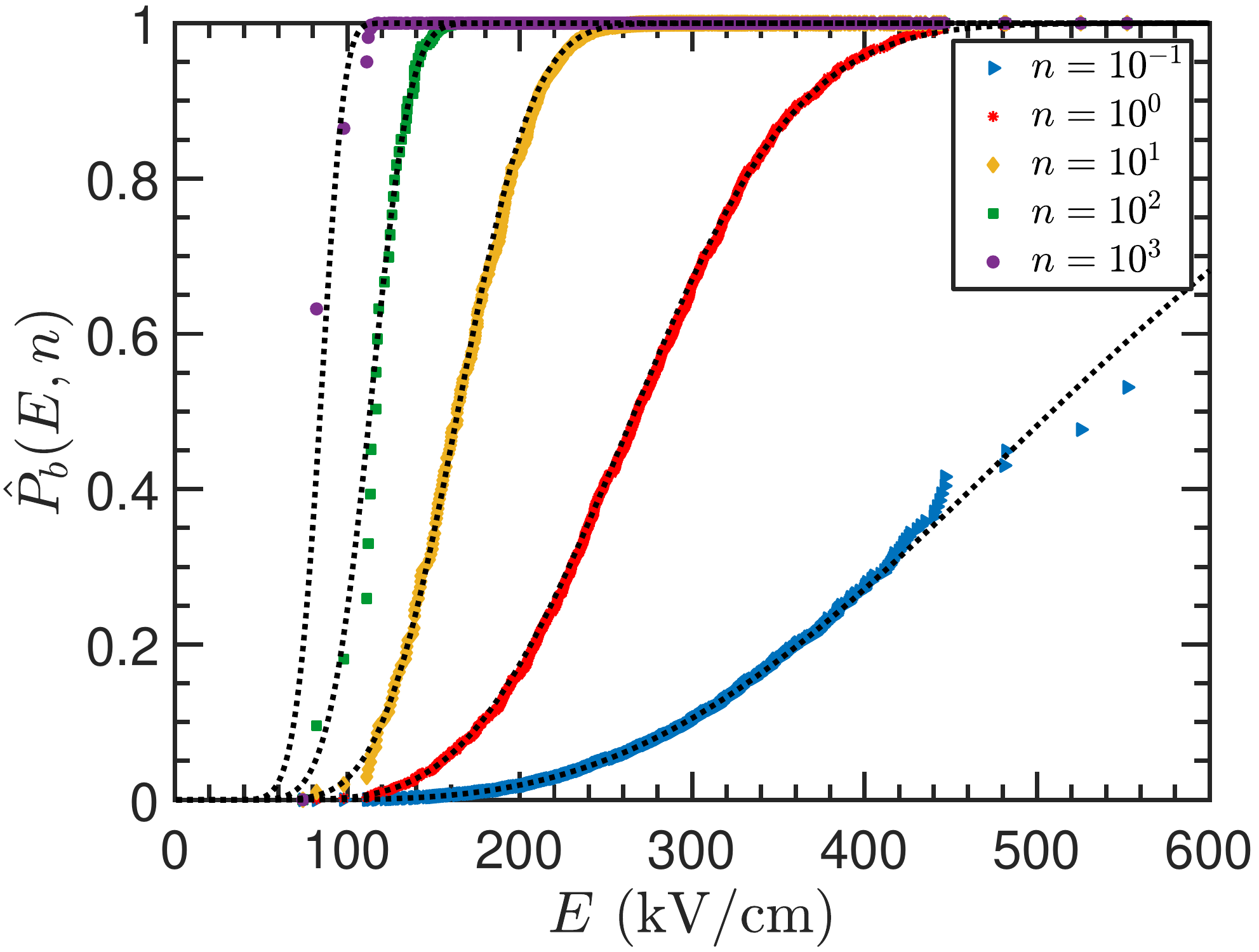}
	\end{adjustbox}
	\caption{The electrode surface area scaling of the empirical cumulative distribution function for different values of the scaling factor, $n$. The $n=1$ reference curve is determined from dataset MP:53 in the SSHV data using the procedure described in Sec.~\ref{sec:discuss_statistical_approach}. The dotted curves are obtained by fitting the reference dataset and applying the scaling factors.}
	\label{fig:scaled_ecdf_MP53}
\end{figure}

The area-scaled empirical cumulative distribution function for several different values of the scaling factors is shown in Fig.~\ref{fig:scaled_ecdf_MP53}.  The $n=1$ data corresponds to the reference surface area of the SSHV electrode while the other values represent electrodes with surface areas that are different than the SSHV electrode by a factor $n$.  From Fig.~\ref{fig:scaled_ecdf_MP53}, the median breakdown field ($\hat{P}_b = 0.5$), or any other percentile, can be immediately read off from the curves to determine how it scales with the electrode surface area.  There, however, are clearly statistical limits to this data-based approach as exemplified by the $n = 10^{-1}$ and $n = 10^3$ curves.  The extent of the limitation is primarily determined by the statistics in the reference dataset.  As a result of this, when the value of $n$ is much larger or smaller than $n=1$, the median, or other chosen percentile, breakdown field is no longer well-constrained by the data points.  However, this would also imply that with sufficient statistics in the reference dataset, the data-based scaling can be carried out to an arbitrarily sized electrode.  Thus, the method discussed here represents a robust, model-independent approach to area scaling, avoiding possibly invalid assumptions regarding the underlying breakdown distribution.

\begin{figure*}[]
	\captionsetup[subfigure]{justification=centering}
	\centering
	\begin{subfigure}[]{0.50\textwidth}
		\includegraphics[width=\textwidth]{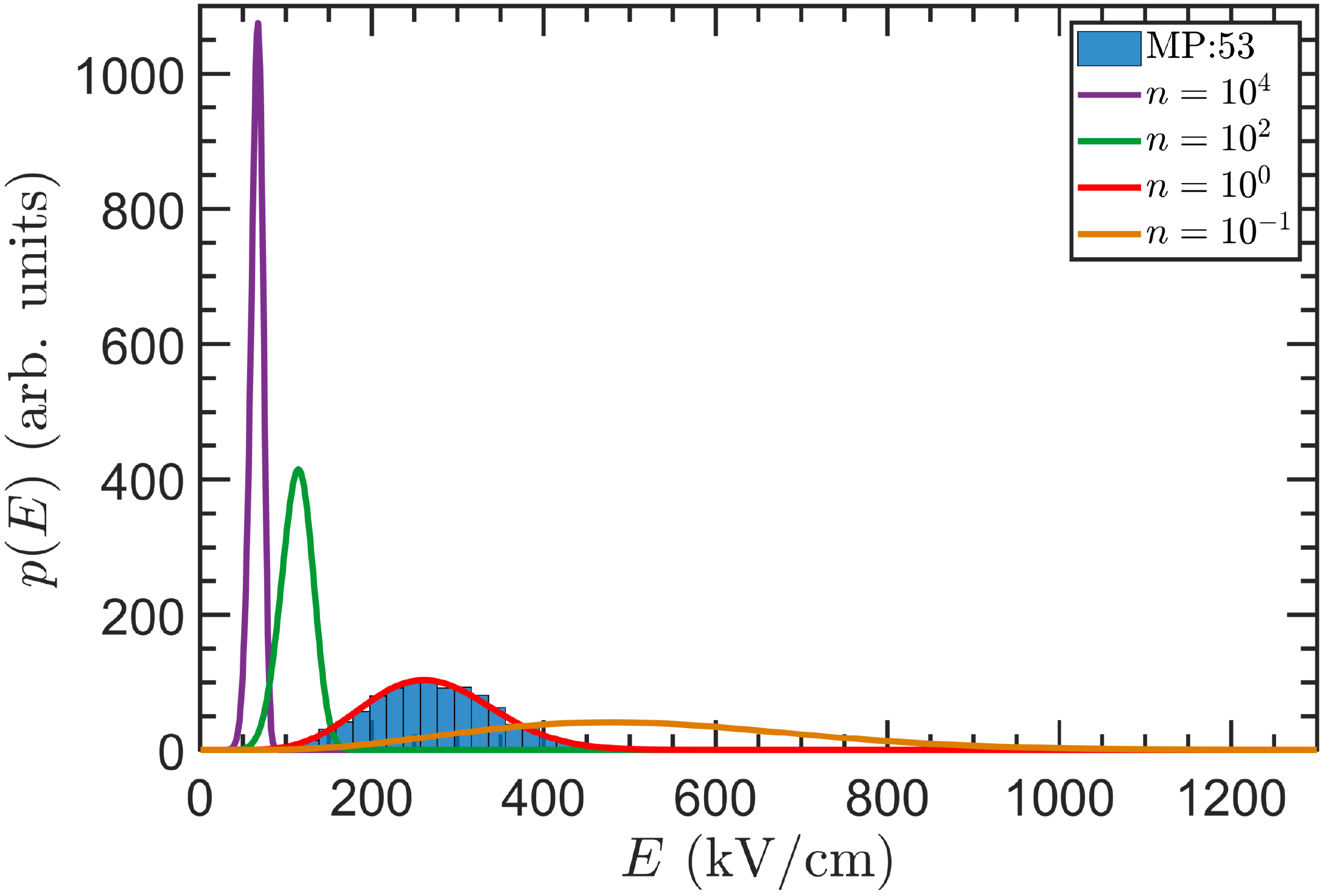}
		\caption{Mechanically-polished electrode}
		\label{fig:MP_pdf_scaling}
	\end{subfigure}
	\hfill
	\begin{subfigure}[]{0.49\textwidth}
		\includegraphics[width=\textwidth]{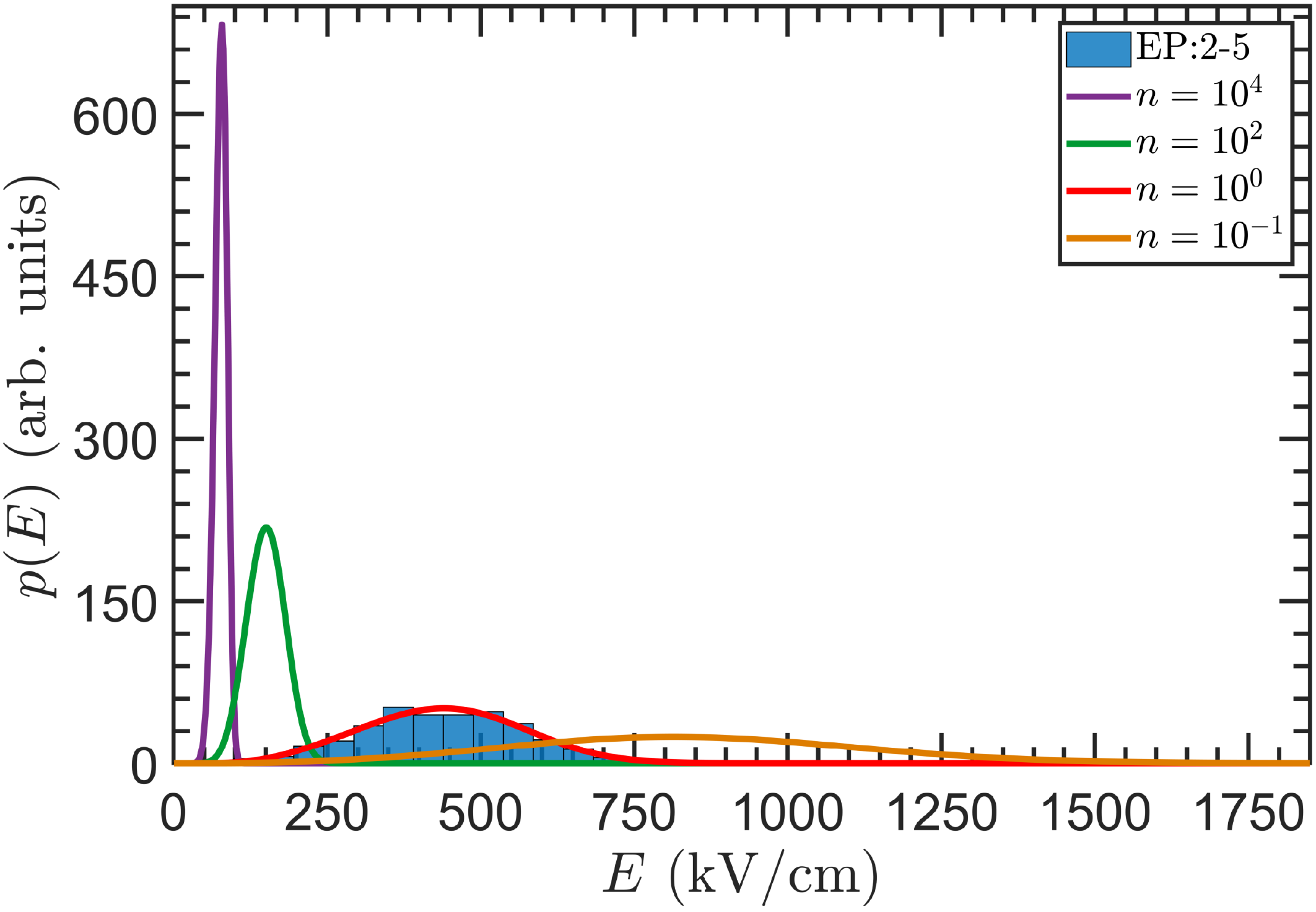}
		\caption{Electropolished electrode}
		\label{fig:EP_pdf_scaling}
	\end{subfigure}
	\caption{The probability density function, $p(E)$, derived from fitting the cumulative hazard function for (a) SSHV dataset MP:53 (mechanically-polished, 10.8~Torr, 1.76~K) and (b) the combined SSHV dataset EP:2-5 (electropolished, 12.8~Torr, 1.81~K) for different scaling factors, $n$.  The experimentally acquired breakdown field distribution for the datasets are also plotted.  The statistics for the curves are given in Table~\ref{tab:pdfstats}.}
	\label{fig:pdf_scaling}
\end{figure*}

\begin{figure*}[]
	\captionsetup[subfigure]{justification=centering}
	\centering
	\begin{subfigure}[]{0.49\textwidth}
		\includegraphics[width=\textwidth]{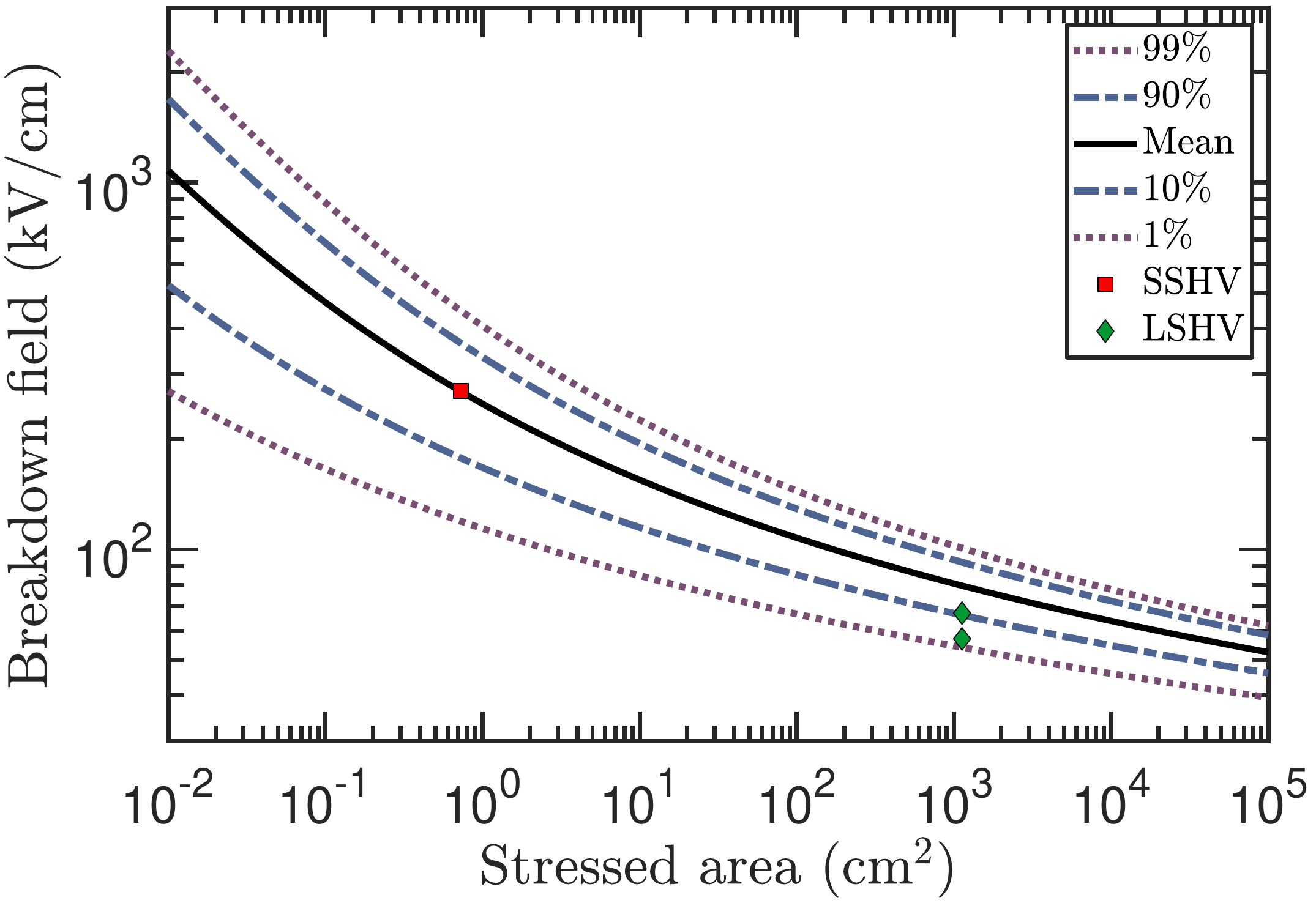}
		\caption{Mechanically-polished electrode}
		\label{fig:MPscaling}
	\end{subfigure}
	\hfill
	\begin{subfigure}[]{0.49\textwidth}
		\includegraphics[width=\textwidth]{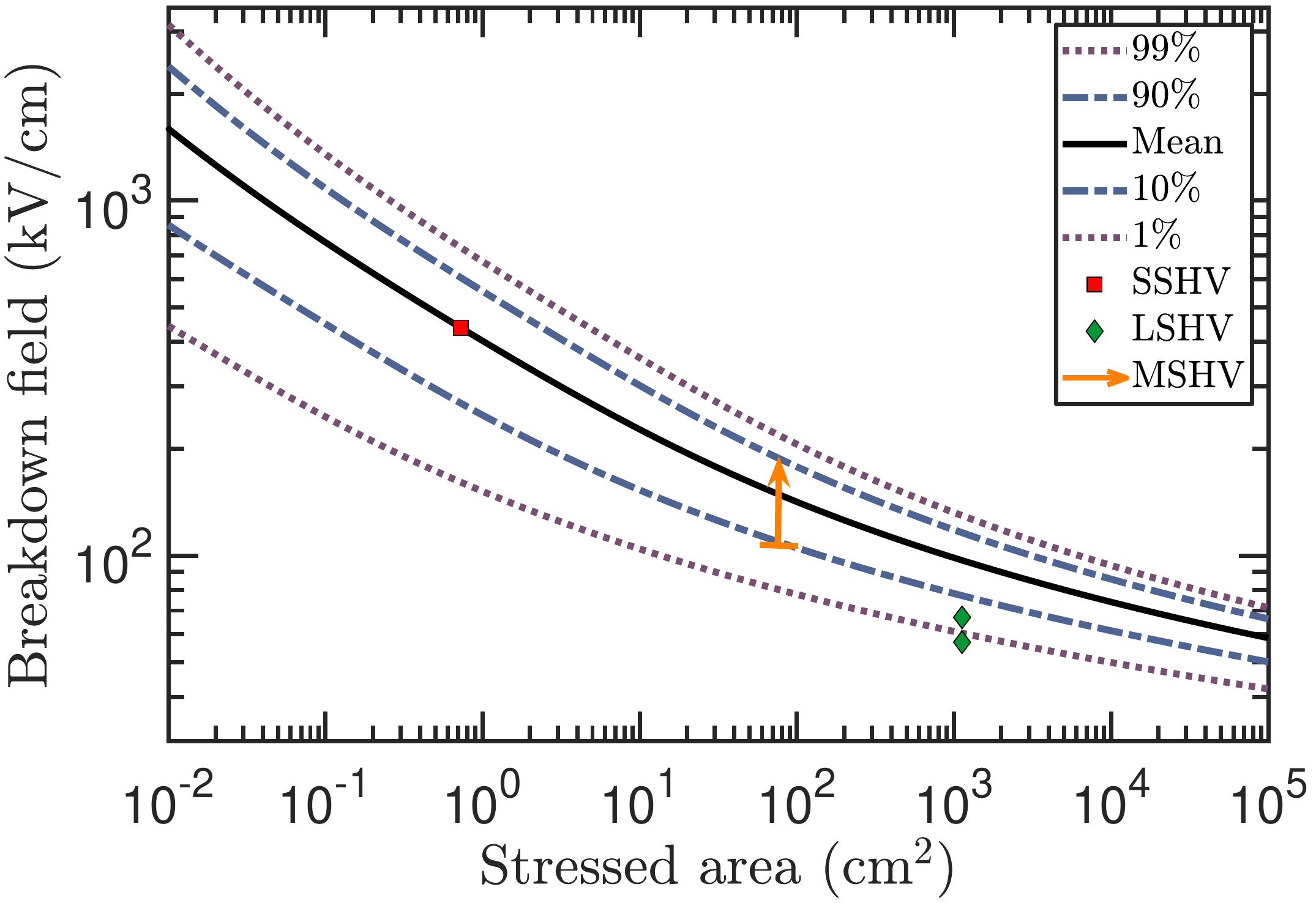}
		\caption{Electropolished electrode}
		\label{fig:EPscaling}
	\end{subfigure}
	\caption{Electrode area scaling of the mean and $p^{\textrm{th}}$ percentile breakdown fields determined from fitting (a) the 10.8 Torr (1.76 K), mechanically-polished electrode SSHV dataset MP:53 and (b) the 12.8 Torr (1.81 K), combined electropolished electrode SSHV datasets EP:2-5 using a Fowler-Nordheim function for the cumulative hazard function.  The mean breakdown fields of the SSHV data used in the scaling are shown along with the mean breakdown fields of the LSHV data of 57 kV/cm and 67 kV/cm at 6.9 Torr and 16.2 Torr, respectively.  The data point at 105 kV/cm from the MSHV apparatus (arrow) represents the highest field achieved using electropolished electrodes, but at which no breakdown was observed. }
	\label{fig:area_scaling}
\end{figure*}

There are several important features in Fig.~\ref{fig:scaled_ecdf_MP53} to highlight.  First, as the area of the electrode is scaled up, the width of the breakdown field distribution becomes narrower.  A comparison of the SSHV and LSHV data shows agreement with this finding.  Conversely, scaling down to smaller areas results in a broader distribution.  This implies that more breakdown statistics are needed to characterize the breakdown field at a given confidence level when the electrode used in the measurement is small.

Figure~\ref{fig:scaled_ecdf_MP53} also shows that the effect of scaling on the tails of the distribution function is dependent on the direction of the scaling.  When the area is scaled up(down) the data points are pushed towards the high(low) field region on the curve.  This suggests that scaling up may present somewhat of a obstacle for reliability analysis, which is typically more concerned with the breakdown field at low percentiles as this allows for safe operational boundaries and parameters of a device to be established in order to reduce its failure risk.  But when the breakdown field distribution is nearly symmetric about the median field, it is possible to use the high percentile breakdown field values (e.g., 90\%, 95\%, 99\%), that are still well-constrained by the data to estimate the low percentile (e.g., 1\%, 5\%, 10\%) breakdown fields.  If the assumption of symmetry cannot be applied to the distribution, the skewness must also be taken into account in a more careful determination.

The change in the distribution width and the median breakdown field with electrode area may be understood without reference to any particular breakdown mechanism in the following way:  An electrode can be viewed as a union of multiple sub-electrodes that we can categorize as one with low-field breakdown initiation, medium-field breakdown initiation, and so on.  Importantly, the division and categorization of these sub-electrodes are made in the statistical, or probabilistic, sense and is in reference to its breakdown initiating probability rather than a geometric one.  These sub-electrodes can also be viewed as sub-groups of electrode surface states.  An overall scaling on the electrode scales each of these sub-groups in the same manner.  For larger electrode sizes, the sub-group leading to low-field breakdown initiation is similarly larger than for a smaller electrode.  As the electric field is ramped up, the contribution to the total probability of breakdown due to the low-field sub-group becomes high enough that a breakdown occurs before reaching the higher fields needed to initiate breakdown in the other sub-groups.  In a more mathematical illustration, the hazard function, $W$, can be expressed as a linear combination of many terms, arranged in order from low-field to high-field breakdown initiation.  These terms are related to the distribution of surface protuberances and their associated field enhancements.  When the area, $S_0$, is scaled up, the leading term dominates and determines the outcome of the particular voltage ramp sequence.  Thus, when the electrode is large, the properties of the observed breakdown distribution is determined primarily by the low-field sub-group (i.e., the ``weakest sub-group"), and this explains the lower breakdown fields and narrower distributions encountered in larger-sized electrodes.  

The above picture implies a type of self-censoring\footnote{Here, we say the data is censored because there is only partial information about the distribution function, that is, the value of the function above some boundary field is unknown.} of the data when going in the reverse direction, that is, scaling from a large-sized reference electrode down to a smaller one.  This is because the predominance of the low-field breakdown initiating sub-group prevents experimental probing of the higher-field sub-groups.  Therefore, no information on these sub-groups is obtained by the experiment, leading to the appearance of a censoring of the empirical distribution function.  This would explain the very peculiar behavior of the $n = 10^{-1}$ curve in Fig.~\ref{fig:scaled_ecdf_MP53} where the distribution function terminates abruptly, signaling an end to the region of accessible information.  As a consequence, area scaling is a non-symmetric operation in the sense that the direction matters just as does the scaling factor.  The implication for experimental studies of electrical breakdown is that more complete information on the breakdown distribution function, or more precisely, the cumulative hazard function, may be obtained by employing smaller-sized electrodes rather larger ones, which on the surface may certainly appear quite counter-intuitive. But such an approach will also have trade-offs that must be considered, and these will be more evident in the discussion that follows.

\begin{table}
	\caption{\label{tab:pdfstats} Statistics for the area-scaled probability density functions for datasets MP:53 and EP:2-5. $n$ is the scaling factor, $\mu$ is the mean breakdown field, FWHM is the full-width at half maximum, and $\gamma$ is the skewness.}
	\begin{ruledtabular}
		\begin{tabular}{lcccc}
			Set & $n$ & $ \mu$ (kV/cm) & FWHM  (kV/cm)&  $\gamma$  \\ \hline
			
			MP &  $10^{-1}$  &  522.8   &  446.7  &  0.441     \\
			MP &  $10^{0}$   &  269.6   &  177.8  &  0.240     \\
			MP &  $10^1$     &  164.1   &  82.5   &  0.033     \\
			MP &  $10^2$     &  112.5   &  43.9   &  -0.142    \\
			MP &  $10^3$     &  83.7    &  26.0   &  -0.281    \\
			MP &  $10^4$     &  65.8    &  16.8   &  -0.389    \\
			
			EP &  $10^{-1}$  &  839.2  &  666.6  &  0.213     \\
			EP &  $10^{0}$   &  438.6  &  327.5  &  0.085     \\
			EP &  $10^1$     &  244.3  &  159.5  &  0.127     \\
			EP &  $10^2$     &  150.2  &  75.4   &  0.023     \\
			EP &  $10^3$     &  103.2  &  40.2   &  -0.144    \\
			EP &  $10^4$     &  76.8   &  23.9   &  -0.282    \\
			
		\end{tabular}
	\end{ruledtabular}
\end{table}	

The statistical limits of the empirical-based scaling method are apparent.  However, by a suitable choice of fitting function for the cumulative hazard function in Eq.~\ref{eq:scaledcdf}, these limits may be remedied.  We showed previously that the estimated cumulative hazard function is well-fitted by a function of the form given by the Fowler-Nordheim field emission equation. In Figs.\ref{fig:MP_pdf_scaling} and \ref{fig:EP_pdf_scaling}, the probability density function, $p(E)$, derived from fitting the associated dataset in each figure is shown. The empirical breakdown field distribution is shown along with the fitted function for $n=1$, and the probability density functions for other values of $n$ are also plotted.  The statistics for the scaled functions are tabulated in Table~\ref{tab:pdfstats}.  The skewness is defined as $\gamma = \mu_3/\mu_{2}^{3/2}$, where $\mu_3$ and $\mu_2$ are the third and second central moments, respectively.  

With increasing $n$, the mean breakdown field decreases and a narrowing of $p(E)$ is observed.  These features are consistent with those displayed by the data-based scaling of the cumulative distribution function shown in Fig.~\ref{fig:scaled_ecdf_MP53}.  But also note the change in the shape of the function and, in particular, its skewness with scaling factor. The function shifts from positive to negative skewness as $n$ increases. This is a rather intriguing feature that is not immediately evident from Fig.~\ref{fig:scaled_ecdf_MP53}, but is simply a consequence of the shape of the cumulative hazard function.  This unique characteristic can provide for a strong test of the validity of our area scaling analysis beyond that of the how the breakdown field scales with electrode area, which often have uncertainties associated with surface conditions and definition of stressed area.

In Fig.~\ref{fig:area_scaling}, the predicted breakdown field as a function of the stressed area derived from analysis of the mechanically-polished dataset MP:53 and the combined electropolished datasets EP:2-5 is shown. The breakdown data acquired in the LSHV apparatus and the limit obtained from our Medium Scale High Voltage (MSHV) system \cite{ITO16} are also plotted in the figure for comparison.  The mean breakdown field measured in the LSHV apparatus is about 22\% and 36\% smaller than the predicted values obtained from the mechanically-polished and electropolished electrode dataset, respectively.  This represents very good agreement between measurements and predictions considering the large scaling factor of about $10^3$ and uncertainties associated with defining the stressed area, the limited statistics in the LSHV dataset, and differences in surface conditions.  Concerning the latter point, note there exists large variations of over 30\% in the mean breakdown field between SSHV mechanically-polished datasets acquired under essentially identical conditions (e.g., $\sim$1.7~K and 10 Torr) as shown in Fig.~\ref{fig:sshv_temperature_pressure_dependence}.

Finally, it is important for us to remark on the fundamental limitations of statistical area scaling.  It is evident that the scaling of the electrode is realized only in the mathematical sense, in which we conceptualize the replication, i.e., making \emph{identical} copies, of the reference electrode $n$ times to produce the surface area of the scaled electrode.  There exists an inherent sampling bias in this operation, so that the statistically scaled electrode is not necessarily a perfect representation of the physical electrode.  For instance, one can conceive of a type of defect that causes a breakdown to occur at low fields.  If this class of defect has a sufficiently low density, it is more likely to be absent in small electrodes as compared to larger ones, implying that the selection bias will tend to give slightly higher predicted breakdown fields for very large electrode areas than what will be observed experimentally.  In the end, the above considerations show that there is no true substitute for experimental measurement.


\subsection{Area scaling in other noble liquids}\label{sec:discuss_scaling_liquids}

In studies of the area effect in liquid argon\cite{Acciarri2014, Auger2016, Tvrznikova2019} (LAr) and xenon\cite{ Tvrznikova2019} (LXe), the breakdown field, $E$, as function of stressed area is often modeled by the expression, $E = C A^{p}$, where $C$ is a constant, $A$ is the stressed area and $p < 0$ is the power index.  This scaling relation can be obtained by assuming that the hazard function follows a two-parameter power-law, or equivalently that the cumulative breakdown distribution function is given by the Weibull function. For LAr, $p$ values between $-0.22$\cite{Auger2016, Tvrznikova2019} and $-0.26$\cite{Acciarri2014} are found from a global fit of experimental data with stressed areas in the range of approximately $10^{-4} - 10^{1}$~cm$^2$.  With LXe, a scarcity of breakdown data exists, but a $p$ value of $-0.13$\cite{ Tvrznikova2019} is found by fitting the results of two experiments\cite{Rebel2014, Tvrznikova2019, TvrznikovaThesis}.  However, the reliability of this fit value is quite questionable given the uncertain experimental conditions for the data point from Ref.\onlinecite{Rebel2014}. In addition, the data in Ref.\onlinecite{Tvrznikova2019, TvrznikovaThesis} show non-stationarity, indicating a possible conditioning effect, which makes the interpretation of the data less straightforward.

The breakdown field area-scaling expression, $E = C A^{p}$, is to an extent problematic. Clearly, in the limiting behavior, the breakdown fields obtained from such an expression are physically unreasonable, that is, it becomes arbitrarily small with increasing electrode surface area.  It is possible, though, to maintain the validity of the expression by restricting the domain of applicability, but such an accommodation brings about other questions.  For instance, what is the scaling behavior outside the range of validity and why is it only valid in the given range?  Is the true scaling law different that what is being assumed?

\begin{figure}[!htb]
	\centering
	\begin{adjustbox}{center}
		\includegraphics[width=\columnwidth]{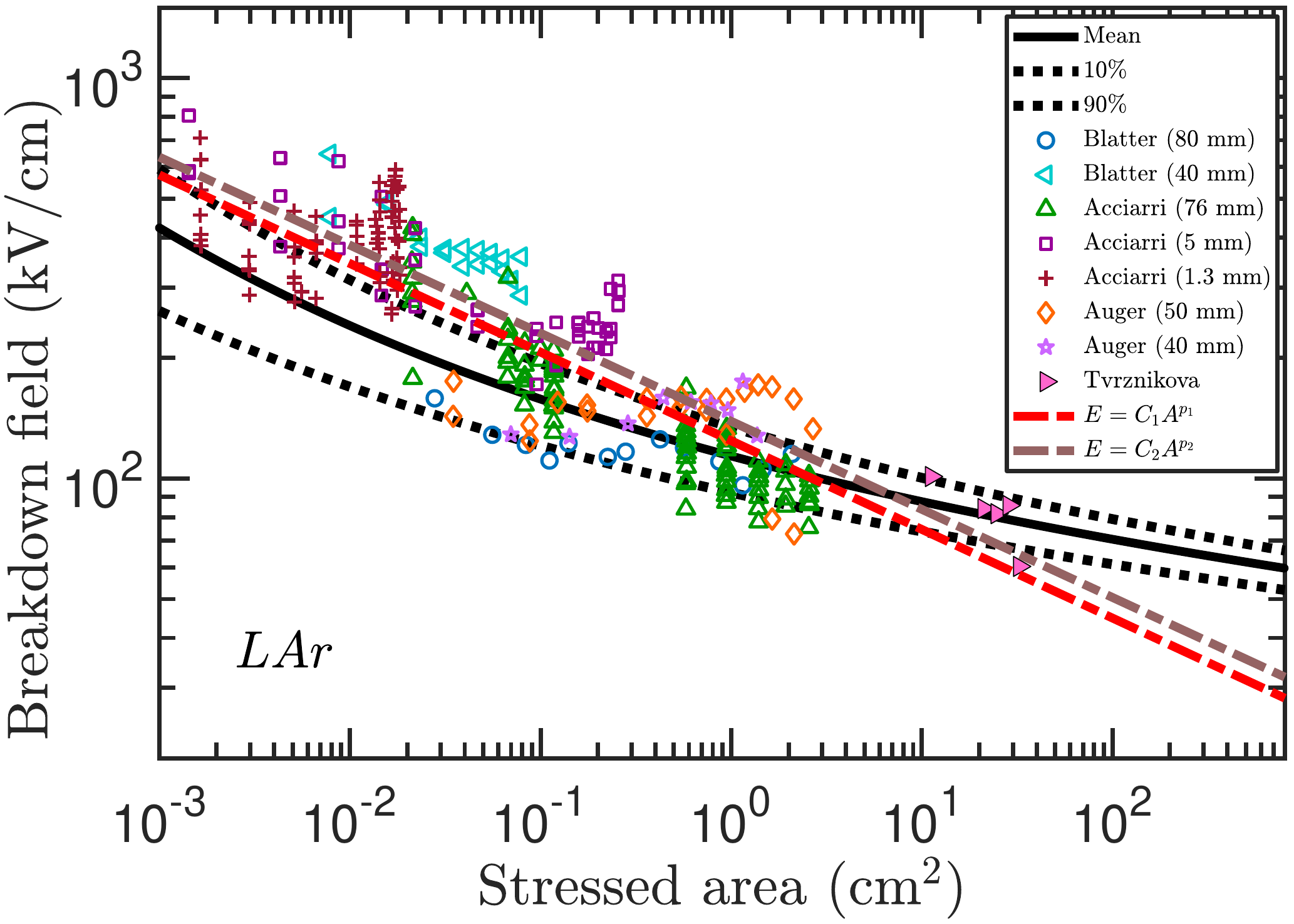}
	\end{adjustbox}
	\caption{LAr breakdown field versus stressed area, defined as the area with an electric field strength greater than 90\% of the maximum field strength. The area scaling curves are determined by applying the analysis method proposed in this work to the breakdown data from Acciarri \emph{et al.}\cite{Acciarri2014} acquired with a stressed area of $\sim 2$~cm$^2$. Data from past LAr experiments and the often assumed scaling curves, $E = CA^{p}$, are also plotted for comparsion.}
	\label{fig:LArscaling}
\end{figure}

To address those questions and further explore the applicability of the proposed analysis method in this work to experimental datasets in other noble liquids, we utilize the LAr breakdown data from Acciarri \emph{et al.}\cite{Acciarri2014}.  Worth noting is that this particular dataset contains only around 170 breakdowns and is also binned.  Therefore, its statistical power is significantly lower than that of our LHe datasets. The data from Acciarri \emph{et al.} were acquired with a sphere-plane electrode geometry where the sphere was 76 mm in diameter and the 90\% stressed area is $\sim 2$~cm$^2$.  The curves representing the mean, 10-percentile, and 90-percentile breakdown field as determined from applying our analysis method to this single dataset is shown in Fig.~\ref{fig:LArscaling}.  Also plotted are the measurements from past studies of LAr breakdown as well the often assumed scaling expression, $E = C A^{p}$, where the values of the parameters are from Auger \emph{et al.}~\cite{Auger2016} and Tvrznikova \emph{et al.} \cite{Tvrznikova2019}.

The 76-mm Acciarri \emph{et al.} data\cite{Acciarri2014} (green triangles), taken for various stressed areas, are well fitted by our scaling curves, determined from data for one single stressed area, as shown in Fig.~\ref{fig:LArscaling}.  These curves also have better agreement with the data of Tvrznikova \emph{et al.} \cite{Tvrznikova2019} as compared to the $E = CA^p$ scaling curves, which overpredict the breakdown fields for the 76-mm data but underpredict for the Tvrznikova \emph{et al.} data.  The Auger \emph{et al.}~\cite{Auger2016} data also show good agreement with our scaling curves  However, the agreement of these curves with the smaller stressed area data is lower, but a likely explanation for this is the difference in surface condition of the electrodes used to acquire those data. Specifically, the 5-mm and 1.3-mm Acciarri \emph{et al.}\cite{Acciarri2014} data were acquired with electrodes with a chrome finish whereas the 76-mm electrode had a mirror-finish.  A comparison of the 5-mm and 76-mm data at around a stressed area of 0.1~cm$^2$ indicates such an effect, which also appears to be present in the Blatter \emph{et al.}~\cite{Blatter2014} data acquired with different electrode sizes. This highlights the point previously made about the problematic nature of fitting data from a group of experiments with variable and sometimes uncertain experimental conditions.


\subsection{Electrode gap spacing dependence}\label{sec:discuss_gap_dependence}

Breakdown in dielectric liquids is shown to be a surface phenomenon that does not involve the bulk properties of the liquid, yet many experiments have been interpreted as showing a electrode gap spacing dependence.  Such a dependence is often expressed in the form $V = k d^s$, where $V$ is the breakdown voltage, $k$ is a constant, $d$ is the gap spacing, and the index $s$ characterizes the strength of the dependence.  However, there is no generally accepted mechanism of breakdown that would form the basis for such a relationship, and some experimenters\cite{SCHWENTERLY74, Gerhold1994} have suggested an alternative relation of $V = V_{0} + ha^{1-j}$.  But inconsistencies in experimental results have thus far prevented any satisfactory resolution to this problem.

The possibility that natural sources of radioactivity can give rise to a gap spacing dependence is inconsistent with experimental observation that show electrons, $\alpha$ particles, and the daughter nuclei from radioactive sources such as $^{241}$Am electroplated directly onto high voltage electrodes do not initiate breakdown for applied fields up to 45~kV/cm in LHe\cite{Ito2012, Phan2020}.  By considering the density of energy deposition from the daughter nucleus of an $^{241}$Am decay, a lower bound on the minimum energy density needed for the initiation process can be determined. The lack of breakdown initiation from these highly ionizing particles show they cannot form a basis to explain any presumed spacing dependence.

Experiments performed by Galand\cite{GALAND68} and Lehmann\cite{LEHMANN70} found that the breakdown voltage in LHe is approximately proportional to the electrode gap spacing up to 2.5~mm.  Results from Mathes\cite{MATHES67} using 12.7-mm diameter steel-sphere electrodes are similar for gap spacings up to 1.25~mm.  However, significant gap dependence for spacings between $10-100$~$\mu$m are indicated by the measurements of Goldschvartz and Blaisse \cite{GB1966} using a 10-mm diameter sphere-plane (tungsten-stainless steel) electrodes.  

Measurements by Schwenterly \emph{et al.}\cite{SCHWENTERLY74} utilized stainless steel sphere-plane electrodes with a diameter of 38 mm and gap spacings between $0.25 - 2.5$~mm.  Their results showed that breakdown fields increase for gap spacings below about 1~mm but become nearly constant for spacings above 1.5~mm.  The results of Gerhold\cite{Gerhold1972} made with spherical electrodes of 50-mm diameter and gaps up to 0.5~mm are consistent with those of Schwenterly \emph{et al.}\cite{SCHWENTERLY74} where for gap spacings above 1.5~mm, the breakdown fields are nearly constant.

Experiments employing larger-sized spherical electrodes or uniform-field electrodes have observed much milder gap spacing effects.  For instance, Meats\cite{MEATS72}, using uniform-field electrodes of 60-mm diameter observed that for spacings less than 1.0~mm, observed a slight dependence of breakdown field on $d$. But at larger spacings, the breakdown voltage varied linearly with a gap spacing up to nearly 4~mm.  Measurements made by Meyerhoff\cite{MEYERHOFF95} with a polished 250-mm diameter brass sphere and brass ground plane for spacings up to 12 mm found that the breakdown voltage increases nearly linearly with spacing up to 200 kV. In contrast, the results of Fallou \emph{et al.}\cite{FALLOU69} indicate a pronounced departure from linearity with $s = 0.5$ up to $d = 10$~mm, implying a ``square-root law" as found in vacuum breakdown.  These measurements, however, were made with small-diameter spheres and had spacings greater than the sphere diameter.

\begin{table}
	\caption{\label{tab:gapspacing}Data from previous experiments on the gap spacing dependence of breakdown in LHe by Goldschvartz and Blaisse\cite{GB1966}, Schwenterly \emph{et al.}\cite{SCHWENTERLY74}, and Meyerhoff\cite{MEYERHOFF95}.  All data were acquired at 4.2~K and SVP and utilized a sphere-plane electrode geometry.  Both the boiling(non-boiling) results from Meyerhoff are shown.}
	\begin{ruledtabular}
		\begin{tabular}{ccc}
			Gap (mm) &  Stressed area (mm$^2$)  &  $\overline{E}$ (kV/cm)   \\ \hline
			 0.010\footnote[7]{Goldschvartz and Blaisse}  &  0.137  & 3760  \\
			 0.020\footnotemark[7]  & 0.266  & 3100  \\
			 0.030\footnotemark[7]   & 0.396  & 2480  \\
			 0.040\footnotemark[7]   & 0.536  & 2280  \\
			 0.050\footnotemark[7]   & 0.688  & 2010  \\
			 0.060\footnotemark[7]  & 0.826  & 1980  \\
			 0.070\footnotemark[7]   & 0.955  & 1850  \\
			 0.080\footnotemark[7]   & 1.077  & 1700  \\
			 0.090\footnotemark[7]   & 1.203  & 1650  \\
			 0.100\footnotemark[7]  & 1.378  & 1500  \\
			 0.25\footnote[19]{Schwenterly \emph{et al.}}  &  12.59  &  800  \\
			 0.50\footnotemark[19] &  25.99  &  655  \\
			 0.75\footnotemark[19]  &  38.42  &  569  \\
			 1.00\footnotemark[19]  &  50.82  &  530  \\
			 1.25\footnotemark[19]  &  66.65  &  500  \\
			 1.50\footnotemark[19]  &  79.49  &  480  \\
			 1.75\footnotemark[19]  &  93.35  &  445  \\
			 2.00\footnotemark[19]  &  108.39  &  425  \\
			 2.25\footnotemark[19]  &  118.98  &  420  \\
			 2.50\footnotemark[19]  &  135.77  &  415  \\
			 3.0\footnote[13]{Meyerhoff} &  987.63  &  173.8(238.1)    \\
			 4.0\footnotemark[13]  &  1392.40  &  176.0(241.1)   \\
			 5.0\footnotemark[13]  &  1761.40  &  163.5(224.0)   \\
			 6.0\footnotemark[13]  &  2196.90  &  151.5(207.5)   \\
			 7.0\footnotemark[13]  &  4156.10  &  161.1(220.7)  \\
			 8.0\footnotemark[13]  &  2670.90  &  161.5(221.2)  \\
			 9.0\footnotemark[13]  &  2783.20  &  159.9(219.1)  \\
			 10.0\footnotemark[13]  &  3102.00  &  158.7(217.4)  \\
			 11.0\footnotemark[13]  &  3226.80  &  155.8(213.4)  \\
			 12.0\footnotemark[13]  &  4156.10  &  152.3(208.7)  \\
		\end{tabular}
	\end{ruledtabular}
\end{table}

\begin{figure}[!htb]
	\centering
	\begin{adjustbox}{center}
		\includegraphics[width=\columnwidth]{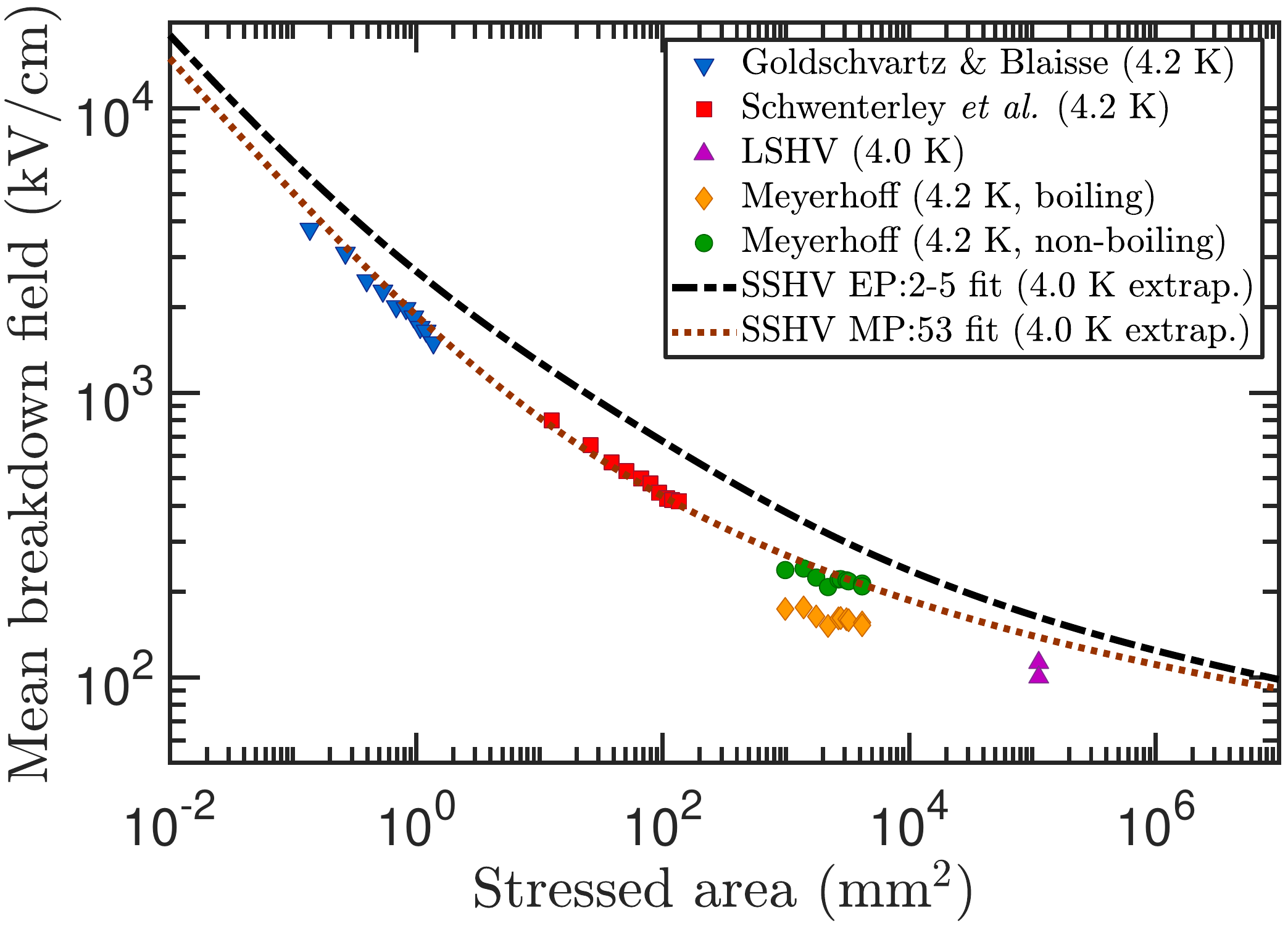}
	\end{adjustbox}
	\caption{The mean breakdown field as a function of the stressed area as measured by Goldschvartz and Blaisse \cite{GB1966}, Schwenterly \emph{et al.}\cite{SCHWENTERLY74}, and Meyerhoff\cite{MEYERHOFF95}. For the Meyerhoff measurements, the breakdown fields in non-boiling LHe are about 37\% higher than the the fields in boiling liquid.  Also shown are the scaling curves determined from analysis of the SSHV data and the measured breakdown fields at 4.0~ K in the LSHV apparatus. }
	\label{fig:gapspacing}
\end{figure}

Although the inconsistencies in experimental results are quite evident, some patterns do emerge.  In particular, the magnitude of the observed effect appears to be associated with the geometry of the electrodes (sphere-sphere vs sphere-plane and uniform-field) as well as the size of the electrodes relative to the gap spacing.  As we previously discussed in Sec.\ref{sec:scale dependence}, these elements have a direct and considerable impact on the characterization of the electrode stressed area.  Hence, it is conceivable that the spacing effect is merely a consequence of an unaccounted for area effect.

To investigate that possibility, we re-analyzed the data from Goldschvartz and Blaisse \cite{GB1966}, Schwenterly \emph{et al.}\cite{SCHWENTERLY74}, and Meyerhoff \cite{MEYERHOFF95}.  The geometries of these experiments (sphere-plane) are modeled in the COMSOL Multiphysics software package \cite{COMSOL} to determine the stressed area, defined to be the region on the electrode surface with a field strength greater than 70\% of the maximum field, for each of the gap spacings measured in their experiments. The choice of this stressed area definition is motivated by the discussion in Sec.~\ref{sec:scale dependence} concerning the SSHV electrode stressed area.  The results of these calculations are shown in Table~\ref{tab:gapspacing} along with the experimental measurements.  In Fig.~\ref{fig:gapspacing}, the measured breakdown fields from these three experiments as well as those from the LSHV apparatus are plotted as a function of the calculated stressed area.  Also shown are the predicted scaling curves at 4.0~K as determined from analysis of the SSHV measurements for the mechanically-polished and electropolished datasets at $1.7-1.8$~K.  The curves at the lower temperatures are extrapolated to 4.0~K by applying a factor that is determined from the data in Fig.~\ref{fig:sshv_temperature_pressure_dependence}.  Both the electropolished and mechanical-polished curves are plotted due to the fact the polish of electrodes used in some of the previous experiments were not clearly specified.

Remarkably, the measurements of Goldschvartz and Blaisse \cite{GB1966}, Schwenterly \emph{et al.}\cite{SCHWENTERLY74}, and Meyerhoff \cite{MEYERHOFF95} closely follow the predicted area scaling behavior. It is particularly important to re-emphasize that these measurements cover a wide range of gap spacings, 10~$\mu$m to 12~mm, that essentially encompasses the full range of gap sizes in all previous experiments exploring this dependence.  Thus, we have a reasonable expectation that the results of other experiments not included in our re-analysis may be explained in similar fashion. Also of note is that measurements made with non-boiling LHe by Meyerhoff are stated as being 37\% higher than those made in boiling liquid.  In Fig.~\ref{fig:gapspacing}, the non-boiling measurements are determined by applying a 1.37 factor to the boiling measurements included in the Meyerhoff paper.  Related to this observation is that although the low breakdown fields from the LSHV measurements in relation to the scaling curves may be explained by a surface quality effect, it may also be suggestive of possible boiling effects present when those measurements were made. Altogether, the re-analysis of the gap spacing measurements from these previous experiments that assumed to demonstrate a gap dependence can be satisfactorily shown to be the consequence of an areal dependence.  This is in keeping with liquid dielectric breakdown being a surface phenomenon.  The finding also lends further support to the validity of our area scaling analysis.


\subsection{Temperature and pressure dependence}\label{sec:discuss_temp_pressure_dependence}

A qualitative understanding of the temperature and pressure dependence of electrical breakdown in LHe may be realized by considering the following simple picture. If electrical breakdown is initiated by the formation of bubbles, then the stability of the bubble against collapse is a necessary condition for a breakdown to develop. The bubble development process is connected to the heterogeneous nucleation process in LHe; the latter may also be tied to the stochastic nature of the breakdown phenomenon. Nevertheless, further consideration of this topic will be deferred to a future study.  When the bubble is created by the injection of energy into the liquid resulting from field emission on the electrode surface, the amount of energy emitted must be sufficient to counteract the opposing forces of surface tension and externally applied pressure that work to prevent the bubble from expanding and ultimately leading to breakdown.  However, only a fraction of the total emitted energy, $W_e$, can go into opposing those two forces because the energy is partitioned into many components to drive different energy processes~\cite{Kattan91} such as ionization, vaporization of the medium, and light production.

Some of the processes are dependent on both the temperature and pressure of the liquid, while others are independent of both quantities.  If we consider the special case of constant temperature, all the pressure independent terms can be combined into the term, $W_{0, T_0}$, and the pressure dependent term can be expressed as $W_{P}(P)$, giving
\begin{eqnarray}
W_{e} = W_{0,T_0} + W_{P}(P).
\label{eq:TPdep3}
\end{eqnarray}
This simplification is motivated by our observation that the breakdown field is primarily dependent on the pressure of the liquid rather than its temperature.  The temperature dependence appears to be a sub-dominant effect, but is certainly not negligible as we discuss below.

The energy injected into the liquid is provided by the field emitted electrons which have a distribution of energies.  The current density is obtained by integration of this energy distribution\cite{Young1959, YoungMuller1959}.  Therefore, the total-energy of emission should be proportional to the emission current.

The total input energy, however, depends on the timescale of the emission process. Experiments that use impulse power (ns to $\mu$s) instead of DC generally observe higher breakdown fields, indicating that the timescale is an important factor to consider.  But without explicit knowledge of this quantity we make a simplifying assumption that this timescale is constant over the pressure and temperature range under consideration here.  As a consequence, the total energy resulting from field emission is taken to be proportional to the emission current.

With that, Eq.~\ref{eq:TPdep3} can be expressed as
\begin{equation}
\begin{aligned}
W_{0,T_0} + W_{P}(P) \propto E^2 \exp\left( \frac{-D\phi^{3/2}}{\beta E} \right).
\end{aligned}
\label{eq:TPdepFN1}
\end{equation}
Defining the parameter in the exponential as $B = D\phi^{3/2}/\beta$, we obtain
\begin{eqnarray}
 W_{0,T_0} + \frac{4 \pi R^3 P}{3} \propto E^2 \exp\left( \frac{-B}{ E} \right),
\label{eq:TPdepFN2}
\end{eqnarray}
where the work due by the external pressure, $P$, on a bubble of radius, $R$, is $4\pi R^3 P/3$ and $E$ is the applied field expressed in units of kV/cm.  Finally, the externally applied pressure is related to the breakdown field by
\begin{eqnarray}
P = A E^2 \exp\left( \frac{-B}{ E} \right) - C_{T_{0}},
\label{eq:TPdepFN3}
\end{eqnarray}
with $A$, $B$, and $C_{T_{0}}$ as positive constants.  

Calculating the numerical values of the constants $A$ and $C_{T_{0}}$ from first principles is very difficult because of the simultaneous, complicated dynamical processes occurring during bubble development, but the constant $B$ is related to the work function and field enhancement factor.  If we use the numerical values of these parameters obtained from fitting $S_{0}\hat{W}$ for SSHV dataset MP:53, we obtain $B = 567$~kV/cm.  However, we will not fix $B$ to this value because the field enhancement factor varies from dataset to dataset and should viewed only as an average effective value.  Notably, at high fields, the exponential factor in Eq.~\ref{eq:TPdepFN2} approaches one and the breakdown field scales as the square root of the applied pressure.  We note that a similar pressure dependence for breakdown in LHe is suggested by Meats~\cite{MEATS72}.

\begin{figure}[!htb]
	\centering
	\begin{adjustbox}{center}
		\includegraphics[width=\columnwidth]{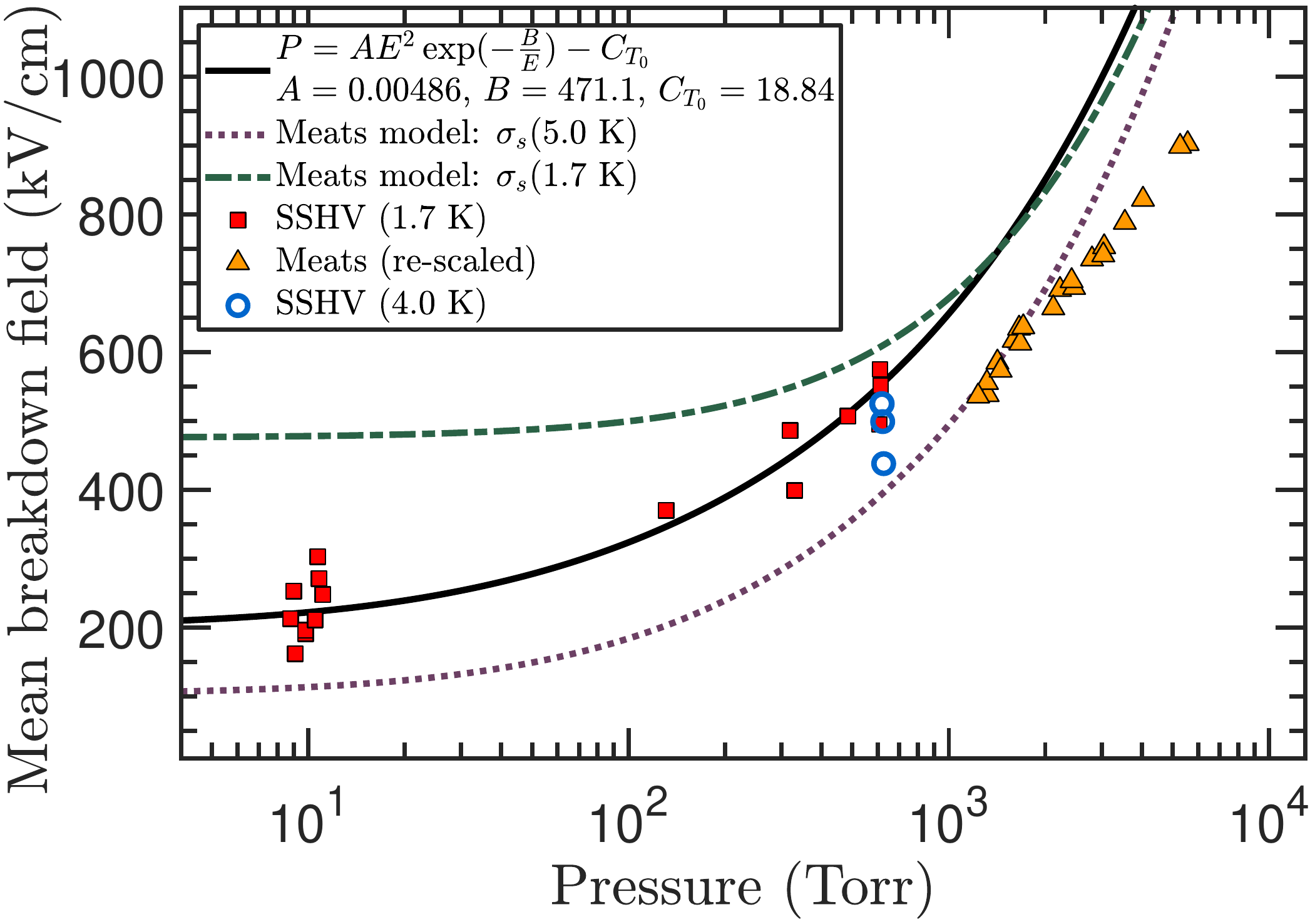}
	\end{adjustbox}
	\caption{Black-line: Fit of breakdown data acquired at $T\approx 1.7$~K with Eq.~\ref{eq:TPdepFN2};  Red-squares: SSHV 1.7~K data points used in the fit; Open-blue-circles: SSHV 4~K SVP data;  Orange-triangles: Pressurized data from Meats; Dotted and dash-dotted lines: Empirical pressure dependence relationship from Meats for two surface tensions. The Meats data points and empirical curves are re-scaled using Fig.~\ref{fig:MPscaling} to account for the difference in electrode surface areas between experiments. }
	\label{fig:fit_const_T_FN_energy}
\end{figure}

Shown in Fig.~\ref{fig:fit_const_T_FN_energy} are the mean breakdown fields for mechanically-polished SSHV datasets that were acquired at a temperature of $\approx 1.7$~K over a range of pressures.  In addition, the breakdown fields for datasets acquired at $\approx 4.0$~K and SVP are also plotted.  An interesting observation is that at the same pressure the typical breakdown field is lower at 4.0~K than it is at 1.7~K.  This does not appear to be merely the result of statistical scatter because similar behavior is observed at other temperatures and pressures (see Figs.~\ref{fig:MP-13-32-33}, \ref{fig:MP-39-40}, \ref{fig:MP-42-43}, and \ref{fig:MP-48-51}) where the lower temperature data at a given pressure has the higher average breakdown field.  This difference could be due to the effects of boiling liquid on measurements acquired at SVP.  But the trend in Fig.~\ref{fig:fit_const_T_FN_energy} is also consistent with the change in the surface tension and latent heat of vaporization between the two temperatures plotted (Table~\ref{tab:LHeProps}).  Both of these quantities decrease between 1.7~K to 4.0~K, and this results in a reduction in the constant $C_{T_{0}}$ in Eq.~\ref{eq:TPdepFN2}.  Physically, this suggests that less energy is required to vaporize the liquid and counteract the surface tension of the bubble for the same externally applied pressure.  Thus, for a given breakdown field, the corresponding pressure increases, which effectively causes the data points in Fig.~\ref{fig:fit_const_T_FN_energy} to shift to the right.

\begin{table}
	\caption{\label{tab:LHeProps}Selected properties of liquid helium at different temperatures\cite{Donnelly1998}: saturated vapor pressure, $P$, surface tension, $\sigma_{s}$, and latent heat of vaporization, $L$.}
	\begin{ruledtabular}
		\begin{tabular}{lccc}
			T (K) & P (Torr) & $\sigma_{s}$ (N/m) & $L$ (J/mol)  \\ \hline
			5.0 & 1470.1 & $1.570\times10^{-5}$ & 47.67     \\			4.6 & 1064.3 & $5.160\times10^{-5}$ & 72.59     \\
			4.2 & 744.3 & $8.954\times10^{-5}$ & 83.19     \\
			4.0 & 612.2 & $1.095\times10^{-4}$ & 87.00      \\
			3.5 & 352.9 & $1.626\times10^{-4}$ & 92.84      \\
			3.0 & 180.4 & $2.161\times10^{-4}$ & 94.11     \\
			2.5 & 76.7 & $2.623\times10^{-4}$ & 92.50      \\
			2.17 & 37.2 & $2.862\times10^{-4}$& 90.75     \\
			1.7 & 8.5 &$3.224\times10^{-4}$ & 91.91      \\
			1.2 & 0.6 & $3.426\times10^{-4}$ & 84.14      \\
			0.7 & $< 0.1$ & $3.508\times10^{-4}$ & 74.35      \\
			0.4 & $< 0.1$ & $3.536\times10^{-4}$ & 68.17   \\
		\end{tabular}
	\end{ruledtabular}
\end{table}	

In addition to the SSHV data, the pressurized data from Meats~\cite{MEATS72} are also plotted in Fig.~\ref{fig:fit_const_T_FN_energy}.  The Meats data were acquired with 6-cm diameter uniform-field electrodes ($S \approx 20$ cm$^2$).  Their electrodes were prepared with metal polish, so the surface condition should be most comparable to the mechanically-polished SSHV electrode.  To account for the electrode surface area difference between the experiments, a scaling factor obtained from Fig.~\ref{fig:MPscaling} is used to re-scale the Meats data.  

The dotted and dashed-dotted curves in Fig.~\ref{fig:fit_const_T_FN_energy} are calculated from the model for pressure dependence proposed by Meats~\cite{MEATS72} in which the breakdown field in V/m is given by:
\begin{eqnarray}
E = \frac{0.20}{\sqrt{\epsilon_0}}\left( P + \frac{2\sigma_s}{5\times10^{-9}}, \right)^{1/2},
\label{eq:TPdepMeats}
\end{eqnarray}
where $P$ is the externally applied pressure in Pa and $\sigma_{s}$ is the surface tension in N/m.  The curve for the Meats model with the surface tension at 5~K provides an adequate fit to both datasets.  However, the 1.7~K surface tension curve has a large discrepancy with our SSHV data.

In Fig.~\ref{fig:fit_const_T_FN_energy}, the solid black curve represents the fit of the 1.7~K SSHV data to Eq.~\ref{eq:TPdepFN3} and has good agreement with the SSHV data as well as the data from Meats, especially considering the uncertainties in electrode conditions between experiments.  The value of the parameter $B$ in the fit corresponds to $\beta \approx 1739$, which is close to the value of $\beta$ obtained from fits of the cumulative hazard functions for datasets MP:53 and EP:2-5. The physical reasonableness of the other two fit parameters is difficult to gauge without independent estimates of their expected values.  Nevertheless, the pressure dependence of the breakdown field is shown to be well described by the simple qualitative picture presented here.

\subsection{Determining breakdown for arbitrary electrode geometry}\label{sec:discuss_application}

Area scaling in Eq.~\ref{eq:scaledcdf} can be extended to the general case in which the electrode has an arbitrary field distribution. For a given potential value, the surface of the electrodes has a distribution of field strengths. If the area that has a field value $E_i$ is given by $S(E_i)$, then its contribution to the survival function is given by $P_s=e^{-S(E_i)W(E_i)}$. The survival function for the entire electrode surface is then given by
\begin{equation}
    P_s = \prod_{\{ E_i \}} e^{-S(E_i)W(E_i)}.
\label{eq:scaledcdf_gen}
\end{equation}
Here, it is understood that $S(E_i)$ is a function of the electrode potential and that a limit in which $E_i$ takes continuous values is assumed. 

\begin{figure}
	\centering
	\begin{adjustbox}{center}
		\includegraphics[width=\columnwidth]{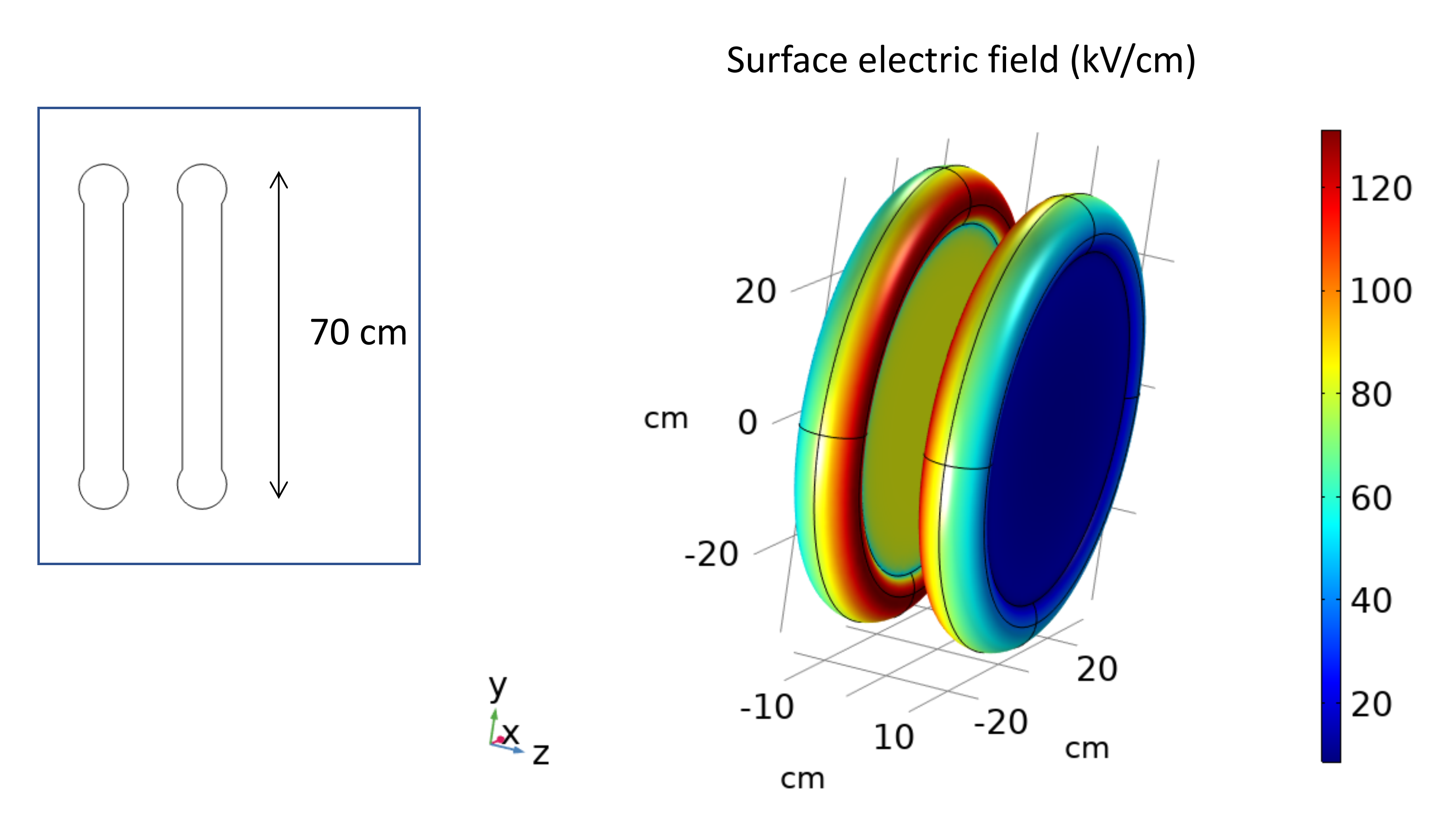}
	\end{adjustbox}
	\caption{An example electrode geometry to illustrate the utility of Eq.~\ref{eq:scaledcdf_gen}. The shape of the two identical electrodes is made of a flat circular disk with a torus attached to the edge. The surface field strengths shown are calculated for a potential difference of 1~MV. Inset: the cross section of the electrodes.}
	\label{fig:Example_field_imange}
\end{figure}

We illustrate how this is done in practice with the simple example electrode geometry shown in Fig.~\ref{fig:Example_field_imange}. The shape of the two identical electrodes is made of a flat circular disk with a torus attached to the edge. The different colors on the surface indicates the field strength at that location.  As seen the part of the torus facing the other electrode has high fields, and the flat part in the middle has a large area with a modest field strength. 

\begin{figure}
	\centering
	\begin{adjustbox}{center}
		\includegraphics[width=\columnwidth]{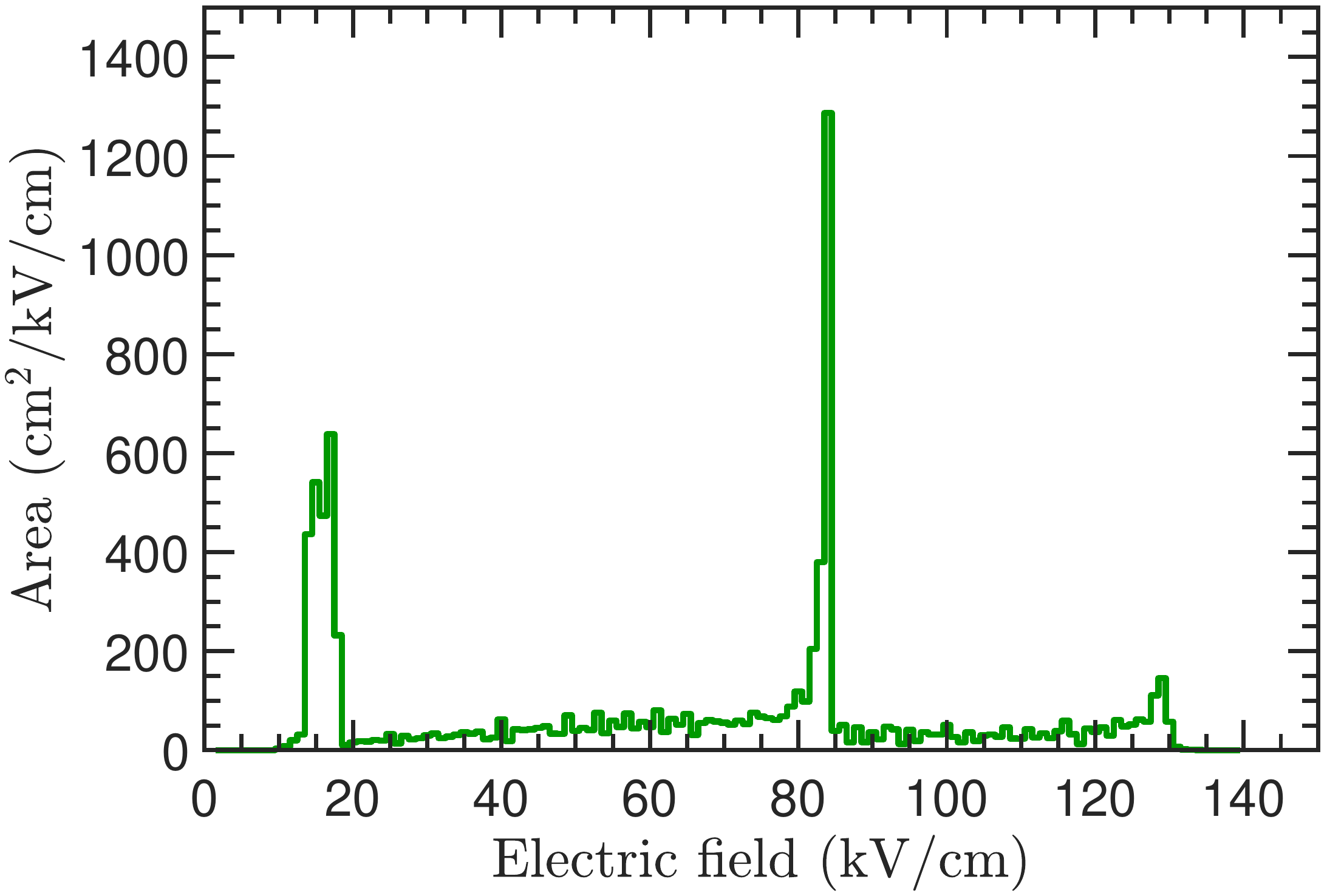}
	\end{adjustbox}
	\caption{The field distribution calculated for the positive electrode shown in Fig.~\ref{fig:Example_field_imange}. The potential difference between the two electrodes is 1~MV. }
	\label{fig:Example_field_distribution}
\end{figure}

Figure~\ref{fig:Example_field_distribution} shows the distribution of the surface field for one of electrodes.  Three distinct peaks are observed. The small peak at $\sim$130~kV/cm corresponds to the high field area on the surface of the torus facing the other electrode. The peak at $\sim$80~kV/cm corresponds to the flat area in the middle facing the other electrode. The peak at $\sim$15~kV/cm is due to the back of the electrode, facing the grounded surface (not shown in Fig.~\ref{fig:Example_field_imange}). For such a geometry it is important to take into account the contribution from the entire electrode surface to breakdown. 

\begin{figure}
	\centering
	\begin{adjustbox}{center}
		\includegraphics[width=\columnwidth]{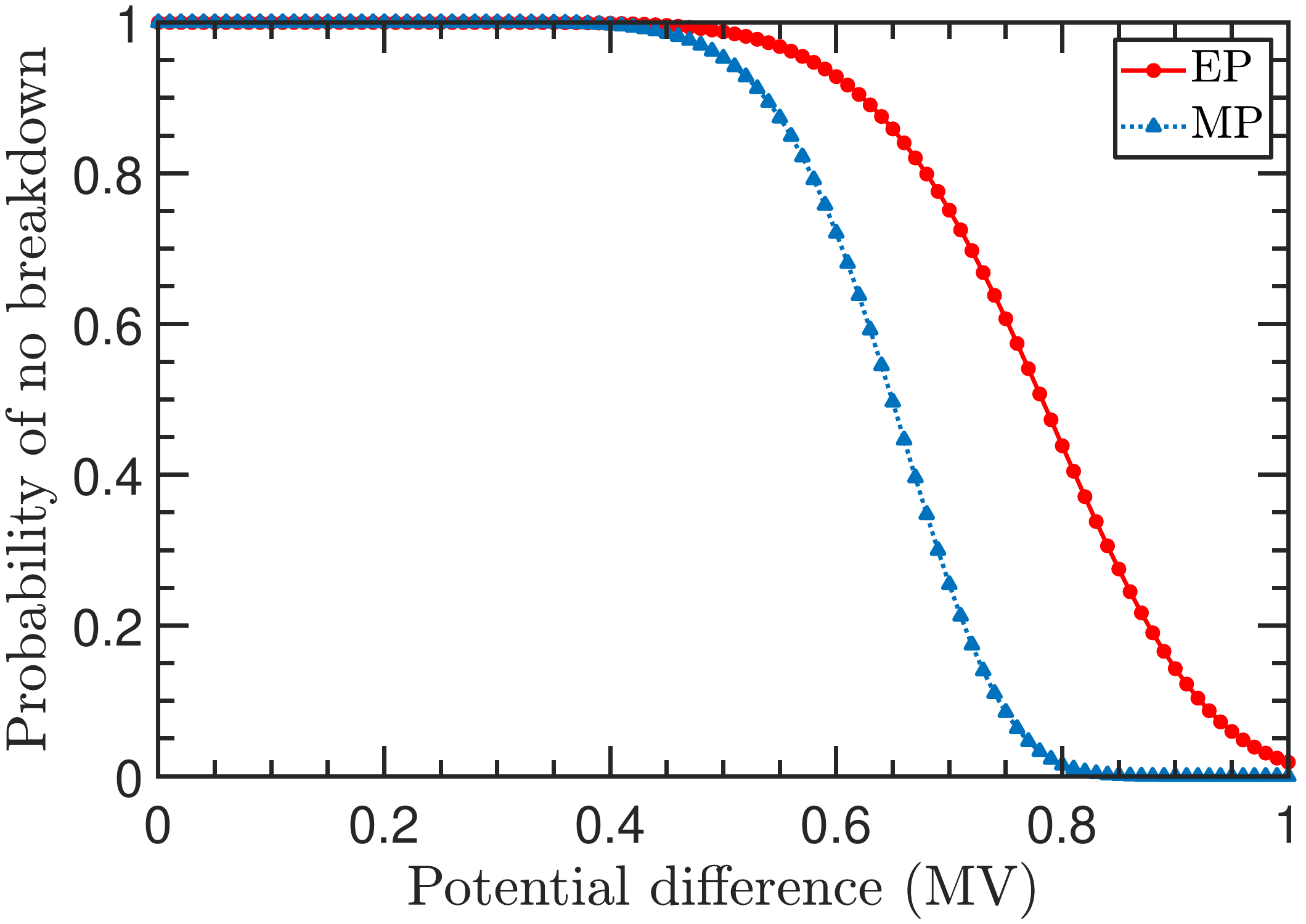}
	\end{adjustbox}
	\caption{The calculated probability of no breakdown for the electrode geometry shown in Fig.~\ref{fig:Example_field_imange} as a function of the potential difference for the case of electropolished (EP) and mechanically-polished (MP) electrodes.}
	\label{fig:Example_surv_prob}
\end{figure}

Figure~\ref{fig:Example_surv_prob} shows the calculated probability of no breakdown (i.e., the survival probability) given by Eq.~\ref{eq:scaledcdf_gen} for this geometry. The value of $W(E_i)$ has been determined by a fit to our SSHV dataset MP:53 for mechanically-polished (MP) electrodes and dataset EP:2-5 for electropolished (EP) electrodes.

For experiments that require large electrodes with a complex shape, an optimization of the electrode shape can be performed to minimize Eq.~\ref{eq:scaledcdf_gen} while meeting other experimental requirements. This inclusive optimization method is different that what is typically done, which is to reduce the so-called local ``hot-spots," or small regions with high field intensities.  The latter approach overlooks the scale dependence of electrical breakdown and the larger-area but lower field regions' contributions to the overall breakdown probability.


\section{Conclusion and prospects}

The electrical breakdown phenomenon in liquid helium has been investigated using several electrodes of different surface quality. The probability distribution of breakdown with respect to field has been analyzed using extreme value theory. This analysis confirms that breakdown is a surface phenomenon directly correlated with Fowler-Nordheim field emission. The statistical theory immediately provides a means of predicting the breakdown behavior of electrodes of different size and shape. The results of several earlier experiments on breakdown in liquid helium are found to be consistent with the current analysis. The implications of the present results to measurements with other noble liquids and to arbitrary shaped electrodes are discussed.

The connection between breakdown and field emission raises the possibility of characterizing the breakdown properties of an electrode in a non-destructive manner such as by relating field emission in vacuum to the breakdown properties in noble liquids. A better understanding of the time-dependent phenomenon, specifically the time scales involved in the breakdown process and the effects of a holding field, is required. We leave exploration of these topics to future studies.

\section{Acknowledgments}
	
This work was supported by the United States Department of Energy,
Office of Science, Office of Nuclear Physics under Contract Numbers 
DE-AC52-06NA25396 and 89233218CNA000001 under proposal
LANLEEDM (LANL), DE-FG02-ER41042 (NCSU), and DE-AC05-00OR22725 (ORNL) through Los Alamos National Laboratory Integrated Contract Order No. 4000129433 with Oak Ridge National Laboratory.  In addition, this work was supported by the National Science Foundation under Grant Nos. 1506451 (Valparaiso) and PHY-1307426 (NCSU). 
We gratefully acknowledge the support provided by the
Physics and AOT Divisions of Los Alamos National Laboratory. We are also
grateful to Vince Cianciolo for his encouragement and support.



\begin{thebibliography}{99}

\bibitem{GER98}
J.~Gerhold, Cryogenics {\bf 38}, 1063 (1998).

\bibitem{BEL15}
A.~A.~Belevtsev, High Temperature {\bf 53}, 770 (2015).

\bibitem{REB14}
B.~Rebel {\it et al.}, JINST {\bf 9}, T08004 (2014).  

\bibitem{GOL94}
R.~Golub and S.~K.~Lamoreaux, Phys. Rep. {\bf 237}, 1 (1994).

\bibitem{GRI09}
M.~G.~D.~van~der~Grinten {\it et al.}, Nucl. Instrum. Methods
Phys. Res. Sect. A {\bf 611}, 129 (2009).  

\bibitem{ITO12}
T.~M.~Ito {\it et al.}, Phys. Rev. A {\bf 85}, 042718 (2012).

\bibitem{ITO16}
T.~M.~Ito {\it et al.}, Rev. Sci. Instrum. {\bf 87}, 045113 (2016).

\bibitem{AHM19}
M.~W.~Ahmed {\it et al.}, JINST {\bf 14,} P11017 (2019).

\bibitem{GUO13}
W.~Guo and D.~N.~McKinsey, Phys. Rev. D {\bf 87}, 115001 (2013).

\bibitem{ITO13}
T.~M.~Ito and G.~M.~Seidel, Phys. Rev. C {\bf 88}, 025805 (2013).

\bibitem{KNA17}
S.~Knapen, T.~Lin, and K.~Zurek, Phys. Rev. D {\bf 95}, 056019 (2017).

\bibitem{HER18}
S.~A.~Hertel, {\it et al.}, arXiv:1810.06283v1.  

\bibitem{CLA18}
S.~M.~Clayton, {\it et al.}, JINST {\bf 13}, P05017 (2018).  

\bibitem{BLA59}
B.~S.~Blaisse, A.~van~den~Boogaart, F.~Ern\'{e},
Bull. Inst. Intern. Froid Anexie, {\bf 1,} 333 (1959).
  
\bibitem{BLA60}
C.~Blank and M.~H.~Edwards, Phys. Rev. {\bf 119,} 50 (1960).

\bibitem{BEL93}
A.~A.~Belevtsev, Nucl. Instrum. Methods Phys. Res., Sect. A {\bf 327,} 18
(1993).

\bibitem{KAT89}
R.~Kattan, A.~Denat, and O.~Lesaint, J. Appl. Phys. {\bf 66,} 4062 (1989).

\bibitem{KEL81}
E.~F.~Kelley and R.~E.~Hebner, Jr., J. Appl. Phys. {\bf 52,} 191 (1981).

\bibitem{FOR90}
E.~O.~Forster, J. Phys. D: Appl. Phys. {\bf 23,} 1506 (1990).

\bibitem{SNSnEDM}
The SNS nEDM experiment (V. Cianciolo and B.~W.~Filippone, spokespersons) [{\tt https://www.phy.ornl.gov/nedm/}].

\bibitem{ITO07}
T.~M.~Ito, J. Phys. Conf. Ser. {\bf 69}, 012037 (2007).

\bibitem{LON06}
J.~C.~Long {\it et al.}, arXiv:physics/0603231  

\bibitem{JANIS}
Janis Research, {\tt https://www.janis.com}

\bibitem{COMSOL}
COMSOL Multiphysics, http://www.comsol.com.

\bibitem{KUP02}
A.~L.~Kupershtokh, E.~I.~Palchikov, D.~I.~Karpov, I.~Vitellas, D.~P.~Agoris, and V.~P.~Charalambakos, J. Phys. D:Appl. Phys. {\bf 35,} 3106 (2002).


\bibitem{DWtest}
J.~Durbin and G.~S.~Watson, Biometrika {\bf 37}, 409 (1950).

\bibitem{LBtest}
G.~Ljung and G.~E.~P.~Box, Biometrika {\bf 66}, 66 (1978).


\bibitem{MAY81}
D.~May and H.~Krauth, ``Influence of the electrode surface conditions on the breakdown of liquid helium”, IEEE Trans. MAG. Vol. 17, pp. 2089-2092, 1981; D.~May, PhD. thesis, TU Karlsruhe, 1981.

\bibitem{CRCHandbook14}
W.~M.~Haynes, CRC Handbook of Chemistry and Physics, 95th ed. (CRC Press, Boca Raton, Fla., 2014).

\bibitem{VIJH76}
A.~K.~Vijh and K.~Dimoff, J. Mater. Sci. {\bf 11}, 150 (1976).

\bibitem{Qui2015}
H.~Qiu, R.~P.~Joshi, A.~Neuber, and J.~Dickens, Semicond. Sci. Technol. {\bf 30}, 105038 (2015).

\bibitem{Tsong1969}
T.~T.~Tsong and E.~W.~M\"{u}ller, Phys. Rev. {\bf 181}, 530 (1969).



\bibitem{MATHES67}
K.~N.~Mathes, IEEE Trans.{\bf EI-2}, 24 (1967). 

\bibitem{GALAND68}
J.~Galand, Compt. Rend. {\bf 266}, 1302 (1968). 

\bibitem{LEHMANN70}
J.~P.~Lehmann, ``Measures dielectriques dans les ftuides cryogeniquesjusqu'a 200 kV-SO Hz," Revue Gen. Elec., 79: IS (1970). 

\bibitem{MEATS72}
R.~J.~Meats, ``Pressurized-helium Breakdown at very Low Temperatures", Proc. of IEE, vol. 119, no. 6, pp. 760-766, 1972.

\bibitem{MEYERHOFF95}
R.~W.~Meyerhoff, Development of a Rigid AC Superconducting Power Transmission Line, In: Timmerhaus K.D. (eds) Advances in Cryogenic Engineering. Advances in Cryogenic Engineering, vol 19. Springer, Boston, MA (1995). 

\bibitem{FALLOU69}
B.~Fallou {\it et al.}, in: Low Temperature and Electric Power, Intern. Inst. Refrigeration, London (March 1969), p. 201. 

\bibitem{SCHWENTERLY74}
S.~W.~Schwenterly {\it et al.}, ``Dielectric strength of liquid helium in millimeter gaps," Conference on Electrical Insulation \& Dielectric Phenomena - Annual Report 1974, Downingtown, PA, USA, 1974, pp. 585-593.

\bibitem{Sharbaugh1955}
A.~H.~Sharbaugh, E.~B.~Cox, R.~W.~Crowe, and P.~L.~Auer, ``The effect of electrode configuration on the electric strength of hexane," 1955 Conference on Electrical Insulation, pp. 16-20, doi: 10.1109/EIC.1955.7533320.

\bibitem{Gallagher75}
T.~J.~Gallagher, {\it Simple dielectric liquids}, (Clarendon Press, Oxford, 1975).

\bibitem{GOSHIMA95}
H.~Goshima {\it et al.}, ``Weibull statistical analysis of area and volume effects on the breakdown strength in liquid nitrogen", IEEE Trans. on Diel. and Electr. Insul., vol. 2, no. 3, pp. 385-393 (1995).

\bibitem{Weber1956}
K.~H.~Weber and H.~S.~Endicott, Trans. Amer. Inst. Elect. Eng. {\bf 75}, 371 (1956).

 \bibitem{WEIBULL51}
W.~Weibull, J. Appl. Mech.{\bf 18}, 293 (1951).

\bibitem{CHOULKOV05}
V.~V.~Choulkov, IEEE T. Dielect. El. In Vol. 12 p.98 (2005). 

\bibitem{GUMBEL58}	
E.~I.~Gumbel, {\it Statistics of Extreme}, Columbia University Press, New York, USA, 1958.


\bibitem{KaplanMeier58}
E.~L.~Kaplan and P.~Meier, ``Nonparametric estimation from incomplete observations", Journal of the American Statistical Association, 53:457-481 (1958).

\bibitem{Nelson69}
W.~Nelson, ``Hazard Plotting for Incomplete Failure Data", Journal of Quality Technology, 1:1, 27-52 (1969).

\bibitem{Nelson72}
W.~Nelson, ``Theory and Applications of Hazard Plotting for Censored Failure Data", Technometrics, 14:4, 945-966 (1972).

\bibitem{Aalen78}
O.~Aalen, ``Nonparametric Inference for a Family of Counting Processes", Ann. Statist. 6, no. 4, 701-726 (1978).



\bibitem{FN1928}
R.~H.~Fowler and L.~Nordheim, Proc. Roy. Soc. A 119, 173-81, 1928.

\bibitem{Good1956}
R.~H.~Good and E.~W.~M\"{u}ller, Field Emission, in ``Handbuch der Physik," Springer Verlag, Berlin, 21, 176-231, 1956.

\bibitem{Wang1997}
J.~W.~Wang and G.~A.~Loew, ``Field emission and RF breakdown in high gradient room temperature linac structures," SLAC-PUB-7684.

\bibitem{Fursey2005}
G.~Fursey, \emph{Field Emission in Vacuum Microelectronics}, (Plenum Publishers, New York, 2005).

\bibitem{Broomall1976}
J.~R.~Broomall, W.~D.~Johnson, and D.~G.~Onn, Phys. Rev. B {\bf 14}, 2819 (1976).

\bibitem{BastaniNejad2015}
M.~BastaniNejad, M.~A.~Mohamed, A.~A.~Elmustafa, P.~Adderley, J.~Clark, S.~Covert, J.~Hansknecht, C.~Hernandez-Garcia, M.~Poelker, R.~Mammei, K.~Surles-Law and P.~Williams, Phys. Rev. ST Accel. Beams \textbf{15}, 083502 (2012).

\bibitem{HalpernGomer1969A}
B.~Halpern and R.~Gomer, J. Chem. Phys. {\bf 51}, 1031 (1969).


\bibitem{Ito2012}
T.~M.~Ito, S.~M.~Clayton, J.~Ramsey, M.~Karcz, C.~Y.~Liu, J.~C.~Long, T.~G.~Reddy, and G.~M.~Seidel, Phys. Rev. A \textbf{85}, 042718 (2012).

\bibitem{Phan2020}
N.~S.~Phan, V.~Cianciolo, S.~M.~Clayton, S.~A.~Currie, R.~Dipert, T.~M.~Ito, S.~W.~T.~MacDonald, C.~M.~O'Shaughnessy, J.~C.~Ramsey and G.~M.~Seidel, \textit{et al.}, Phys. Rev. C \textbf{102} 3, 035503 (2020).


\bibitem{Acciarri2014}
R.~Acciarri {\it et al.}, JINST {\bf 9} P11001 (2014).

\bibitem{Auger2016}
M.~Auger {\it et al.}, JINST {\bf 11} P03017 (2016).

\bibitem{Tvrznikova2019}
L.~Tvrznikova {\it et al.}, JINST {\bf 14} P12018 (2019).

\bibitem{TvrznikovaThesis}
L.~Tvrznikova, \emph{Sub-GeV dark matter searches and electric field studies for the LUX and LZ experiments}, Ph.D. thesis, Yale University (2019).

\bibitem{Blatter2014}
A.~Blatter {\it et al.}, JINST {\bf 9} P04006 (2014).

\bibitem{Rebel2014}
B.~Rebel \emph{et al.}, JINST {\bf 9} T08004 (2014).


\bibitem{GB1966}
J.~M.~Goldschvartz and B.~S.~Blaisse, Br. J. Appl. Phys. {\bf 17} 1083 (1966).

\bibitem{Gerhold1994}
J.~Gerhold, M.~Hubmann, and E.~Telser, Cryogenics {\bf 34} 579 (1994).

\bibitem{Gerhold1972}
J.~Gerhold, Cryogenics {\bf 12} 370 (1972).

\bibitem{Davidson2011}
A.~J.~Davidson, \emph{HighVoltage Breakdown in the Ramsey Cell of the CryoEDM
experiment - An experimental study of some relevant parameters}, Ph.D. Thesis, University of Sussex (2011).

\bibitem{Gerhold1989}
J.~Gerhold, IEEE Transactions on Electrical Insulation 24 2 155-166 (1989).

\bibitem{Hara1993}
M.~Hara \emph{et al.}, Proceedings 8th International symposium on High Voltage Engineering, 2 559-562 (1993).

\bibitem{Chigusa1999}
S.~Chigusa \emph{et al.},IEEE Transactions on Dielectrics and Electrical Insulation, 6 3 385-392 (1999).

\bibitem{Fallou1970}
B.~Fallou \emph{et al.}, Cryogenics, 10 2 142-146 (1970).

\bibitem{Burnier1970}
P.~H.~Burnier \emph{et al.}, Advances in Cryogenic engineering, 15 76-84 (1970).

\bibitem{Yoshino1982}
K.~Yoshino, K.~Ohseko, M.~Shiraishi, M.~Terauchi, and Y.~Inuishi, J. Electrostatics {\bf 12}, (1982).



\bibitem{Kattan91}
R.~Kattan, A.~Denat, N.~Bonifaci, ``Formation of vapor bubbles in non-polar liquids initiated  by  current pulses",  IEEE Transactions on  Electrical Insulation, Vol. 26, No. 4, 1991, pp. 656-662.


\bibitem{Young1959}
R.~D.~Young, Phys. Rev. {\bf 113}, 110 (1959).

\bibitem{YoungMuller1959}
R.~D.~Young and E.~W.~M\"{u}ller, Phys. Rev. {\bf 113}, 115 (1959).

\bibitem{Donnelly1998}
R.~J.~Donnelly and C.~F.~Barenghi, J. Phys. Chem. Ref. Data {\bf 27}, 1217 (1998).


\end{thebibliography}
\end{document}